\documentclass[12pt]{article}

\usepackage{
    amssymb,
    amsmath,
    amsfonts,
    calc,
    eurosym,
    geometry,
    ulem,
    graphicx,
    caption,
    color,
    setspace,
    sectsty,
    comment,
    footmisc,
    caption,
    pdflscape,
    subcaption,
    subfiles,
    titling,
    array,
    hyperref,
    booktabs,
    longtable,
    float,
    makecell,
    bbm,
    threeparttable}

\usepackage{pgfplots} 

\usepackage[
    backend=biber,
    style=nature,
    date=year,
    doi=true,
    isbn=false,
    url=false,
    eprint=false
]{biblatex}

\AtEveryBibitem{%
  \clearfield{note}%
}
\AtEveryCitekey{\clearlist{publisher}}
\AtEveryBibitem{\clearlist{publisher}}

\usepackage{pgf,tikz}
\usetikzlibrary{arrows, automata}
\usetikzlibrary{shapes.geometric,positioning}
\usetikzlibrary{positioning,calc}

\usepackage{siunitx}
\newcolumntype{d}{S[input-symbols = ()]}

\newcolumntype{L}[1]{>{\raggedright\arraybackslash}m{#1}}
\newcolumntype{C}[1]{>{\centering\arraybackslash}m{#1}}
\newcolumntype{R}[1]{>{\raggedleft\arraybackslash}m{#1}}
\subsubsectionfont{\normalfont\itshape}

\usepackage{mathtools}

\usepackage{letltxmacro}
\LetLtxMacro\orgvdots\vdots
\LetLtxMacro\orgddots\ddots

\makeatletter
\DeclareRobustCommand\vdots{%
  \mathpalette\@vdots{}%
}
\newcommand*{\@vdots}[2]{%
  \sbox0{$#1\cdotp\cdotp\cdotp\m@th$}%
  \sbox2{$#1.\m@th$}%
  \vbox{%
    \dimen@=\wd0 %
    \advance\dimen@ -3\ht2 %
    \kern.5\dimen@
    \dimen@=\wd2 %
    \advance\dimen@ -\ht2 %
    \dimen2=\wd0 %
    \advance\dimen2 -\dimen@
    \vbox to \dimen2{%
      \offinterlineskip
      \copy2 \vfill\copy2 \vfill\copy2 %
    }%
  }%
}
\DeclareRobustCommand\ddots{%
  \mathinner{%
    \mathpalette\@ddots{}%
    \mkern\thinmuskip
  }%
}
\newcommand*{\@ddots}[2]{%
  \sbox0{$#1\cdotp\cdotp\cdotp\m@th$}%
  \sbox2{$#1.\m@th$}%
  \vbox{%
    \dimen@=\wd0 %
    \advance\dimen@ -3\ht2 %
    \kern.5\dimen@
    \dimen@=\wd2 %
    \advance\dimen@ -\ht2 %
    \dimen2=\wd0 %
    \advance\dimen2 -\dimen@
    \vbox to \dimen2{%
      \offinterlineskip
      \hbox{$#1\mathpunct{.}\m@th$}%
      \vfill
      \hbox{$#1\mathpunct{\kern\wd2}\mathpunct{.}\m@th$}%
      \vfill
      \hbox{$#1\mathpunct{\kern\wd2}\mathpunct{\kern\wd2}\mathpunct{.}\m@th$}%
    }%
  }%
}
\makeatother

\def\spacingset#1{\renewcommand{\baselinestretch}%
{#1}\small\normalsize} \spacingset{1.2}

\usepackage[noblocks]{authblk}

\begin{document}
\begin{titlepage}
\title{Inferring infectiousness: a joint model of the within-host viral kinetics of SARS-CoV-2}
\author[1,2]{Christopher B. Boyer\thanks{Corresponding author: \href{mailto:boyerc5@ccf.org}{boyerc5@ccf.org}}}
\author[3,4]{Stephen M. Kissler}
\author[5]{Seran Hakki}
\author[5]{Jakob Jonnerby}
\author[5]{Ajit Lalvani}
\author[6,7,8]{Marc Lipsitch}
\affil[1]{Department of Quantitative Health Sciences, Cleveland Clinic, Cleveland, OH, USA}
\affil[2]{Department of Medicine, Cleveland Clinic Lerner College of Medicine of Case Western Reserve University, Cleveland, OH, USA}
\affil[3]{Department of Computer Science, University of Colorado Boulder, Boulder, CO, USA}
\affil[4]{Department of Epidemiology, Colorado School of Public Health, Aurora, CO, USA}
\affil[5]{NIHR Health Protection Research Unit in Respiratory Infections, National Heart and Lung Institute, Imperial College London, London, UK}
\affil[6]{Division of Infectious Diseases and Geographic Medicine, Department of Medicine, Stanford University, Stanford, CA, USA}
\affil[7]{Department of Biology, Stanford University, Stanford, CA, USA}
\affil[8]{Center for International Security and Cooperation, Freeman-Spogli Institute, Stanford University, Stanford, CA, USA}

\date{\today}
\maketitle
\clearpage
\newpage

\begin{abstract}
During an infectious disease outbreak, providing accurate answers to policy questions about transmission requires a detailed model of the natural history of infectiousness. Unfortunately, direct measures of infectiousness are generally unavailable. Instead, we often rely on indirect proxies, such as viral load measured by PCR or antigen tests, viral culture to detect replication-competent virus, or symptom onset, each of which reflects different aspects of viral dynamics or host response. However, these proxies vary in terms of the ease of collection, scalability, and their relationship to viral shedding and therefore underlying infectiousness. Here, we use data from five prospective, densely sampled cohorts with longitudinal data on multiple proxies of viral shedding for approximately 2,000 infections to develop a Bayesian joint model for the within-host viral kinetics of SARS-CoV-2 infection. Modeling the joint distribution allows us to infer the trajectory of infectious virus shedding — the most direct correlate of infectiousness — for individuals who contribute only PCR data, and to compute derived quantities that are inaccessible from any single proxy alone. These include the population-level probability and expected duration of ongoing infectiousness as a function of time since diagnosis, stratified by variant, vaccination status, and infection history; the residual risk of releasing an individual from isolation; and personalized, real-time estimates of infectiousness that are sequentially updated as new test results become available.
\noindent \\
\vspace{0in} 
\bigskip
\end{abstract}
\setcounter{page}{0}
\thispagestyle{empty}
\end{titlepage}
\pagebreak \newpage
\spacingset{1.45} 
\section{Introduction} \label{sec:introduction}
Accurate information about the timing and intensity of infectiousness within a host is crucial for designing effective public health policies and building accurate transmission models. However, direct measurement and quantification of infectiousness \textemdash defined as the instantaneous probability of onward transmission to a susceptible host given contact \textemdash is challenging. Transmission events are rare and highly contingent on behavioral and environmental factors that can be hard to fully enumerate or control. Nonetheless, a necessary component of transmission is the production of replication competent virus by an infected individual~\cite{watson2015characterisation,marks2021transmission}. As a result, researchers often rely on proxies of viral shedding, such as self-reported symptoms, antigen and polymerase chain reaction (PCR) test results, or viral culture data, to estimate infectiousness.

These proxies vary widely in their ease of collection, scalability, and correlation with actual viral shedding~\cite{puhach2022sars,cevik2021sars}. Symptom-based assessments are subjective and can be influenced by individual variability and reporting biases. Antigen tests offer rapid results but may lack sensitivity, especially in asymptomatic individuals or at different stages of infection~\cite{mina2021covid}. PCR tests are highly sensitive but can detect non-viable viral RNA for weeks after acute infection, potentially overestimating the period of infectiousness~\cite{binnicker2020challenges}. Viral culture is considered the gold standard for assessing infectivity but is labor-intensive and time-consuming, especially for viruses with long replication cycles such as respiratory syncytial virus (RSV) where plaque formation can take several days to a week. It therefore may not be feasible for large-scale or rapid assessments.

Existing approaches to modeling within-host viral kinetics treat each proxy independently, e.g., fitting PCR kinetics to PCR data, culture kinetics to culture data, and so on. This precludes translating a routine diagnostic result (e.g., a Ct value) into a probabilistic statement about the unobserved quantity of interest: how long the individual is likely to remain infectious. The reliance on single proxies or sparsely collected data can lead to incomplete or inaccurate representations of an individual's infectiousness over time~\cite{he2020temporal,jones2021estimating}. This limitation hampers our ability to answer critical policy questions, such as determining optimal isolation periods or prioritizing testing strategies~\cite{quilty2021quarantine,larremore2021test}. It also is inconsistent with how proxy data are collected during an acute infection in the real world, where often only one high quality measurement is available and supplemented by symptoms or rapid tests. There is therefore a clear need for a comprehensive approach that integrates multiple proxies to more accurately model within-host viral kinetics.

In this study, we address this gap by developing a Bayesian joint model for the within-host viral kinetics of acute infection that links multiple proxies of viral shedding through a shared latent trajectory of infectiousness. The model and the inferential framework are general --- applicable to any acute viral infection for which longitudinal proxy data are available. However, we demonstrate the approach using SARS-CoV-2 infection as an empirical example, drawing on data from five prospective, densely sampled cohorts comprising approximately 2,000 infections with longitudinal measurements of qPCR, viral culture, lateral flow antigen tests, and symptom diaries.

By modeling the joint distribution, we obtain several capabilities that are unavailable via single-proxy analyses. First, the model imputes the latent infectious virus trajectory for all individuals, including the approximately 1,750 who contributed only PCR data and for whom culture results are unavailable. Second, because the imputed trajectories propagate the full posterior uncertainty in the relationship between viral RNA and infectious virus, we can compute the probability and expected duration of ongoing infectiousness at any point during infection --- a quantity that previous analyses could only estimate for the small subset of individuals with serial culture data. Third, by stratifying over covariate profiles, we quantify how vaccination, variant, and prior infection modulate the duration of infectiousness, revealing, for example, that boosted individuals infected with Omicron have a longer tail of infectious virus shedding than immunologically na\"ive individuals with pre-Alpha despite lower peak viral loads. Finally, by conditioning on an individual's accumulating test history via Bayesian filtering, the model can provide personalized, real-time estimates of residual infectiousness that can directly inform clinical isolation and discharge decisions.

In the sections that follow, we first review the proxies for infectiousness that the model integrates (Section~\ref{sec:proxies}), then specify the model (Section~\ref{sec:model}), describe the Bayesian estimation strategy and prior specification (Section~\ref{sec:priors}), and apply the model to the five SARS-CoV-2 cohorts (Section~\ref{sec:results}). We conclude with a discussion of the model's implications for outbreak response and its applicability to other pathogens (Section~\ref{sec:discussion}).

\section{Proxies for infectiousness} \label{sec:proxies}
Our goal is to characterize individual infectiousness over time (purple curve in panel A of Figure~\ref{fig:overview}). We take the shedding of viable virus as the most direct, measurable, indicator of infectiousness (red curve in panel A of Figure~\ref{fig:overview}). However, we recognize that measuring viable virus shedding (e.g., via culture) is costly and slow, so we seek a statistical framework that allows us to translate more common proxies such as the outcomes of common diagnostic tests like RT-qPCR (blue curve) or LFDs (green) or reported symptoms (yellow) into an estimate of viral shedding. We review the most common proxies for viral shedding below.

\subsection{Infectious virus}
The gold standard diagnostic approach to determining whether replication competent virus is present is recovery via cell culture. This is achieved by inoculating susceptible cell lines with clinical specimen and measuring cytopathic effects via light microscopy, with infection confirmed either via RT-PCR of the supernatant of infected cells or by immunostaining for viral proteins. In the case of SARS-CoV-2, several cell lines can be used for isolation including a line from African green monkey kidney cells (Vero E6) and human lines from colorectal adenocarcinoma cells (Caco-2), lung adenocarcinoma cells (A549), and hepatocellular carcinoma cells (Huh7). These lines typically express either angiotensin converting enzyme 2 (ACE2; the receptor required for virus entry) or transmembrane protease 2 (TMPRSS2; another receptor important for virus entry).

While simple culture provides a qualitative assessment of the presence of infectious virus in an inoculated specimen, other methods are available that directly quantify the number of infectious virions~\cite{wolk2020sars,bullard2020predicting}. To produce viral titres, the specimen is serially diluted and used for inoculation. In a focus-forming assay, a monolayer of cells is inoculated with the clinical specimen and the cells are fixed one day after infection and immunostained with virus-specific antibodies creating visible groups of infected cells (foci) that can be counted. In a plaque-forming assay, a monolayer of cells is inoculated and plates are fixed 2--3 days post-infection and then stained with crystal violet, forming plaques in the presence of infectious virus that can be counted. Finally, in a 50\% tissue culture infectious dose (TCID$_{50}$) assay, serial dilutions of the specimen are inoculated onto cell monolayers and observed for cytopathic effect over several days; the titer is defined as the dilution at which 50\% of inoculated wells show cytopathic effect.

The advantages of viral culture assays are that they directly detect replication competent virus and therefore come the closest to measuring infectious viral shedding. The disadvantages are that cultures are time-consuming and require specially trained laboratory personnel to properly handle and store. Results can vary substantially across cell lines or laboratories. In the case of an emerging pathogen like SARS-CoV-2, cultures had to be performed under biosafety level 3 laboratory conditions. For these reasons, cultures are often expensive and not routinely performed in the clinical setting. Therefore high quality cultures collected longitudinally within the same infection are rare.

\subsection{Viral nucleic acids}
In the clinical setting, laboratory diagnosis of viral infection generally relies instead on demonstration of viral RNA via a virus-specific RT-PCR performed on the clinical specimen. The amount of viral RNA in the sample can be quantified by calibrating the cycle threshold (Ct) value, representing the number of amplification cycles necessary to detect a signal, to reference samples containing a prescribed number of viral copies per milliliter or swab. Following previous literature, we refer to this measure of viral RNA concentration as viral load~\cite{puhach2022infectious}. Viral load is correlated, but not synonymous, with infectious virus.

RT-PCR is highly sensitive and specific, and thus the gold standard for clinical diagnosis, but it does not distinguish between replication competent and residual RNA~\cite{binnicker2020challenges}. For determining individual infectiousness, this creates challenges particularly in the late stages of infection when destroyed or inactivated virus creates a surfeit of residual RNA leading to positive results even when virions are not themselves infectious. Unlike viral culture, RT-PCR assays can be standardized and scaled for mass testing in laboratory settings. They are therefore more commonly used in research, including in longitudinal studies. However, outside certain mass testing, quarantining, or occupational health environments, repeated testing with RT-PCR in the community, as a means for individuals to monitor infectiousness, is not available.

\subsection{Viral antigens}
Diagnostics that can be used in a household or community setting are more restricted. However, since the COVID-19 pandemic, there has been a surge in availability of antigen-detecting rapid diagnostic tests. These tests detect viral proteins from lysed samples by forcing them across a lateral flow field containing conjugate antibody, hereafter referred to as a lateral flow device (LFD). For SARS-CoV-2 tests, the nucleocapsid proteins are typical target antigens of LFDs. In their basic form, LFDs provide a qualitative assessment of whether viral antigen is present but there is often no set procedure for quantifying the amount or concentration, although some studies have attempted to develop grading systems based on the intensity of the test line~\cite{toptan2021evaluation,routsias2021diagnostic} and a handful of companies offer digital lateral flow reader that provide quantitative results~\cite{joung2022rapid}. 

The advantages of LFDs are their speed and simplicity, allowing them to be used at home for more biologically-informed and individualized screening and isolation decisions. Although they can be inaccurate relative to RT-PCR when used for clinical diagnosis, LFDs generally correlate well with the presence of infectious virus~\cite{pickering2021comparative}, though this correlation is weaker early in infection when antigen levels have not yet reached the detection threshold despite active viral replication~\cite{hakki2022onset}. LFD results can also vary by manufacturer, whether they were self-swabbed or taken by a professional, and the type of sample (nasal versus oral). There is also some evidence that performance may vary across variants of SARS-CoV-2, although the evidence is mixed~\cite{eyre2023performance}.

\subsection{Symptoms}
Symptom-based assessment is perhaps the oldest and most widely accessible approach to inferring infectiousness. Common symptoms of acute respiratory viral infections include cough, sore throat, congestion, fever, fatigue, and myalgia. These symptoms are broadly understood to reflect the host innate and adaptive immune response to infection rather than direct viral cytopathology, which is why symptom onset typically lags behind the initial rise in viral shedding by one to several days~\cite{puhach2022infectious} but prior to peak~\cite{hakki2022onset}. The temporal relationship between viral kinetics and symptom onset is of direct public health relevance: if symptoms reliably coincide with or precede peak infectiousness, symptom-based isolation may be an effective strategy for reducing onward transmission~\cite{fraser2004factors}. However, if symptoms substantially lag or fail to appear, as in asymptomatic or pre-symptomatic infections, symptom-based strategies will miss a considerable amount of potential infectiousness.

The advantages of symptom-based assessments are their simplicity and universality --- they require no laboratory infrastructure and can be self-reported. For these reasons, symptom-based screening and isolation policies have been the backbone of pandemic response since well before the availability of modern diagnostics. The disadvantages are substantial. Symptom presentation is subjective and varies considerably across individuals, with some infections remaining entirely asymptomatic. For SARS-CoV-2, estimates of the asymptomatic fraction have ranged from 20\% to over 40\%~\cite{oran2020prevalence,buitrago2020occurrence}, and even among symptomatic individuals the time from infection to symptom onset is variable. With a novel pathogen, symptom presentation can also vary as population immunity accrues. For instance, during the initial waves of the COVID-19 pandemic, symptoms were often due to viral cytopathology (e.g., anosmia, Covid toes) arising later in infection and making symptom-based isolation ineffective. However, later, as prior exposure through vaccination or infection became more common, symptoms were more immune-mediated and arose earlier. Despite these limitations, the temporal relationship between symptoms and viral shedding contains information about the immune response that is not captured by virological proxies alone, motivating its inclusion as a component of a joint model.

\section{The model} \label{sec:model}
We propose a generative model of within-host viral kinetics during an acute infection that can be linked, through an observational process, to multiple measures of viral shedding. Consider longitudinal observations from $i = 1, \ldots, N$ infected individuals at times $t$ since inoculation. The viral kinetics of person $i$ at time $t$ are given by $\{(V_{it}, R_{it}, Y_{it}, \mathbf{Z}_{it}, \mathbf{X}_i, S_{i})\}$ where $V_{it}$ is infectious virus, $R_{it}$ is viral RNA copies, $Y_{it}$ is an indicator of symptom onset, $\mathbf{Z}_{it}$ is a vector of observed test or biomarker measurements, $\mathbf{X}_i$ is a vector of individual covariates (e.g. sex, age, immune history), and $\mathbf{S}_i$ is a vector of test characteristics, such as swab type and gene target. The vector $\mathbf{Z}_{it}$ contains proxy measurements for viral shedding including viral culture ($V^*_{it}$), RT-PCR ($R^*_{it}$), and lateral flow tests for viral antigens ($L^*_{it}$), such that $\mathbf{Z}_{it} = (V^*_{it}, R^*_{it}, L^*_{it})$. We use the superscript $*$ to denote observed rather than true value and, for notational convenience, hereafter we suppress the indexes $i$ for individuals. 

\begin{landscape}
\begin{figure}[p]
    \centering
    \includegraphics[width=\linewidth]{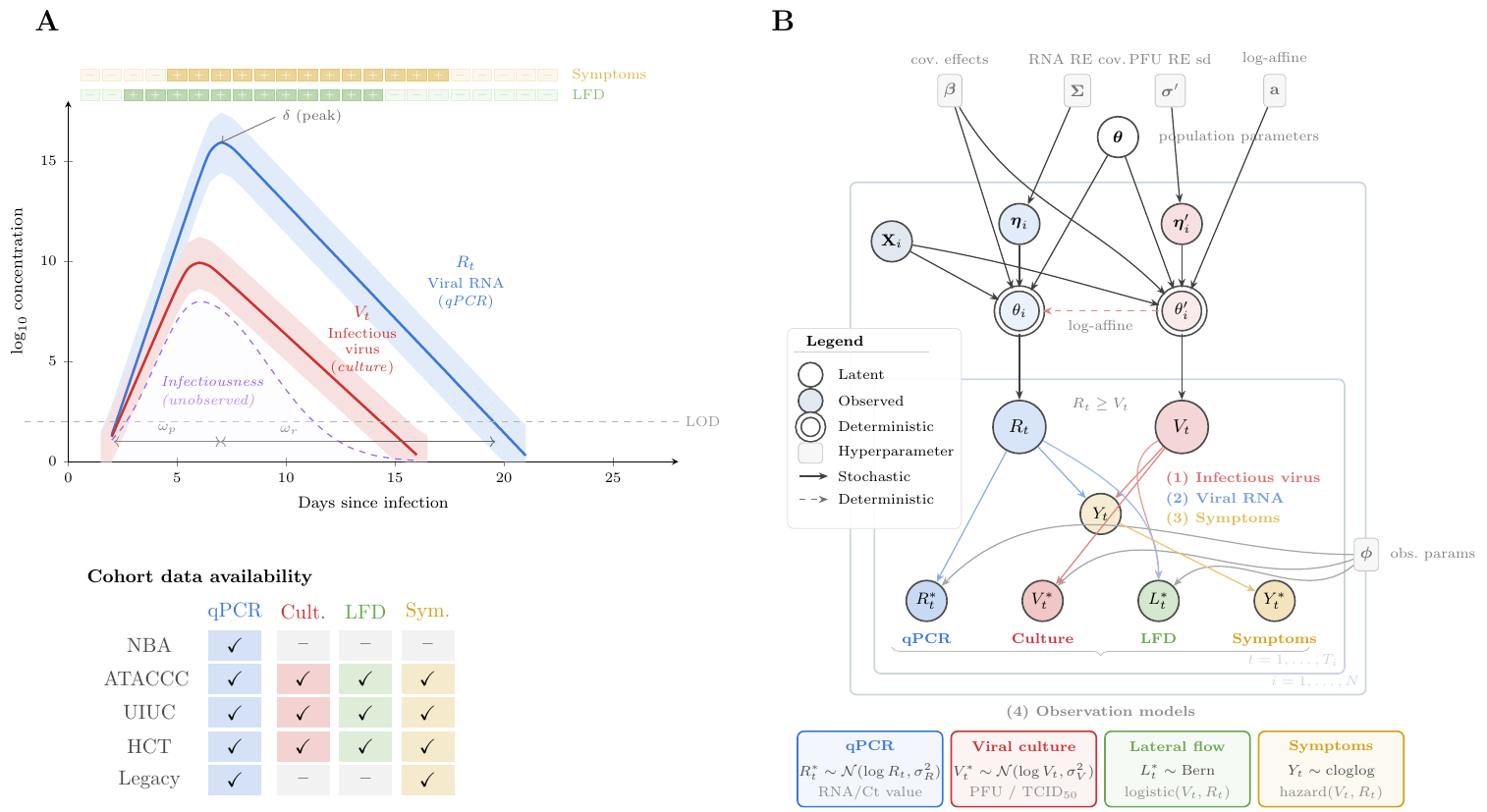}
    \caption{Overview of the joint model for within-host viral kinetics. \textbf{(A)}~Latent trajectories of viral RNA ($R_t$, blue) and infectious virus ($V_t$, red), with a conceptual infectiousness curve (dashed, purple, unobserved) and the periods of LFD positivity and symptoms during the infectious episode. The data availability table shows which assays are available in each cohort. \textbf{(B)}~Plate diagram showing probabilistic dependencies. Population parameters~($\boldsymbol{\theta}$) are modified by covariates~($\boldsymbol{\beta}$, $\mathbf{X}_i$) and random effects~($\boldsymbol{\eta}_i$, $\boldsymbol{\eta}'_i$) to produce individual trajectory parameters; RNA parameters~($\theta_i$) are mapped to PFU parameters~($\theta'_i$) via a log-affine transformation~($\mathbf{a}$). Circles denote random variables (shaded~= observed), double circles deterministic functions, and rounded rectangles hyperparameters.}
    \label{fig:overview}
\end{figure}
\end{landscape}

Our goal is to characterize the joint distribution, 
$$
f(V_t, R_t, Y_t, \mathbf{Z}_t | \mathbf{X}, t, S; \theta)
$$
defining the within-host trajectories of viral shedding and test or biomarker results over time, where $\theta$ is a vector of parameters (we use $\theta$ generically throughout to represent as yet to be defined model parameters). To specify the model, we factorize the joint distribution into a product of conditional distributions. A natural factorization is:

\begin{align*}
    f(&V_t, R_t, Y_t, \mathbf{Z}_t | \mathbf{X}, S, t; \theta) = \\ & \quad \underbrace{f(V_t | \mathbf{X}, t; \theta)}_{\shortstack{infectious \\ virus}} \times \underbrace{f(R_t | V_t, \mathbf{X}, t; \theta)}_{\shortstack{viral RNA}} \times \underbrace{f(Y_t | V_t, R_t, \mathbf{X}, t; \theta)}_{\shortstack{natural history \\ of symptoms}} \times \underbrace{f(\mathbf{Z}_t | V_t, R_t, \mathbf{X}, t, S; \theta)}_{\shortstack{observation \\ model}}
\end{align*}
where the first term describes the trajectory of infectious virus over time, the second term describes the total concentration of viral RNA over time conditional on the amount of infectious virus, the third term describes the onset of symptoms conditional on the amount of infectious virus and viral RNA, and the fourth term relates these to observed biomarker values and test results. Figure~\ref{fig:overview} provides a schematic overview of the model structure, showing the hierarchical dependencies among population parameters, individual random effects, latent viral trajectories, and observed biomarker measurements.

We posit models for each of these components in turn below.

\subsection{Infectious virus}
A commonly used mechanistic model describing the dynamics of viral proliferation during acute infection is the target cell limited model~\cite{perelson2002modelling,baccam2006kinetics}. In its simplest form, the model tracks the number of susceptible cells ($T$), infected cells ($I$), and free virions ($V$), described by the following system of ordinary differential equations:
\begin{align*}
    \dfrac{dT}{dt}&= -b V T \\
    \dfrac{dI}{dt}&= b V T - d I \\
    \dfrac{dV}{dt}&= p I - c V
\end{align*}
Target cells become infected at rate $b$ upon contact with free virus. Infected cells produce virus at rate $p$ and die at per capita rate $d$, while free virus is cleared at rate $c$. Extensions of this model include the addition of an eclipse phase, in which infected cells do not yet produce virus, and the inclusion of the immune response. An example numerical solution for the trajectory of viral particles over time is shown in Figure~\ref{fig:tcl_1} of the Supplement. We note the following salient features: an initial exponential expansion of the infected cell population and number of free virus particles (proliferation), followed by a peak and subsequent exponential decline due to immune response and depletion of susceptible target cells (clearance).

Due to the complexity of the human immune response and the challenges of measuring the initial number of target or infected cells, it is often difficult to specify the target cell model in practice. Therefore, we instead posit a semi-mechanistic model which retains many of the phenomenological features of the model, but is agnostic to the true biological structure of the response. Namely, we assume that the number of infectious viral particles follows a piecewise exponential function, defined as
\begin{equation*}
    \log V_t = g(t; \theta),
\end{equation*}
where
\begin{equation*}
    g(t; \theta) = \begin{cases}
   \dfrac{\delta}{\omega_p} (t - (t_p - \omega_p)) & \text{if } t \leq t_p \\
    \delta - \dfrac{\delta}{\omega_r} (t - t_p) & \text{if } t > t_p,
\end{cases}
\end{equation*}
with parameters representing peak viral concentration ($\delta$), proliferation duration ($\omega_p$), clearance duration ($\omega_r$), and time of peak ($t_p$), all of which can potentially vary with covariates $\mathbf{X}$. Panel A of Figure~\ref{fig:overview} illustrates the piecewise exponential trajectory and its parameterization.

\paragraph{Smooth trajectory approximation.} The piecewise exponential function has a non-differentiable kink at the peak, which can create challenges for gradient-based sampling. We therefore introduce a smooth approximation based on the log-sum-exp envelope of the two exponential arms. Defining the proliferation rate $a = \delta / \omega_p$ and the clearance rate $b = \delta / \omega_r$, the smooth trajectory is
\begin{equation*}
    g_s(t; \theta) = \delta + \log \frac{a + b}{b \exp(-a(t - t_p)) + a \exp(b(t - t_p))}
\end{equation*}
which serves as a continuously differentiable envelope of the two exponential arms and converges to the piecewise exponential in the limit of sharp rates. The smooth formulation improves the efficiency of Hamiltonian Monte Carlo sampling by providing well-defined gradients everywhere.

\paragraph{Limitations.} A notable limitation of our smoothed piecewise exponential model above is that it assumes a single growth and decay phase that may not capture more complex viral dynamics. For instance, real trajectories of many viral infections exhibit a plateau, i.e., a sustained period with high viral load, which is not well captured by a single peak model. We introduce a generalization of the model above that includes an approximate plateau around the peak via a flat-top duration parameter in Section~\ref{sec:supp_flattop} of the supplement. However, we note, in our application, the simpler form without a plateau was preferred. Another common feature of viral kinetics is biphasic decay, which can arise from the emergence of an adaptive immune response or the presence of multiple compartments with different clearance rates (see for instance Figure~\ref{fig:tcl_2} of the Supplement). Our model does not capture this feature, but it is also possible to extend the model to include a second clearance phase by adding an additional exponential component to the piecewise function. Ultimately, the choice of model complexity should be guided by what is known about the biology of the infection, with a preference for simpler models when they provide an adequate fit to the observed trajectories.

\subsection{Viral RNA copies}
Many tests, such as RT-PCR, detect or quantify viral RNA rather than infectious virus. That is, they do not distinguish between replication-competent (infectious) virus and residual (non-infectious) viral RNA which could be present due to production of nonviable viral particles during infection or may linger in a nonviable but not fully degraded state after the infection has been cleared by the immune system. In the target cell limited model, we could track the amount of viral RNA via additional compartments. For instance, we could add compartments tracking the number of  free floating RNA ($R$) via 
\begin{equation*}
    \dfrac{dR}{dt} = q p I - e R 
\end{equation*} 
where $q$ is the amount of viral RNA per infectious virion and $e$ is the rate of degradation and clearance of viral RNA. However, we do not have direct measurements of the number of viral RNA copies, and the relationship between the number of infectious virus and the number of viral RNA copies is not well understood. Therefore, we again posit a semi-mechanistic model. Because quantitative RT-PCR data are available for all individuals whereas viral culture data are limited to a subset, we directly parameterize the viral RNA trajectory and derive the infectious virus trajectory from it. The RNA trajectory follows the same functional form (smooth piecewise exponential, as described above):
\begin{equation*}
    \log R_t = g_s(t; \theta),
\end{equation*}
where $\theta = (\delta, \omega_p, \omega_r, t_p)$ denotes the RNA kinetic parameters. The infectious virus trajectory uses the same functional form with parameters $\theta' = (\delta^\prime, \omega^\prime_p, \omega^\prime_r, t^\prime_p)$ derived from the RNA parameters via log-affine transformations:
\begin{align*}
    \log \delta^\prime &= a_{0,\delta} + a_{1,\delta} \log \delta \\
    \log \omega^\prime_{p} &= a_{0,\omega_p} + a_{1,\omega_p} \log \omega_{p} \\
    \log \omega^\prime_{r} &= a_{0,\omega_r} + a_{1,\omega_r} \log \omega_{r} \\
    t^\prime_{p} &= a_{0, t_p} + a_{1, t_p} \cdot t_{p} 
\end{align*}
where $\delta^\prime$, $\omega^\prime_{p}$, $\omega^\prime_{r}$, and $t^\prime_{p}$ are the PFU parameters expressed as log-affine transformations of the corresponding RNA parameters (equivalently, $\delta^\prime = e^{a_{0,\delta}} \delta^{a_{1,\delta}}$). This form ensures that the transformed parameters are strictly positive, nests the proportional model (when $a_1 = 1$), and is interpretable: $a_1$ represents the elasticity of each PFU parameter with respect to its RNA counterpart. We note that this log-affine form can be motivated as an approximation to the convolution integral that arises from tracking RNA as a separate ODE compartment; this relationship is derived formally in Section~\ref{sec:supp_logaffine} of the Supplement.

\subsection{Viral antigens}
\label{sec:antigens}
Lateral flow devices (LFDs) detect viral nucleocapsid protein (antigen) rather than RNA or infectious virus directly. Antigen accumulates in proportion to viral replication and is cleared by the host, so its concentration $A_t$ follows the production--clearance dynamics
\begin{equation*}
    \frac{dA}{dt} = \alpha_1 V_t - \alpha_2 A_t, \qquad A_{t_0} = 0,
\end{equation*}
where $\alpha_1$ governs production from infectious virus and $\alpha_2$ governs clearance. Because the solution $A_t = \alpha_1 \int_0^t e^{-\alpha_2(t-s)} V_s\,ds$ is a convolution of $V_t$ with an exponential kernel, antigen accumulates with a lag during viral proliferation and persists after viral clearance begins. This produces an asymmetry: at the same instantaneous viral load, antigen concentration is lower during the rising phase than during the falling phase---consistent with empirical observations that LFD sensitivity is reduced early in infection despite high viral loads~\cite{hakki2022onset} (see Supplementary Figure~\ref{fig:antigen_schematic}).

In principle, one could derive an antigen trajectory via a second log-affine transformation analogous to the RNA-to-PFU mapping above (Section~\ref{sec:supp_logaffine}). However, unlike RNA (quantified continuously via Ct values) and infectious virus (quantified via PFU assays), we lack direct quantitative measurements of antigen concentration---only dichotomous LFD outcomes are available. With only binary data, the trajectory shape parameters of a piecewise exponential antigen curve would be poorly identified. Instead of modeling the full antigen trajectory, we capture its key dynamic feature---the proliferation--clearance asymmetry---by augmenting the LFD observation model with a post-peak indicator interaction, which allows the relationship between viral load and LFD sensitivity to differ between the rising and falling phases of infection (Section~\ref{sec:supp_lfd_deriv}).

\subsection{Symptom onset}
The relationship between infectious virus and the onset of symptoms or symptom profile is not well understood. Therefore, we posit a statistical model for the onset of symptoms. Biologically, symptoms are a manifestation of the infection or the immune response, for which the number of infectious virus and the number of viral RNA copies are at least a proxy (and one we have information about). We model symptom onset via a discrete-time hazard model with a complementary log-log link, which corresponds to the discretization of a continuous-time proportional hazards model. Specifically, for individual $i$ at time $t$, conditional on not yet having developed symptoms,
\begin{align*}
    P&(Y_{it} = 1 \mid Y_{i,t-1} = 0, V_{it}, R_{it}) = \\
    &\qquad 1 - \exp\!\left(-\exp\!\left(\zeta_0 + \zeta_1 \frac{\log V_{it}}{s} + \zeta_2 \frac{\log R_{it}}{s} + \zeta_3 \mathbbm{1}(t \geq t_p) + \zeta_4 \mathbbm{1}(t \geq t_p) \frac{\log R_{it}}{s} + u_i\right)\right)
\end{align*}
where $\zeta_0$ is the baseline log-hazard, $\zeta_1$ and $\zeta_2$ (both constrained positive) capture the effect of infectious virus and viral RNA on symptom onset, $s$ is a fixed normalizing constant (set to the prior mean peak viral load, which keeps the linear predictor on the unit scale), and $u_i \sim N(0, \sigma_u)$ is an individual-specific random effect capturing heterogeneity in symptom susceptibility beyond what is explained by viral dynamics. The post-peak indicator $\mathbbm{1}(t \geq t_p)$ and its interaction with log RNA allow the hazard to differ between the proliferation and clearance phases, reflecting the biological observation that immune effectors (cytokines, interferons, T-cell responses) accumulate with a lag relative to viral load, so at a given viral load the hazard of symptom onset is expected to be higher during clearance than during proliferation. On the log-hazard scale, $\exp(\zeta_1/s)$ and $\exp(\zeta_2/s)$ have the interpretation of hazard ratios for a unit increase in $\log V_t$ and $\log R_t$, respectively.

\subsection{Observation models}

\paragraph{Binary viral culture.} A number of viral culture assays seek to directly detect and quantify the number of infectious virus particles $V_t$. The simplest isolates the virus in cell culture by inoculating a sample onto a monolayer of susceptible cells and observing whether a cytopathic effect occurs. We model the probability of a positive viral culture result from this test as a function of the number of infectious virus particles via the logistic saturation model
\begin{equation*}
    V^{*}_{t,\text{culture}} \sim \text{Bernoulli}(\text{logit}^{-1}(\pi_0 + \pi_1 \log V_t)),
\end{equation*}
where $\pi_0$ represents where saturation occurs and $\pi_1$ captures the steepness of the curve. This model allows for a non-zero probability of a positive culture even at low levels of infectious virus, which could occur due to stochasticity in the culture process or contamination, and for a non-zero probability of a negative culture even at high levels of infectious virus, which could occur due to sample degradation or processing errors.
\paragraph{Quantitative viral culture (TCID50).} Alternatively, the number of infectious virus particles can be directly quantified by 50\% tissue culture infectious dose (TCID50), plaque forming units (PFU) assays, or focus-forming assays. For tests based on TCID50, we derive a mechanistic observation model from the kinetics of viral growth in culture. After inoculation with a sample containing $V_t$ infectious particles, the virus replicates exponentially in culture at rate $r$, and cytopathic effect (CPE) is detected when the viral population exceeds a threshold $V^{\dagger}$. Solving for the time to detection gives
\begin{equation*}
    d^{*} = a - b \log V_t + \varepsilon, \quad \varepsilon \sim \text{Normal}(0, \sigma_{\text{TCID50}})
\end{equation*}
where $a = (\log V^{\dagger} - \log c_0)/r$ absorbs the detection threshold and proportionality constant, $b = 1/r > 0$ is the inverse culture growth rate, and $\varepsilon$ captures stochastic variation in culture conditions. Since cultures are examined at daily intervals, the observed day of positivity $D \in \{2, 3, 4, 5, \text{neg}\}$ follows an interval-censored normal model:
\begin{equation*}
    P(D = d \mid V_t) = \begin{cases}
        \Phi\!\left(\frac{2 - \mu}{\sigma_{\text{TCID50}}}\right) & \text{if } d = 2 \\[4pt]
        \Phi\!\left(\frac{d - \mu}{\sigma_{\text{TCID50}}}\right) - \Phi\!\left(\frac{d - 1 - \mu}{\sigma_{\text{TCID50}}}\right) & \text{if } d \in \{3, 4, 5\} \\[4pt]
        1 - \Phi\!\left(\frac{5 - \mu}{\sigma_{\text{TCID50}}}\right) & \text{if } d = \text{neg}
    \end{cases}
\end{equation*}
where $\mu = a - b \log V_t$.

\paragraph{Quantitative viral culture (PFU/FFA).} The plaque forming units (PFU) assay and focus-forming assays seek to directly characterize the amount of infectious virus in a sample by counting plaques or foci. We model the concentration of plaques or foci as a function of the true number of infectious virus particles via the log-normal model
\begin{align*}
    \log V^{*}_{t,\text{FFA}} &= \log V_t + \varepsilon_{\text{FFA}} \\
    \varepsilon_{\text{FFA}} &\sim \text{Normal}(0, \sigma_{\text{FFA}})_{lod} \\
    \log V^{*}_{t,\text{PFU}} &= \log V_t + \varepsilon_{\text{PFU}} \\
    \varepsilon_{\text{PFU}} &\sim \text{Normal}(0, \sigma_{\text{PFU}})
\end{align*}
where we assume errors are homoscedastic and censored at the level of detection for each assay.

\paragraph{Qualitative RT-PCR.} Much like simple viral isolation, qualitative RT-PCR tests provide information on the presence or absence of viral RNA in a sample, but not direct quantification. We model the probability of a positive RT-PCR test as a function of the number of viral RNA copies via the logistic model
\begin{equation*}
    R^{*}_{t,\text{PCR}} \sim \text{Bernoulli}(\text{logit}^{-1}(\pi_0 + \pi_1 \log R_t)).
\end{equation*}
\paragraph{Quantitative RT-PCR (qPCR).} By contrast, quantitative RT-PCR tests provide a cycle threshold (Ct) value, which is inversely related to the concentration of viral RNA in the clinical sample. Through calibration using an external standard with a defined number of RNA copies, the Ct value can be transformed into an estimate of the number of viral RNA copies via a characteristic curve. We model the number of viral RNA copies as a function of the Ct value via the log-normal model
\begin{align*}
    \log R^{*}_{t,\text{qPCR}} &= \log R_t + \varepsilon_{\text{qPCR}} \\
    \varepsilon_{\text{qPCR}} &\sim \text{Normal}(0, \sigma_{\text{qPCR}})_{lod}
\end{align*}
where, as with the PFU and FFA assays previously, we assume errors are homoscedastic and censored at the level of detection for each assay.

\paragraph{Test error mixture.} Beyond simple measurement error, systematic test failures can occur due to sample quality, processing errors, or contamination. We model these through a mixture framework that accounts for both false positive and false negative results. With probability $\lambda_{\text{fp}}$ (false positive), a positive result is spurious and the observed value is drawn from an exponential distribution centered near the limit of detection; with probability $\lambda_{\text{fn}}$ (false negative), a truly positive sample yields a below-LOD reading. The resulting mixture for quantitative RT-PCR is
\begin{equation*}
    \log R^{*}_{t,\text{qPCR}} \sim (1 - \lambda_{\text{fp}}) (1 - \lambda_{\text{fn}}) \cdot \text{Normal}(\log R_t, \sigma_{\text{qPCR}})_{lod} + \lambda_{\text{fp}} \cdot \text{Exp}(1/\mu)
\end{equation*}
where below-LOD observations incur an additional factor of $(1 - \lambda_{\text{fn}})$ and $\mu$ is the mean of the false positive error distribution, set so that 90\% of the distribution falls within 1 Ct unit of the limit of detection. An analogous mixture is applied to quantitative viral culture results (PFU/FFA), with the same false positive and false negative probabilities.

\paragraph{Lateral flow device (LFD).} Lateral flow tests detect the presence of viral nucleocapsid proteins (antigens) in a sample; the relationship between antigen concentration and infectious virus is discussed in Section~\ref{sec:antigens}. Rather than model the latent antigen trajectory explicitly, we capture the key features---dependence on both infectious virus and viral RNA, as well as the proliferation--clearance asymmetry---through a logistic model with a post-peak indicator interaction:
\begin{equation*}
    L_t^* \sim \text{Bernoulli}\!\left(\text{logit}^{-1}\!\left(\gamma_0 + \gamma_1 \log R_t + \gamma_2 \log V_t + \gamma_3 \mathbbm{1}(t \geq t_p) + \gamma_4 \mathbbm{1}(t \geq t_p) \log R_t\right)\right)
\end{equation*}
where $t_p$ is the individual's estimated RNA peak time and $\mathbbm{1}(t \geq t_p)$ is a post-peak indicator. The interaction term $\gamma_4 \mathbbm{1}(t \geq t_p) \log R_t$ allows the slope of LFD sensitivity with respect to viral load to differ between the proliferation and clearance phases, while $\gamma_3$ captures a level shift. Together, these terms accommodate the asymmetry identified in Section~\ref{sec:antigens}: at the same instantaneous viral load, LFD sensitivity may be lower during the rising phase than during the falling phase, consistent with the antigen lag observed by~\cite{hakki2022onset}. Setting $\gamma_3 = \gamma_4 = 0$ recovers the phase-symmetric model. An alternative derivative-based parameterization using $\dot{g}_s(t)$ is discussed in Section~\ref{sec:supp_lfd_deriv}.

\paragraph{Symptom diaries.} The gold standard for symptom data is a daily diary where participants record the presence or absence and severity across a range of symptoms. We define symptom onset as the first day on which any COVID-19-related symptom is reported; however, this definition varies across studies (e.g., first any symptom versus first respiratory symptom versus patient-reported onset date), which introduces heterogeneity that our model absorbs through the individual-level random effect $u_i$ but cannot fully resolve. In some cohorts (HCT, UIUC), we have full daily diaries and construct the aggregate symptom onset indicator $Y_{it}$ from individual symptom scores. In others (ATACCC, Legacy), only the overall time to symptom onset is recorded, in which case we set $Y_{it} = 0$ for all days prior to onset and $Y_{it} = 1$ at the recorded onset day; observations after onset do not contribute to the hazard likelihood. In either case, only at-risk person-days (prior to onset or, for asymptomatic individuals, all observed days) contribute to the likelihood.

\subsection{Covariate effects} 
It is well established that viral shedding varies with individual characteristics such as age, variant, and prior vaccination or infection history~\cite{hay2022quantifying}. Therefore, we allow the parameters of the piecewise exponential model for the number of infectious virus particles and the number of viral RNA copies, as well as the parameters of the symptom onset and observation models, to vary with individual covariates $\mathbf{X}$. In particular, we allow the parameters of $g(t; \theta)$ and $g(t; \theta^\prime)$ to vary with $\mathbf{X}$ via
\begin{align*}
    \delta(\mathbf{X}) &= \delta_0 \exp(\beta_\delta \mathbf{X}) \\
    \omega_{p}(\mathbf{X}) &= \omega_{p_0} \exp(\beta_{\omega_p} \mathbf{X}) \\
    \omega_{r}(\mathbf{X}) &= \omega_{r_0} \exp(\beta_{\omega_r} \mathbf{X})
\end{align*}
and 
\begin{align*}
    \log \delta^\prime(\mathbf{X}) &= a_{0,\delta} + a_{1,\delta} \log \delta(\mathbf{X}) + \beta^\prime_\delta \mathbf{X} \\
    \log \omega^\prime_{p}(\mathbf{X}) &= a_{0,\omega_p} + a_{1,\omega_p} \log \omega_{p}(\mathbf{X}) + \beta^\prime_{\omega_p} \mathbf{X} \\
    \log \omega^\prime_{r}(\mathbf{X}) &= a_{0,\omega_r} + a_{1,\omega_r} \log \omega_{r}(\mathbf{X}) + \beta^\prime_{\omega_r} \mathbf{X}
\end{align*}
as well as symptom onset hazard via

\begin{align*}
    P&(Y_t = 1 \mid Y_{t-1}=0, V_t, R_t, \mathbf{X}) = \\
    &\quad 1 - \exp\!\left(-\exp\!\left(\zeta_0 + \zeta_1 \frac{\log V_t}{s} + \zeta_2 \frac{\log R_t}{s} + \zeta_3 \mathbbm{1}(t \geq t_p) + \zeta_4 \mathbbm{1}(t \geq t_p) \frac{\log R_t}{s} + u_i + \boldsymbol{\beta}_{Y} \mathbf{X}\right)\right)
\end{align*}
where $s$ is a fixed scale factor (set to the prior mean of the peak viral load) that keeps the linear predictor $O(1)$ and prevents numerical overflow in the double-exponential cloglog link.

In our data example below, we include in $\mathbf{X}$: age (categorized as 0 to 30 years old, 30 to 50 years old, or 50+); variant (categorized as Pre-Alpha, Alpha, Delta, Omicron, BA.4/BA.5, or Other); a binary indicator of previous infection; and vaccination history (categorized as Unvaccinated, Vaccinated boosted, Vaccinated unboosted, Vaccinated unreported, Unreported, or Boosted unreported primary). For categorical variables, we use indicator coding with the first category as the reference level, in which case the reference model is for an unvaccinated and immunologically naive 0 to 30 year old infected with Pre-Alpha wildtype or variant.

\subsection{Individual and setting-specific random effects}
There is often residual variation in viral trajectories at the individual level beyond that which can be explained by the covariates in $\mathbf{X}$. This could be due to heterogeneity in inoculating dose, differences in immune function, or other complex interactions between host and pathogen characteristics. We model residual variation in infectious virus and viral RNA at the individual level through the inclusion of individual-specific random effects for peak height, proliferation duration, clearance duration, and timing of peak, i.e. 
\begin{align*}
    \delta_{0,i} &= \delta_0 \exp(\tilde\delta_{0,i}) & \delta^\prime_{0,i} &= \delta^\prime_0 \exp(\tilde\delta^\prime_{0,i}) \\
    \omega_{p_0,i} &= \omega_{p_0} \exp(\tilde\omega_{p_0,i}) & \omega^\prime_{p_0,i} &= \omega^\prime_{p_0} \exp(\tilde\omega^\prime_{p_0,i}) \\
    \omega_{r_0,i}  &= \omega_{r_0} \exp(\tilde\omega_{r_0,i}) & \omega^\prime_{r_0,i}  &= \omega^\prime_{r_0} \exp(\tilde\omega^\prime_{r_0,i})
\end{align*}
where the random effects on the RNA trajectory may be either independent or correlated.

\paragraph{Independent random effects.} Under independent effects,
\begin{equation*}
\tilde\delta_{0,i} \sim N(0, \sigma_{\tilde\delta}), \quad
\tilde\omega_{p_0,i} \sim N(0, \sigma_{\tilde\omega_p}), \quad
\tilde\omega_{r_0,i} \sim N(0, \sigma_{\tilde\omega_r}), \quad
t_{p,i} \sim N(0, \sigma_{t_p})
\end{equation*}
and analogously for the PFU-specific random effects $(\tilde\delta^\prime_{0,i}, \tilde\omega^\prime_{p_0,i}, \tilde\omega^\prime_{r_0,i}, t^\prime_{p,i})$, all mutually independent.

\paragraph{Correlated random effects.} For the RNA trajectory, the four individual effects $(t_{p,i}, \tilde\delta_{0,i}, \tilde\omega_{p_0,i}, \tilde\omega_{r_0,i})$ may exhibit biologically plausible correlations --- for example, individuals with higher peak viral load may also have longer clearance times. To capture this, we model the RNA individual effects as draws from a multivariate normal:
\begin{equation*}
    \boldsymbol{\eta}_i = (t_{p,i}, \tilde\delta_{0,i}, \tilde\omega_{p_0,i}, \tilde\omega_{r_0,i})^\top \sim \text{MVN}(\mathbf{0}, \boldsymbol{\Sigma})
\end{equation*}
where $\boldsymbol{\Sigma} = \text{diag}(\boldsymbol{\sigma}) \cdot \boldsymbol{\Omega} \cdot \text{diag}(\boldsymbol{\sigma})$ and $\boldsymbol{\sigma} = (\sigma_{t_p}, \sigma_{\tilde\delta}, \sigma_{\tilde\omega_p}, \sigma_{\tilde\omega_r})^\top$ collects the marginal standard deviations. To facilitate efficient posterior sampling, we employ a non-centered Cholesky parameterization:
\begin{equation*}
    \boldsymbol{\eta}_i = \text{diag}(\boldsymbol{\sigma}) \cdot \mathbf{L} \cdot \mathbf{z}_i, \qquad \mathbf{z}_i \sim N(\mathbf{0}, \mathbf{I}_4)
\end{equation*}
where $\mathbf{L}$ is the lower-triangular Cholesky factor of $\boldsymbol{\Omega}$, so that $\boldsymbol{\Omega} = \mathbf{L}\mathbf{L}^\top$. We place an LKJ($\nu$) prior on the correlation matrix~\cite{lewandowski2009generating} (see Section~\ref{sec:priors} for choice of $\nu$), which provides shrinkage toward the identity (i.e., toward independence), balanced against the data evidence for non-zero correlations. The PFU individual effects remain independent to limit model complexity, since the infectious virus trajectory parameters are derived from the RNA parameters via the log-affine transformation.

When synthesizing results across settings, additional variability may be present due to differences in measurement, such as the type of test used, the swab/sample type, administrator, or the gene target, or differences in characteristics of participants, the pathogen, or other outbreak dynamics. We model this setting-specific variation through the inclusion of setting-specific random effects for all trajectory and observation model parameters.

\subsection{Missing data} 
We limit our sample to individuals with complete covariate information; however, some individuals do not have complete data for all tests or biomarkers at each time point. Let $C_t$ be a missingness indicator ($C_t = 1$ if the value is missing, $0$ if observed) and let $\mathbf{D}_t$ denote all observed outcome data at time $t$, stacking the biomarker measurements $\mathbf{Z}_{t}$ and symptom indicators $Y_{t}$ across time points $t$. We partition $\mathbf{D}_t$ into the observed components $\mathbf{D}^{\text{obs}}_t$ and the missing components $\mathbf{D}^{\text{mis}}_t$ and define $\mathbf{H}_t = (V_t, R_t, \mathbf{X}, t, S)$ for notational convenience

The full-data likelihood can be written as
\begin{equation*}
    f(\mathbf{D}^{\text{obs}}_t, \mathbf{D}^{\text{mis}}_t, C_t \mid \mathbf{H}_t; \theta, \phi) = f(\mathbf{D}^{\text{obs}}_t, \mathbf{D}^{\text{mis}}_t \mid \mathbf{H}_t; \theta) \cdot f(C_t \mid \mathbf{D}^{\text{obs}}_t, \mathbf{D}^{\text{mis}}_t, \mathbf{H}_t; \phi)
\end{equation*}
where $\theta$ governs the data-generating model and $\phi$ governs the missingness mechanism. We assume that the missingness mechanism is \emph{missing at random} (MAR) conditional on the observed data and covariates~\cite{rubin1976inference}, i.e.,
\begin{equation*}
    f(C_t \mid \mathbf{D}^{\text{obs}}_t, \mathbf{D}^{\text{mis}}_t, \mathbf{H}_t; \phi) = f(C_t \mid \mathbf{D}^{\text{obs}}_t, \mathbf{H}_t; \phi).
\end{equation*}
Under MAR with distinct parameters $(\theta, \phi)$, the observed-data likelihood for $\theta$ is obtained by integrating out the missing values:
\begin{equation*}
    f(\mathbf{D}^{\text{obs}}_t, C_t \mid \mathbf{H}_t; \theta, \phi) = f(C_t \mid \mathbf{D}^{\text{obs}}_t, \mathbf{H}_t; \phi) \int f(\mathbf{D}^{\text{obs}}_t, \mathbf{D}^{\text{mis}}_t \mid \mathbf{H}_t; \theta)\, d\mathbf{D}^{\text{mis}}_t
\end{equation*}
so that the missingness mechanism factors out and inference for $\theta$ depends only on $f(\mathbf{D}^{\text{obs}}_t \mid \mathbf{H}_t; \theta) = \int f(\mathbf{D}^{\text{obs}}_t, \mathbf{D}^{\text{mis}}_t \mid \mathbf{H}_t; \theta)\, d\mathbf{D}^{\text{mis}}_t$. The missingness mechanism is therefore \emph{ignorable} for likelihood-based inference about $\theta$~\cite{little2019statistical}.

In the Bayesian framework, we handle the missing values by treating them as additional latent variables to be sampled alongside all other parameters. That is, rather than analytically integrating out $\mathbf{D}^{\text{mis}}_t$, the MCMC sampler draws from the joint posterior
\begin{equation*}
    p(\theta, \mathbf{D}^{\text{mis}} \mid \mathbf{D}^{\text{obs}}) \propto f(\mathbf{D}^{\text{obs}}, \mathbf{D}^{\text{mis}} \mid \theta) \cdot \pi(\theta)
\end{equation*}
where each missing value $D^{\text{mis}}_{t}$ is imputed at every MCMC iteration from its full conditional distribution given the current values of all other parameters and latent quantities. This is equivalent to multiple imputation performed jointly with parameter estimation, and yields valid posterior inference under the MAR assumption~\cite{gelman2013bayesian}.

\section{Estimation}

The model described in Section~\ref{sec:model} specifies a joint generative process for all observed biomarkers and symptom data, conditional on latent viral trajectories, individual random effects, and population-level parameters. Up to this point, the likelihood is well-defined and the model could, in principle, be fit under either a maximum likelihood or Bayesian framework. In this section, we motivate the choice of Bayesian inference and describe the prior distributions used. For notational compactness, let $\mathbf{D}_i^{\text{obs}}$ denote all observed outcome data for individual $i$, stacking the biomarker measurements $\mathbf{Z}_{it}$ and symptom indicators $Y_{it}$ across time points $t$ conditioning on covariates $\mathbf{X}_i$ and test characteristics $S_i$ throughout. By the ignorability result established in Section~\ref{sec:model}, only the observed components contribute to the likelihood.

\subsection{Maximum likelihood}

Maximum likelihood estimation (MLE) seeks the parameter vector $\hat\theta$ that maximizes the marginal likelihood $\mathcal{L}(\theta) = \prod_{i=1}^{N} \int f(\mathbf{D}_i^{\text{obs}} \mid \boldsymbol{\eta}_i, \theta)\, f(\boldsymbol{\eta}_i \mid \theta)\, d\boldsymbol{\eta}_i$, where $\boldsymbol{\eta}_i$ generically collects all individual-level random effects for subject $i$ --- encompassing the RNA trajectory effects, PFU trajectory effects, and symptom onset effect defined in Section~\ref{sec:model}. This integration over individual random effects is intractable in closed form for all but the simplest hierarchical models, and the high dimensionality of the random effects space (thousands of correlated individual effects across multiple trajectory components) makes numerical approximations such as the Laplace approximation or adaptive Gaussian quadrature unreliable.

Even if the marginal likelihood could be evaluated, several features of the model make MLE problematic. First, the observation model parameters --- particularly the false positive rate $\lambda_{\text{fp}}$ and false negative rate $\lambda_{\text{fn}}$ --- are only weakly identified by the data: distinguishing a genuine low-level detection from a false positive requires information that is largely latent, and the likelihood surface is correspondingly flat in these directions. Second, the log-affine transformation parameters $(a_{0,\cdot}, a_{1,\cdot})$ linking RNA to PFU trajectories are informed only by the subset of individuals with paired quantitative culture data ($N = 275$ of ${\sim}2{,}000$), creating a tension between the global RNA trajectory (informed by all subjects) and the transformation (informed by few). Without regularization, these parameters can drift to implausible values. Third, the correlation matrix $\boldsymbol{\Omega}$ for the RNA individual effects has $\binom{4}{2} = 6$ free parameters, and the Cholesky factor must remain positive definite; unconstrained maximization near the boundary of the positive definite cone is numerically fragile.

\subsection{Bayesian inference}

We adopt a Bayesian approach in which inference is based on the posterior distribution
\begin{equation*}
    p(\theta, \boldsymbol{\eta}_{1:N} \mid \mathbf{D}^{\text{obs}}) \propto \prod_{i=1}^{N} f(\mathbf{D}_i^{\text{obs}} \mid \boldsymbol{\eta}_i, \theta)\, f(\boldsymbol{\eta}_i \mid \theta) \cdot \pi(\theta)
\end{equation*}
where $\pi(\theta)$ is the prior distribution over population-level parameters. The Bayesian framework offers several advantages for this problem.

\paragraph{Partial pooling across heterogeneous data sources.} The five cohorts contributing data to this analysis differ substantially in their sampling designs, assay panels, populations, and circulating variants (Table~\ref{tab:cohorts}). Rather than fitting separate models to each cohort or pooling all data under a common likelihood with no acknowledgment of between-study heterogeneity, the hierarchical structure of the model enables \emph{partial pooling}: cohort-specific parameters (setting-level random effects on trajectory and observation model parameters) are shrunk toward the population mean in proportion to the amount of information each cohort contributes. This is particularly important for the PFU trajectory, where only three of five cohorts provide quantitative viral culture data, and for symptom onset, where cohorts differ in whether full daily diaries or only onset dates are recorded. The posterior automatically balances cohort-specific evidence against the population prior, borrowing strength where data are sparse without forcing homogeneity where the data support heterogeneity.

\paragraph{Regularization of weakly identified parameters.} Several parameters in the model are only weakly constrained by the data. The false positive rate $\lambda_{\text{fp}}$ and false negative rate $\lambda_{\text{fn}}$ are identified primarily through rare events (a spurious positive near the limit of detection, or a missing detection during high viral load); the TCID50 observation model parameters $(a, b, \sigma_{\text{TCID50}})$ are informed by a single cohort with interval-censored culture results; and the binary culture model intercept $\pi_0$ is identified by relatively few observations. Informative priors on these parameters regularize inference by concentrating posterior mass in scientifically plausible regions, preventing the sampler from exploring pathological configurations that are technically consistent with the data but biologically implausible. By encoding prior knowledge --- for instance, that false positive rates should be low, or that PFU individual-effect variances should be modest --- the prior acts as a soft constraint that stabilizes the posterior geometry without introducing hard boundaries.

\paragraph{Full uncertainty quantification.} The posterior distribution provides a coherent, joint measure of uncertainty over all model parameters and latent quantities. This is particularly valuable for derived quantities such as the probability of a positive culture result as a function of time since infection, or the expected duration of infectiousness under different testing and isolation policies, which depend non-linearly on multiple parameters. Point estimates from MLE would require delta-method or bootstrap approximations for such quantities, whereas the posterior predictive distribution provides exact finite-sample uncertainty propagation.

\subsection{Priors} \label{sec:priors}

We now describe the prior distributions for all model parameters. Our general strategy is to use weakly informative priors that encode plausible parameter ranges based on the existing literature on SARS-CoV-2 viral kinetics, while remaining sufficiently diffuse to let the data dominate in well-identified directions. For variance components and effect sizes, we use regularizing priors that gently shrink toward zero, reflecting a prior expectation that most effects are modest.

\paragraph{Population-level kinetic parameters.} The population-level peak viral load, proliferation time, and clearance time are parameterized on the positive real line using a non-centered log-normal parameterization. For each parameter $\mu \in \{\delta_0, \omega_{p_0}, \omega_{r_0}\}$:
\begin{equation*}
    \mu = m_\mu \exp(s_\mu \cdot z_\mu), \quad z_\mu \sim N(0, 1)
\end{equation*}
where $m_\mu$ is the prior median and $s_\mu$ controls the prior dispersion. This implies $\log \mu \sim N(\log m_\mu, s_\mu^2)$, so that the prior 95\% interval is $[m \exp(-1.96\, s),\; m \exp(1.96\, s)]$. The specific values are:

\begin{table}[ht]
\centering
\caption{Prior distributions for population-level kinetic parameters. Each parameter follows a log-normal distribution parameterized by its median $m$ and log-scale standard deviation $s$. The implied 95\% prior interval is computed as $[m \exp(-1.96\, s),\; m \exp(1.96\, s)]$.}
\label{tab:priors-kinetics}
\begin{tabular}{lccc}
\toprule
Parameter & Prior median ($m$) & Log-scale SD ($s$) & Implied 95\% CI \\
\midrule
Peak RNA, $\delta_0$ (log copies/mL) & 17 & 0.7 & (4.3, 67.0) \\
Proliferation time, $\omega_{p_0}$ (days) & 8 & 0.7 & (2.0, 31.5) \\
Clearance time, $\omega_{r_0}$ (days) & 15 & 0.7 & (3.8, 59.1) \\
\bottomrule
\end{tabular}
\end{table}

These priors are centered on values broadly consistent with the existing literature~\cite{hay2022quantifying,kissler2021viral,ke2022daily} and are wide enough to accommodate substantial variation across variants and host characteristics.

\paragraph{Log-affine transformation parameters.} The intercepts and elasticities linking RNA kinetic parameters to PFU kinetic parameters receive priors
\begin{equation*}
    a_{0,\cdot} \sim N(-1, 1), \qquad a_{1,\cdot} \sim N(1, 0.5)\, T[0, \infty)
\end{equation*}
The intercept prior is centered near $-1$ (reflecting the expectation that the PFU peak is somewhat lower than the RNA peak on the log scale), while the elasticity prior is centered at 1 (i.e., proportional scaling) with positivity enforced by truncation. These priors provide mild regularization while permitting substantial departures from proportionality.

\paragraph{Observation error.} The observation standard deviations for quantitative RT-qPCR and quantitative viral culture assays receive half-normal priors:
\begin{equation*}
    \sigma_{\text{qPCR}},\, \sigma_{\text{PFU}} \sim N^+(0, 5)
\end{equation*}
This prior is weakly informative, placing most of its mass below $\sigma = 10$ (log copies/mL or log PFU/mL), which is far wider than any plausible measurement error. 

\paragraph{False positive and false negative rates.} We use Beta priors parameterized by their mean and concentration:
\begin{equation*}
    \lambda_{\text{fp}} \sim \text{Beta}_\mu(0.02, 100), \qquad \lambda_{\text{fn}} \sim \text{Beta}_\mu(0.01, 100)
\end{equation*}
where $\text{Beta}_\mu(\mu, \kappa)$ denotes the Beta distribution with mean $\mu$ and concentration $\kappa$ (i.e., $\alpha = \mu\kappa$, $\beta = (1-\mu)\kappa$). These priors encode the expectation that both error rates are low (1--2\%) with moderate certainty, while still allowing the data to push the rates higher if warranted. The implied 95\% prior intervals are approximately $(0.006, 0.039)$ for $\lambda_{\text{fp}}$ and $(0.001, 0.025)$ for $\lambda_{\text{fn}}$.

\paragraph{Individual random effect standard deviations.} The marginal standard deviations of the RNA individual effects receive half-normal priors:
\begin{equation*}
    \sigma_{t_p},\, \sigma_{\tilde\delta},\, \sigma_{\tilde\omega_p},\, \sigma_{\tilde\omega_r} \sim N^+(0, 1)
\end{equation*}
On the log scale, a standard deviation of 1 corresponds to a multiplicative factor of $e^{\pm 1} \approx (0.37, 2.72)$ for an individual at $\pm 1$ SD from the population mean, which is a reasonable range for between-individual variation in viral kinetics. For the PFU individual effects, we use a tighter prior:
\begin{equation*}
    \sigma_{\tilde\delta^\prime},\, \sigma_{\tilde\omega^\prime_p},\, \sigma_{\tilde\omega^\prime_r},\, \sigma_{t^\prime_p} \sim N^+(0, 0.3)
\end{equation*}
reflecting the expectation that PFU-specific residual variation (beyond that inherited from the RNA trajectory via the log-affine transformation) should be modest. This tighter prior was motivated by preliminary fits in which wide PFU RE priors led to poorly identified variance components and pathological funnel geometry in the posterior~\cite{betancourt2017conceptual}.

\paragraph{Correlation matrix.} The LKJ prior~\cite{lewandowski2009generating} on the RNA individual-effect correlation matrix is set at concentration $\nu = 4$:
\begin{equation*}
    \boldsymbol{\Omega} \sim \text{LKJ}(4)
\end{equation*}
An LKJ($\nu$) prior with $\nu > 1$ shrinks the correlation matrix toward the identity, with stronger shrinkage for larger $\nu$. At $\nu = 4$, the marginal prior on each pairwise correlation is approximately $\text{Beta}(5, 5)$ rescaled to $(-1, 1)$, concentrating mass near zero with 95\% of prior mass in approximately $(-0.6, 0.6)$. This provides moderate regularization against overfitting the off-diagonal correlations while still permitting the data to identify substantial correlations where they exist.

\paragraph{Covariate and source effects.} All regression coefficients for covariate effects on kinetic parameters, observation model parameters, and symptom onset receive independent normal priors centered at zero:
\begin{equation*}
    \beta_{\cdot} \sim N(0, 1), \qquad \beta^\prime_{\cdot} \sim N(0, 1)
\end{equation*}
On the log scale, this corresponds to a prior 95\% interval of $(\exp(-2), \exp(2)) \approx (0.14, 7.4)$ for the multiplicative effect of a covariate, which is generous. Source-level random effects similarly receive $N(0, 1)$ priors.

\paragraph{Symptom onset model.} The discrete-time cloglog hazard parameters receive priors
\begin{align*}
    \zeta_0 &\sim N(-3, 1) \\
    \zeta_1,\, \zeta_2 &\sim N^+(0, 0.5) \\
    \zeta_3,\, \zeta_4 &\sim N(0, 1) \\
    \sigma_u &\sim N^+(0, 1)
\end{align*}
The intercept prior centers the baseline daily hazard at $\exp(-3) \approx 0.05$ (i.e., a ${\sim}5\%$ daily probability of symptom onset absent viral load effects), consistent with the observation that symptoms typically develop over several days from infection. The viral load effect coefficients $\zeta_1$ and $\zeta_2$ are constrained to be positive (more virus implies higher hazard of symptom onset) with a moderately tight prior reflecting the expectation that these effects are positive but not overwhelmingly large. The post-peak indicator coefficients $\zeta_3$ and $\zeta_4$ receive standard normal priors, allowing the data to determine the direction and magnitude of the immune activation lag.

\paragraph{TCID50 observation model.} The interval-censored normal model for TCID50 culture results receives priors based on the expected kinetics of viral growth in culture:
\begin{align*}
    a &\sim N(8, 3) & \text{(days to detection at low inoculum)} \\
    \log b &\sim N(-0.5, 1) & \text{(log inverse culture growth rate)} \\
    \log \sigma_{\text{TCID50}} &\sim N(0, 1) & \text{(log observation scale)}
\end{align*}

\paragraph{LFD and binary culture.} The logistic model intercept for the LFD test is parameterized via a non-centered offset from a prior baseline positivity rate of 1\%:
\begin{equation*}
    \gamma_0 = \text{logit}(0.01) + z_{\gamma_0}, \quad z_{\gamma_0} \sim N(0, 1)
\end{equation*}
which centers the intercept near $\text{logit}(0.01) \approx -4.6$, corresponding to a low false positive rate for the antigen test in the absence of viral load. The slope coefficients $\gamma_1$, $\gamma_2$, and the indicator interaction coefficients $\gamma_3$, $\gamma_4$ each receive standard normal priors. The binary culture model coefficients $(\pi_0, \pi_1)$ similarly receive $N(0, 1)$ priors.

All individual-level random effects are specified with non-centered parameterizations to improve posterior geometry, as described in Section~\ref{sec:computation}.

\section{Empirical example: SARS-CoV-2}
\subsection{The data} \label{sec:data}
We include data from publicly available longitudinal datasets from acute SARS-CoV-2 infections where participants were swabbed or tested repeatedly using multiple assays that measure viral shedding. We only include studies where there was a decent chance that the proliferation phase was adequately captured, such as from contact tracing or occupational health cohorts where surveillance testing was performed independent of symptoms. All studies were approved by institutional review boards and obtained written informed consent where applicable.

The Assessment of Transmission and Contagiousness of COVID-19 in Contacts (ATACCC) study~\cite{singanayagam2022community,hakki2022onset} was a longitudinal, prospective cohort study of community contacts of newly diagnosed, PCR-confirmed SARS-CoV-2 index cases in the United Kingdom spanning two separate enrollment periods: ATACCC1 enrolled contacts from Sept 13, 2020, to March 31, 2021, during the SARS-CoV-2 pre-alpha and alpha variant waves; and ATACCC2 enrolled contacts from May 24, 2021, to Oct 28, 2021, during the delta variant wave. Participants consisted of household and non-household exposed contacts aged 5 years or older who provided informed consent and agreed to complete symptom diary and self-swabbing of the upper respiratory tract for up to 20 days. We use data from 57 well-documented infections which included the growth phase. 

The UIUC cohort study~\cite{ke2022daily,Ranoa2022mitigation} was a longitudinal, prospective cohort at the University of Illinois at Urbana-Champaign. During the fall of 2020 and spring of 2021, all faculty, staff and students were required to undergo at least twice weekly quantitative PCR with reverse transcription (RT-qPCR) testing for SARS-CoV-2. Participants were enrolled if they reported a negative RT-qPCR test result in the past 7 days and were either (1) within 24 h of a positive RT-qPCR result or (2) within 5 days of exposure to someone with a confirmed positive RT-qPCR result and nasal and saliva samples were collected daily for up to 14 days. Participants also completed a daily online symptom survey. We use data from 60 well-documented infections which included the growth phase.

The NBA occupational cohort study~\cite{kissler2021viral,hay2022quantifying} was a longitudinal, prospective cohort study among players, staff, and affiliates of the National Basketball Association who were infected with SARS-CoV-2. Between March 11, 2020, and July 28, 2022, the NBA conducted regular surveillance for SARS-CoV-2 infection as part of an occupational health program. This included frequent viral testing (often daily during high community COVID-19 prevalence) using a variety of platforms, but primarily via nucleic acid amplification tests, as well as clinical assessment including case diagnosis and symptom tracking. To assess viral concentration, RT-qPCR tests were conducted when possible, using anterior nares and oropharyngeal swabs. Data on participant age and vaccination status were collected where possible. Viral lineages were assigned using whole-genome sequencing, when feasible. This resulted in a longitudinal dataset of 424,401 SARS-CoV-2 tests with clinical COVID-19 history and demographic information for 3021 individuals.

The Crick Legacy cohort~\cite{russell2024combined} was a longitudinal, prospective cohort study of staff at the Francis Crick Institute in London, United Kingdom, who were infected with SARS-CoV-2 during the Delta and Omicron waves between July 2021 and April 2022. Staff underwent regular occupational surveillance testing via RT-qPCR, and infections were identified through routine screening or following symptom onset. Upon identification, participants provided serial upper respiratory tract swabs over the course of their infection, with symptom onset dates recorded. All participants in the Legacy cohort were vaccinated (either two or three doses). We use data from 157 infections with adequate sampling of the growth phase. Because Ct values were generated on a different platform than the NBA cohort, we apply a calibration offset ($-2$ on the natural-log scale, corresponding to an approximately 7.4-fold adjustment) to account for differences in extraction and amplification efficiency, estimated from paired samples processed on both platforms.

The SARS-CoV-2 Human Challenge Characterisation Study~\cite{killingley2022safety} was a single-center, phase 1, open-label, human challenge trial. Healthy adults aged 18 to 30 years who were at low absolute risk of hospitalization or death and with no evidence of previous SARS-CoV-2 infection or vaccination were recruited between March and July 2021 and inoculated intranasally with 10 TCID50 of a wild-type SARS-CoV-2 virus. Participants were housed in a quarantine unit and prospectively followed for the length of infection and completed twice daily nasal and throat swabs as well as a symptom diary. We use data from 18 individuals in whom inoculation produced a well-tolerated infection. 

Table~\ref{tab:cohorts} summarizes the characteristics of the five cohorts included in this analysis, including the number of infections, age distribution, recurrence status, variant distribution, vaccination history, and available laboratory measurements. Covariate information was harmonized across cohorts, in some cases by collapsing detailed categories into broader groups (e.g., age categories, vaccination history) to ensure comparability. In many cases, covariate information is suboptimal due to incomplete reporting in the original studies, but they nonetheless provide a sense for the possibilities of modeling viral kinetics. 

\begin{landscape}
\begin{table}[p]
    \centering
    \caption{Summary of cohort characteristics and available laboratory measurements. Individual characteristics include age group, recurrence status, SARS-CoV-2 variant, and vaccination history. Laboratory measurements include Ct-derived viral RNA concentrations, viral cultures, lateral flow device (LFD) results, and symptom diary data.}
    \small
    \begin{tabular}{lcccccccccc}
     \toprule
     & \multicolumn{2}{c}{NBA} & \multicolumn{2}{c}{ATACCC} & \multicolumn{2}{c}{UIUC} & \multicolumn{2}{c}{Legacy} & \multicolumn{2}{c}{HCT} \\
      & N & \%& N & \% & N & \% & N & \% & N & \% \\
     \midrule
     \textit{Individual characteristics} &  &  &  &  &  &  &  &  &  &  \\
     Age: [18,30) & 818 & 41.1 & 16 & 28.1 & 37 & 61.7 & 67 & 42.7 & 18 & 100.0 \\
     Age: [30,50) & 876 & 44.0 & 32 & 56.1 & 17 & 28.3 & 60 & 38.2 & 0 & 0.0 \\
     Age: [50,100) & 295 & 14.8 & 9 & 15.8 & 6 & 10.0 & 30 & 19.1 & 0 & 0.0 \\
     Recurrence: No & 1,796 & 90.3 & 57 & 100.0 & 60 & 100.0 & 120 & 76.4 & 18 & 100.0 \\
     Recurrence: Yes & 193 & 9.7 & 0 & 0.0 & 0 & 0.0 & 37 & 23.6 & 0 & 0.0 \\
     Variant: Pre-Alpha & 191 & 9.5 & 13 & 22.8 & 43 & 71.7 & 0 & 0.0 & 18 & 100.0 \\
     Variant: Alpha & 49 & 2.5 & 12 & 21.0 & 16 & 26.7 & 0 & 0.0 & 0 & 0.0 \\
     Variant: Delta & 191 & 9.5 & 25 & 43.8 & 0 & 0.0 & 34 & 21.7 & 0 & 0.0 \\
     Variant: Omicron & 1,400 & 70.4 & 0 & 0.0 & 0 & 0.0 & 123 & 78.3 & 0 & 0.0 \\
     Variant: BA.4/BA.5 & 71 & 3.6 & 0 & 0.0 & 0 & 0.0 & 0 & 0.0 & 0 & 0.0 \\
     Variant: Other & 278 & 14.0 & 0 & 0.0 & 1 & 1.6 & 0 & 0.0 & 0 & 0.0\\
     History: Unvaccinated & 94 & 4.7 & 24 & 53.4 & 60 & 100.0 & 0 & 0.0 & 18 & 100.0 \\
     History: Vaccinated boosted & 982 & 49.4 & 0 & 0.0 & 0 & 0.0 & 108 & 68.8 & 0 & 0.0 \\
     History: Vaccinated unboosted & 269 & 13.5 & 21 & 46.6 & 0 & 0.0 & 49 & 31.2 & 0 & 0.0\\
     History: Vaccinated unreported & 5 & 0.3 & 0 & 0.0 & 0 & 0.0 & 0 & 0.0 & 0 & 0.0\\
     History: Unreported & 579 & 29.1 & 0 & 0.0 & 0 & 0.0 & 0 & 0.0 & 0 & 0.0\\
     History: Boosted unreported primary & 20 & 1.2 & 0 &0.0 & 0 & 0.0 & 0 & 0.0 & 0 & 0.0\\
     \midrule
     \textit{Lab measurements} &  &  &  &  &  &  &  &  &  &  \\
        Ct values & 21,463 & & 638 & & 934 & & 698 & & 684 \\
        Viral cultures & 0 & & 638 & & 934 & & 0 & & 684 \\
        LFD values & 0 & & 638 & & 934 & & 0 & & 684 \\
        Symptom diaries & 0 & & 638 & & 934 & & 698 & & 684 \\
     \midrule
     Individuals & 1,989 &  & 57 & & 60 & & 157 & & 18 \\
     \bottomrule
     \end{tabular}
    \label{tab:cohorts}
\end{table}
\end{landscape}

\section{Computation} \label{sec:computation}
All models are implemented in Stan~\cite{carpenter2017stan} and fit using the No-U-Turn Sampler (NUTS)~\cite{hoffman2014no}, a variant of Hamiltonian Monte Carlo~\cite{betancourt2017conceptual}, via the \texttt{cmdstanr} interface to CmdStan~\cite{gabry2024cmdstanr}. We run 4 parallel chains, each with 1{,}000 warmup iterations and 4{,}000 sampling iterations, yielding 16{,}000 posterior draws in total. We set the target acceptance probability to $\delta_{\text{adapt}} = 0.95$ (the NUTS \texttt{adapt\_delta} parameter) and maximum tree depth to 12 to reduce the incidence of divergent transitions, which can signal numerical instability in regions of high curvature. 

Several computational strategies are employed to improve sampling efficiency. First, all individual-level random effects use a non-centered parameterization (NCP)~\cite{papaspiliopoulos2007general}, in which standard normal auxiliary variables $\mathbf{z}_i$ are sampled and then scaled by the relevant standard deviations in the transformed parameters block. This eliminates the funnel geometries that arise from centered parameterizations of hierarchical models~\cite{neal2003slice,betancourt2016diagnosing} and that, in preliminary runs, led to severe convergence failures (E-BFMI $<$ 0.07, $\hat{R}$ up to 2.46). For the correlated RNA random effects, the NCP is extended via the Cholesky factorization of the correlation matrix (Section~\ref{sec:model}), which parameterizes the correlation structure through its lower-triangular factor $\mathbf{L}$ and avoids direct sampling of the positive-definite matrix $\boldsymbol{\Omega}$. Second, the smooth trajectory function (Section~\ref{sec:model}) provides continuously differentiable log-densities, which are better suited to the leapfrog integrator used by HMC than the piecewise function with its non-differentiable kink at the peak. Third, population-level kinetic parameters (peak height, proliferation time, clearance time) are parameterized on a non-centered log scale with prior mean and coefficient of variation supplied as data, which standardizes the posterior geometry and reduces sensitivity to the prior scale. 

Convergence is assessed using the split-$\hat{R}$ statistic~\cite{vehtari2021rank}, with all parameters required to satisfy $\hat{R} < 1.05$, and effective sample sizes (both bulk and tail ESS) required to exceed 100 per parameter. We additionally monitor the number of divergent transitions, the energy Bayesian fraction of missing information (E-BFMI)~\cite{betancourt2016diagnosing}, and inspect trace plots for key parameters~\cite{gabry2019visualization}. Model comparison is performed using Pareto-smoothed importance sampling leave-one-out cross-validation (PSIS-LOO)~\cite{vehtari2024pareto,vehtari2017practical} computed from the observation-level log-likelihoods stored in the generated quantities block, supplemented by the widely applicable information criterion (WAIC)~\cite{watanabe2010asymptotic}. Posterior predictive checks are conducted by drawing replicated datasets from the fitted model and comparing their summary statistics to the observed data~\cite{gabry2019visualization}. A simulation-based calibration study (parameter recovery)~\cite{talts2018validating} is performed by generating data from the prior predictive distribution, fitting the model, and assessing whether the true parameters are recovered within the posterior credible intervals.

\section{Model checking} \label{sec:model_checking}

\subsection{Convergence diagnostics}
The final model was fit in approximately 16.5 hours (wall time) on a 16-core workstation. All four chains converged without divergent transitions (0 of 64{,}000 transitions) and without maximum tree depth exceedances. The energy Bayesian fraction of missing information (E-BFMI) ranged from 0.68 to 0.70 across chains, well above the 0.3 threshold that would suggest problematic posterior geometry. The maximum split-$\hat{R}$ across all parameters was 1.03 (for the RNA correlation matrix element $L_{\Omega,\text{rna}}[4,3]$), with only 17 parameters exhibiting $\hat{R} > 1.01$, all of which are elements of the RNA individual-effect correlation matrix. No PFU-related parameters showed convergence concerns. The minimum bulk effective sample size was 183 ($L_{\Omega,\text{rna}}[4,1]$) and the minimum tail ESS was 328 ($L_{\Omega,\text{rna}}[4,3]$), both adequate for reliable posterior inference. Trace plots for key parameters are shown in Figure~\ref{fig:trace}.

\subsection{Posterior predictive checks}
We assess model fit by comparing posterior predictive draws to the observed data across all four assay types. For viral RNA, we generate replicated qPCR observations at each measured time point by drawing from $\text{Normal}(\hat{R}_t, \hat{\sigma}_{\text{qPCR}})$ and compare the distribution of replicated values to the observed Ct-derived log RNA concentrations, both in aggregate and stratified by cohort. For infectious virus, we compare replicated PFU/TCID$_{50}$ values to observed culture results, again both in aggregate and stratified by cohort. For lateral flow tests, we compare predicted positivity rates (averaged across posterior draws) to observed positivity rates in a calibration plot (overall and by cohort). When the LFD calibration is stratified by infection phase (pre-peak versus post-peak RNA), the model reproduces the asymmetry reported by~\cite{hakki2022onset}: predicted LFD sensitivity is lower during the proliferation phase than during clearance at comparable predicted positivity levels, consistent with the lag between viral replication and antigen accumulation to the LFD detection threshold (Supplementary Figure~\ref{fig:ppc_lfd_phase}). For symptoms, we compare the predicted cumulative incidence under the cloglog hazard model to the observed Kaplan--Meier curve of symptom onset, overall and stratified by cohort. In all cases, the observed data fall within the 95\% posterior predictive intervals, indicating adequate model fit across assay types and cohorts (Figures~\ref{fig:ppc_rna}--\ref{fig:ppc_sym}; cohort-stratified versions in Supplementary Figures).

\subsection{Cross-validation and information criteria}
We compute PSIS-LOO and WAIC from the observation-level log-likelihoods. The estimated expected log pointwise predictive density under LOO is $\widehat{\text{elpd}}_{\text{LOO}} = {-45{,}743}$ (SE = 208), with an effective number of parameters $p_{\text{LOO}} = 4{,}736$ (SE = 56). The WAIC estimate is $\widehat{\text{elpd}}_{\text{WAIC}} = {-45{,}320}$ (SE = 209), with $p_{\text{WAIC}} = 4{,}313$ (SE = 62). The close agreement between LOO and WAIC provides reassurance that neither method is substantially biased for this model.

Of the 24{,}715 observations, 1{,}707 (6.9\%) have Pareto shape parameter $\hat{k} > 0.7$ and 658 (2.7\%) have $\hat{k} > 1$, with 258 observations yielding $\hat{k} = \infty$. Investigation reveals that these problematic observations are overwhelmingly symptom-only time points --- observations where only symptom status is recorded with no accompanying RNA, PFU, or LFD data. All 258 infinite-$\hat{k}$ values are symptom-only observations, and symptom-only time points are enriched 24.5-fold among high-$\hat{k}$ observations relative to their prevalence in the dataset. This pattern is expected: when an observation contributes only a symptom log-likelihood term, leaving it out under LOO removes the sole data point constraining the latent trajectory at that time, causing importance sampling weights to degenerate. The ATACCC cohort is disproportionately affected (25.1\% of ATACCC observations have $\hat{k} \geq 1$) because it has the most multi-modal data per observation. These high Pareto $k$ values reflect a structural limitation of PSIS-LOO for mixed-outcome models with sparse symptom-only observations rather than model misspecification. Moment-matching or $K$-fold cross-validation would be required for reliable LOO estimates of these observations; we defer this to future work.

\subsection{Parameter recovery}
We assess the model's ability to recover known parameter values through a simulation study. We generate a synthetic dataset from the model's prior predictive distribution using a 50\% random subsample of the observed covariate and missingness structure, fit the model to the simulated data using the same MCMC configuration (with 2{,}000 sampling iterations per chain for computational efficiency), and compare the recovered posterior intervals to the true generating values. Of 34 parameters assessed, 26 (76.5\%) are covered by the 95\% posterior credible interval of the recovery fit. The eight non-covered parameters fall into three categories. First, the observation error parameters $\sigma_{\text{pfu}}$ and $\hat{\lambda}_{\text{fp}}$ show upward bias ($+0.48$ and $2\times$ overestimation, respectively), reflecting the difficulty of separating measurement error from process variation in limited PFU data. Second, $\hat{\lambda}_{\text{fn}}$ is dominated by its prior near zero, as expected given the rarity of false negatives in qPCR. Third, the simple culture model parameters ($\alpha_{\text{cult},1}$) are not identifiable from the small number of qualitative culture observations available. These limitations are consistent with the data constraints of the respective assay types and do not compromise inference on the primary kinetic parameters, all of which are well recovered (Figure~\ref{fig:recovery}).

\section{Results} \label{sec:results}

\subsection{Individual trajectory fits}

\begin{landscape}
\begin{figure}[p]
    \centering
    \includegraphics[width=\linewidth]{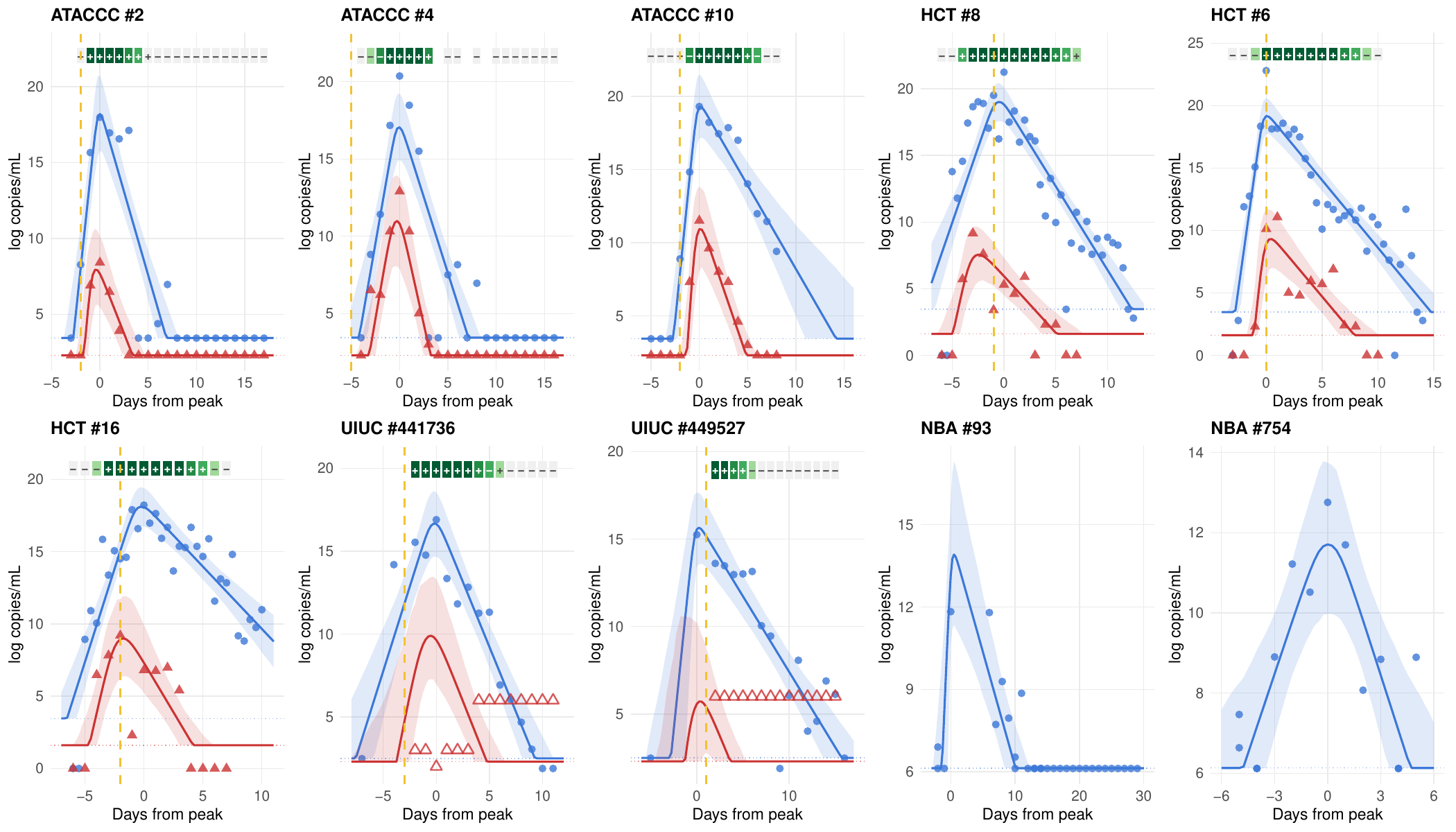}
    \caption{Example individual trajectory fits from the ATACCC, UIUC, HCT, Legacy, and NBA cohorts. For each individual, the posterior median RNA trajectory (blue line) and 95\% credible interval (blue ribbon) are shown with observed qPCR values (blue points). Where available, the PFU trajectory (red line/ribbon) is overlaid with culture measurements: filled triangles denote PFU or FFA assays (ATACCC, HCT) and open triangles denote TCID$_{50}$ assays (UIUC). LFD predictions are shown as shaded tiles (scale runs from dark green to light green as probability increases) overlaid with $+$ or $-$ symbols where $+$ denotes a positive observed LFD result and $-$ denotes a negative observed LFD result. Symptom onset is marked with dashed vertical line.}
    \label{fig:trajectories}
\end{figure}
\end{landscape}

The model produces well-calibrated individual-level trajectory fits across all five cohorts. Figure~\ref{fig:trajectories} shows example fits for selected individuals from the ATACCC, UIUC, HCT, Legacy, and NBA cohorts, illustrating the model's ability to capture diverse kinetic profiles. For each individual, the posterior median RNA trajectory (with 95\% credible interval) closely tracks the observed qPCR values, and where available, the PFU trajectory aligns with culture-based measurements. The model appropriately handles heterogeneous data availability: for NBA participants, who contribute only qPCR data, the infectious virus trajectory is inferred entirely from the RNA-to-PFU transformation and the hierarchical prior. For Legacy participants, who contribute qPCR and symptom onset data but no culture or LFD results, the model leverages the joint structure to impute infectious virus dynamics. For ATACCC and UIUC participants, who contribute culture, LFD, and symptom data in addition to qPCR, the model jointly fits all assay types, with predicted LFD positivity probabilities (shown as shaded tiles) tracking observed LFD results and the estimated symptom onset timing consistent with reported symptom diaries.

Complete individual-level trajectory fits for all participants in each cohort are provided in Figures~\ref{fig:fit_ataccc}--\ref{fig:fit_legacy}. The individual-level random effects allow the model to accommodate substantial heterogeneity in viral kinetics. Some individuals exhibit rapid proliferation and clearance (short, sharp peaks), while others show prolonged shedding with gradual clearance. The correlated random effect structure captures the tendency for individuals with higher peak viral loads to also have longer clearance times (posterior median $\rho_{d_p, \omega_r} = 0.21$, 90\% CrI: 0.15--0.28).

\subsection{Population-level viral kinetics}

\begin{landscape}
\begin{figure}[p]
    \centering
    \includegraphics[width=\linewidth]{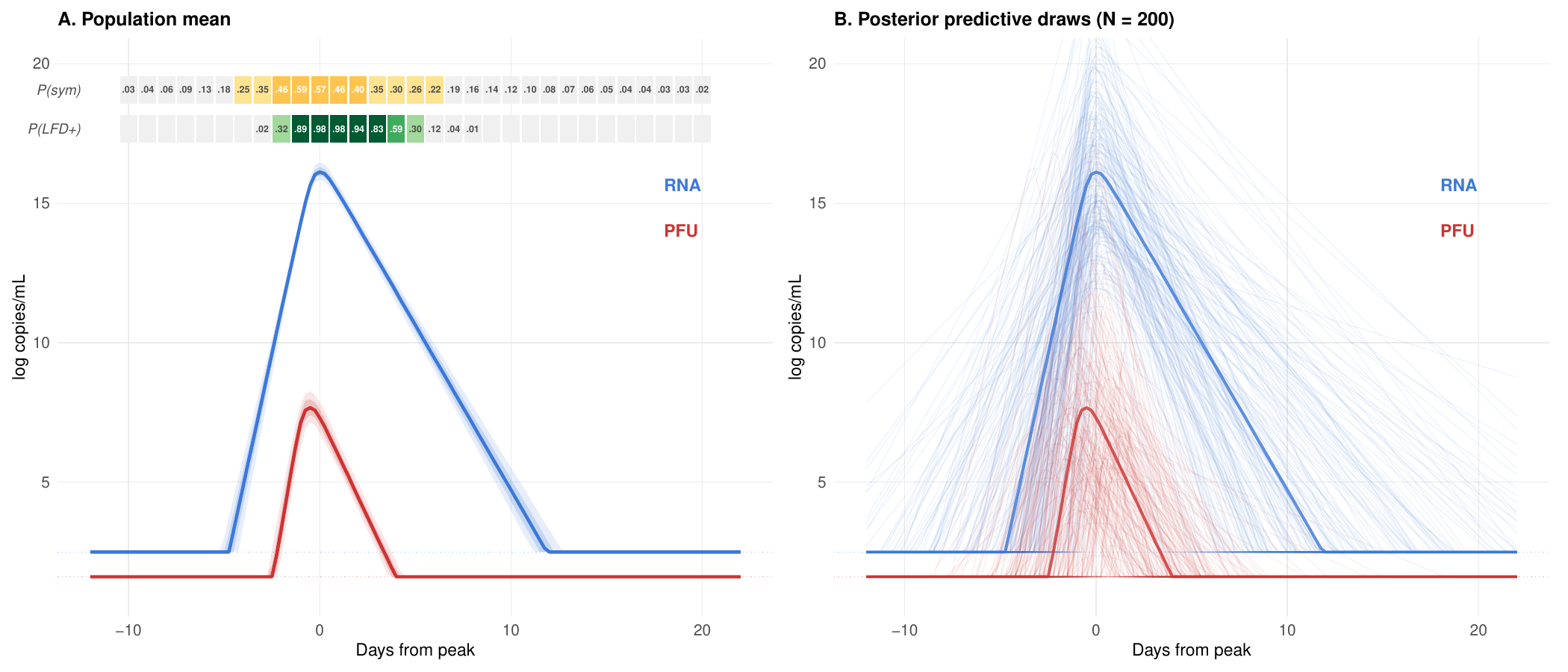}
    \caption{Population-level viral kinetics. \textbf{(A)} Population-mean trajectories for viral RNA (blue) and infectious virus (red) with 50\% (dark shading) and 80\% (light shading) credible intervals, computed from 200 posterior draws. Trajectories are clamped at the assay-specific limits of detection (dashed horizontal lines). LFD positivity probability and daily symptom onset hazard are shown as shaded tiles above the trajectories. \textbf{(B)} Spaghetti plot of 200 individual posterior trajectory draws, each generated by sampling new random effects from the estimated hierarchical distribution, illustrating inter-individual variability in kinetic profiles.}
    \label{fig:population}
\end{figure}
\end{landscape}

Figure~\ref{fig:population} summarizes the population-level viral kinetics implied by the posterior distribution. Panel A shows the population-mean trajectory computed from the median posterior parameters, displaying the smooth piecewise-exponential shape for both viral RNA and infectious virus (PFU). The median peak RNA concentration is 16.2 log copies/mL (90\% CrI: 15.7--16.7), reached at time $t_p = 0$ by convention. The RNA proliferation duration is 5.3 days (90\% CrI: 4.6--6.2 days) and the clearance duration is 13.7 days (90\% CrI: 12.4--15.2 days), yielding an asymmetric trajectory with clearance approximately 2.6 times slower than proliferation. The 50\% and 80\% credible intervals, reflecting uncertainty in the population parameters, show modest uncertainty in the peak height but wider bands during the proliferation and late clearance phases.

The infectious virus trajectory, derived from the RNA trajectory via the log-affine transformation, peaks at a lower level and clears more rapidly. The PFU peak elasticities with respect to RNA are close to unity ($a_{1,\delta} = 1.07$, 90\% CrI: 0.65--1.50), indicating an approximately proportional relationship on the log scale. The clearance elasticity is sub-proportional ($a_{1,\omega_r} = 0.66$, 90\% CrI: 0.44--0.88), meaning that infectious virus clears faster than viral RNA --- consistent with the biological expectation that replication-competent virus is eliminated more rapidly than residual RNA. This asymmetry has direct practical consequences: because infectious virus clears in approximately $0.66 \times 13.7 \approx 9$ days versus 13.7 days for RNA, a substantial ``false-positive window'' of $\sim$5 days exists during which an individual would test positive by PCR but is no longer shedding replication-competent virus. The 2.6:1 clearance-to-proliferation ratio further implies that the post-peak tail of viral shedding --- during which isolation and testing decisions are made --- is considerably longer than the pre-peak phase during which most infections are detected.

The superimposed LFD positivity and symptom hazard probabilities illustrate the temporal relationship between viral dynamics and detectable outcomes. LFD positivity rises rapidly during the proliferation phase, peaking near the time of peak viral load, and declines as viral shedding falls below the antigen detection threshold. The daily symptom onset hazard is highest near and shortly after the peak, consistent with the cloglog hazard model's dependence on both RNA and PFU levels.

Panel B shows 200 posterior predictive draws of individual latent trajectories, each generated by sampling new individual random effects from the posterior distribution of the hierarchical variance--covariance structure. The spaghetti plot illustrates the substantial inter-individual variation in trajectory shape, timing, and magnitude. Some drawn trajectories peak early with high viral loads, while others show delayed, lower peaks with prolonged clearance --- reflecting the estimated individual-level standard deviations of 0.24 (peak timing), 0.16 (peak height), 0.63 (proliferation), and 0.51 (clearance) on the log scale.

\subsection{Covariate effects on viral kinetics}

\begin{landscape}
\begin{figure}[p]
    \centering
    \includegraphics[width=\linewidth]{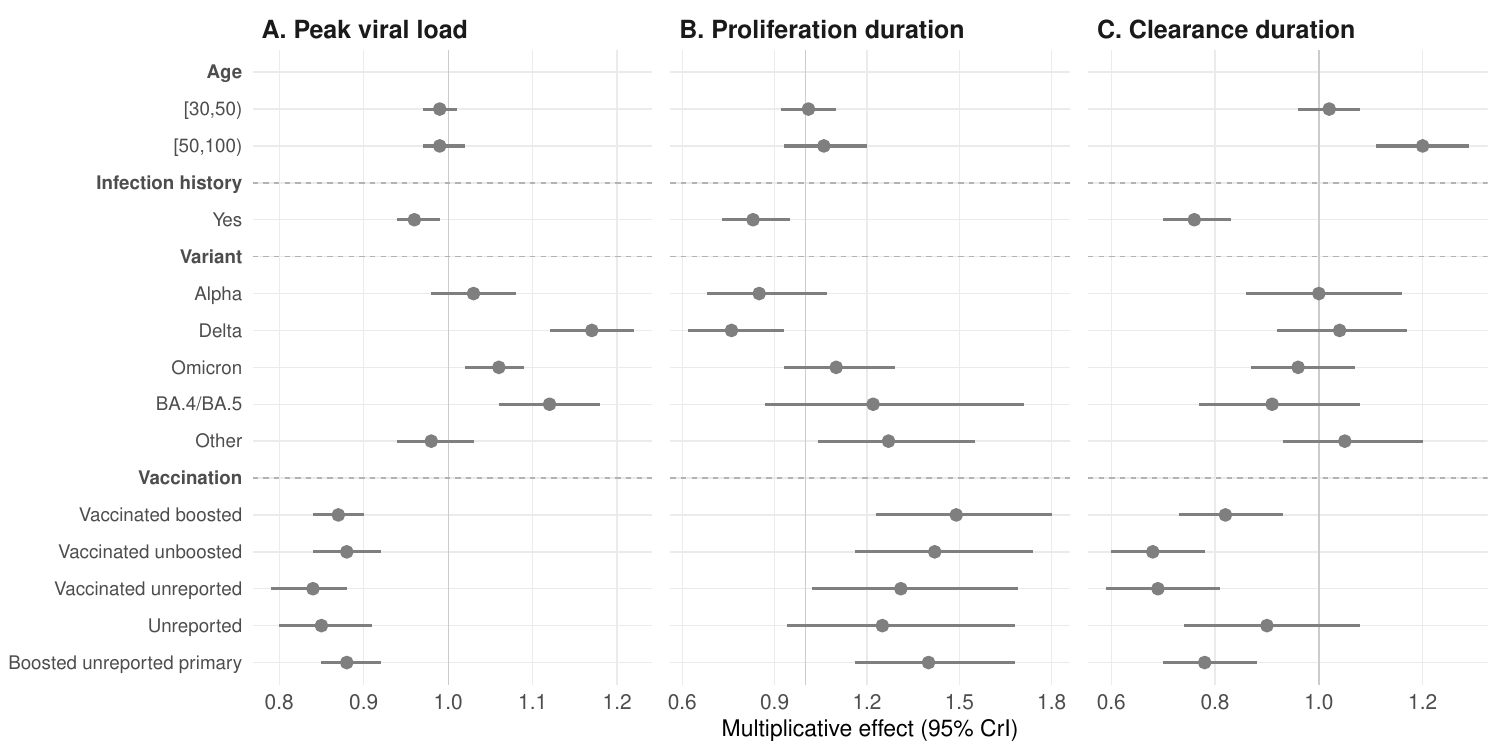}
    \caption{Estimated covariate effects on RNA kinetic parameters. Points and horizontal lines show the posterior median and 95\% credible interval for $\exp(\beta)$, representing the multiplicative fold change relative to the reference category (unvaccinated, immunologically na\"ive, 18--30 year old with pre-Alpha variant). Effects are grouped by covariate type and shown for peak viral load, proliferation duration, and clearance duration.}
    \label{fig:covariates}
\end{figure}
\end{landscape}

Figure~\ref{fig:covariates} presents the estimated multiplicative effects of covariates on the three RNA kinetic parameters: peak viral load, proliferation duration, and clearance duration. All effects are expressed as $\exp(\beta)$, representing the fold change relative to the reference category (unvaccinated, immunologically na\"ive, 18--30 year old infected with pre-Alpha SARS-CoV-2).

\paragraph{Variant effects.} The Delta variant is associated with a 17\% higher peak viral load ($\exp(\hat\beta) = 1.17$, 95\% CrI: 1.12--1.22), a 34\% shorter proliferation phase (0.66, 0.57--0.75), and no significant change in clearance duration. The Omicron variant shows a modest 6\% increase in peak load (1.06, 1.02--1.08) paired with 10\% faster clearance (0.90, 0.84--0.96). BA.4/BA.5 subvariants exhibit a 12\% higher peak (1.12, 1.07--1.18) with 13\% shorter clearance (0.87, 0.78--0.97). The Alpha variant shows no significant effect on peak load but a 21\% reduction in proliferation time (0.79, 0.68--0.91).

\paragraph{Vaccination effects.} Prior vaccination is consistently associated with reduced peak viral loads and altered kinetic timing. Boosted individuals show a 13\% reduction in peak RNA (0.87, 0.84--0.90) but a 44\% longer proliferation phase (1.44, 1.27--1.64), suggesting that vaccine-induced immunity blunts peak shedding while delaying the time to peak. Unboosted vaccinated individuals show a similar 12\% peak reduction (0.88, 0.85--0.92) with a more modest 17\% increase in proliferation time (1.17, 1.02--1.35). Both boosted and unboosted vaccinated individuals clear virus 14--26\% faster than unvaccinated individuals. 

\paragraph{Age and infection history.} Age effects are primarily seen in clearance duration: individuals aged 50 and older clear virus 19\% more slowly (1.19, 1.13--1.26) than those under 30, with no significant effect on peak load or proliferation time. Prior infection (recurrence) is associated with a 4\% lower peak (0.96, 0.93--0.98), a 14\% shorter proliferation phase (0.86, 0.77--0.95), and dramatically faster clearance (0.74, 0.70--0.79), consistent with effective anamnestic immunity from prior exposure.

\paragraph{Combined implications for infectiousness.} These covariate effects do not operate in isolation. A boosted individual infected with Omicron, for instance, would experience a 13\% lower peak and 10\% faster clearance compared to an unvaccinated pre-Alpha infection --- but also a 44\% longer proliferation phase. These offsetting effects produce a distinctively different infectiousness profile: lower but more prolonged, with implications for optimal isolation duration that cannot be determined from any single covariate effect in isolation. Section~\ref{sec:policy} exploits the joint posterior to compute these composite profiles and their consequences for testing and isolation policy.

\subsection{RNA-to-PFU transformation and observation model parameters}
The RNA observation standard deviation is estimated at $\hat\sigma_{\text{qPCR}} = 2.24$ log copies/mL (90\% CrI: 2.21--2.27), reflecting substantial within-individual measurement variability across qPCR platforms and time points. The PFU observation standard deviation is $\hat\sigma_{\text{pfu}} = 2.22$ log PFU/mL (90\% CrI: 2.01--2.47). The estimated false positive rate for qPCR is $\hat\lambda_{\text{fp}} = 0.019$ (90\% CrI: 0.016--0.023), indicating that approximately 2\% of positive qPCR results near the limit of detection may be spurious. The false negative rate is estimated near zero.

The symptom onset cloglog model estimates a baseline log-hazard intercept of $\hat\zeta_0 = -1.47$ (90\% CrI: $-1.90$ to $-1.02$), with positive effects of both infectious virus ($\hat\zeta_1 = 0.71$, 90\% CrI: 0.23--1.31) and viral RNA ($\hat\zeta_2 = 1.22$, 90\% CrI: 0.59--1.85) on the daily hazard of symptom onset. The individual-level symptom random effect standard deviation is $\hat\sigma_u = 0.83$ (90\% CrI: 0.58--1.13), indicating substantial unmeasured heterogeneity in symptom susceptibility.

\subsection{Correlation structure of individual random effects}

\begin{landscape}
\begin{figure}[p]
    \centering
    \includegraphics[width=\linewidth]{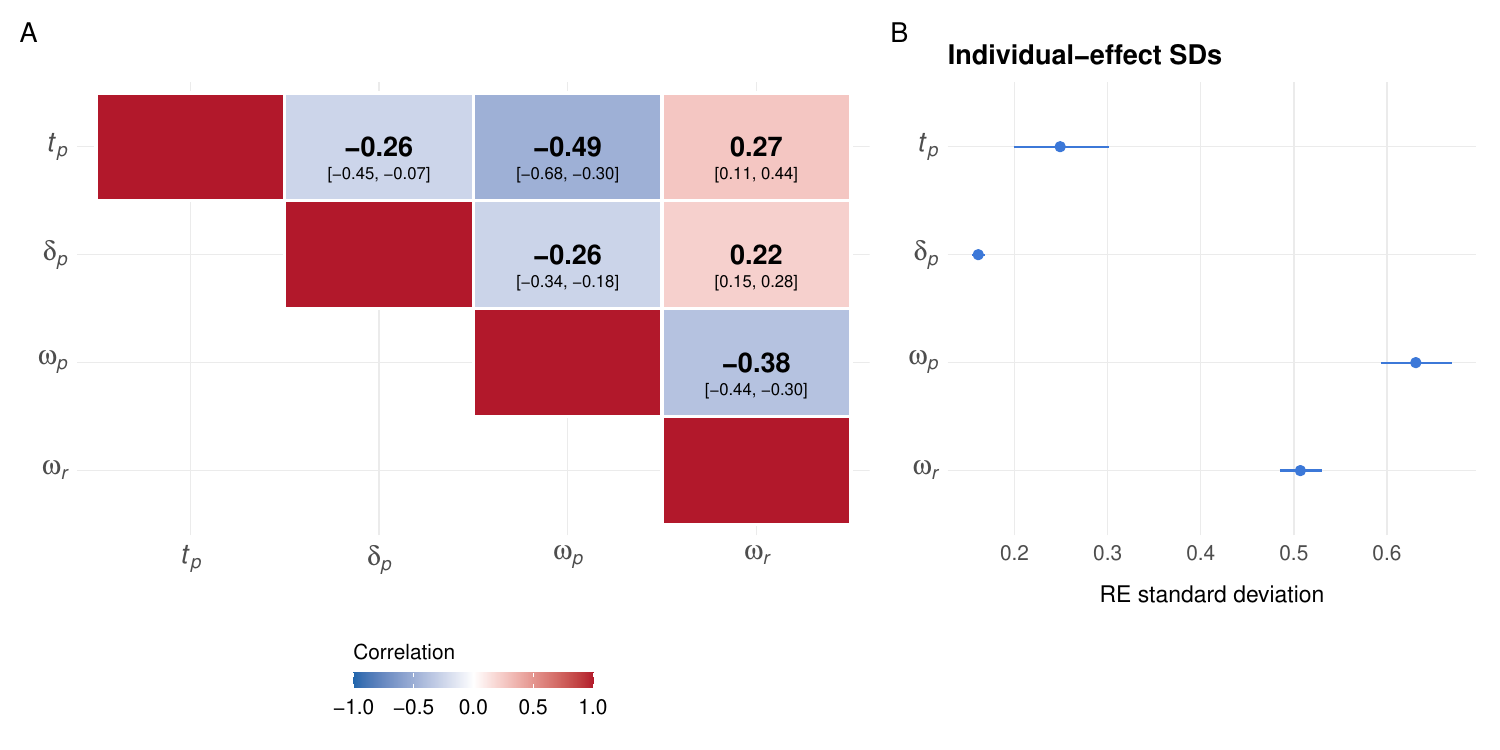}
    \caption{Posterior correlation structure of viral RNA individual random effects. The upper triangle shows the posterior median pairwise correlation between peak timing ($t_p$), peak height ($\delta$), proliferation duration ($\omega_p$), and clearance duration ($\omega_r$). The sidebar shows the estimated marginal standard deviations of each random effect.}
    \label{fig:correlations}
\end{figure}
\end{landscape}

Figure~\ref{fig:correlations} presents the estimated correlation matrix of the RNA individual random effects. Several biologically interpretable patterns emerge. Peak timing and proliferation rate show a strong negative correlation ($\hat\rho = -0.54$, 90\% CrI: $-0.72$ to $-0.35$): individuals who reach peak viral load earlier tend to have faster proliferation. Peak timing and clearance rate are positively correlated ($\hat\rho = 0.28$, 90\% CrI: 0.12--0.44): earlier peaks are associated with slower clearance. Peak height and proliferation rate are negatively correlated ($\hat\rho = -0.25$, 90\% CrI: $-0.33$ to $-0.17$): higher peaks are reached with faster proliferation. Peak height and clearance rate show a modest positive correlation ($\hat\rho = 0.21$, 90\% CrI: 0.15--0.28), and the strongest within-kinetic correlation is between proliferation and clearance rates ($\hat\rho = -0.38$, 90\% CrI: $-0.44$ to $-0.30$), indicating that faster proliferators also tend to clear faster.

These correlations have implications for predicting individual trajectories: an early observation of rapid proliferation, for instance, is informative about both the expected peak height and the likely clearance trajectory, even before those phases are observed. The negative correlation between proliferation rate and peak timing ($\hat\rho = -0.54$) also has a subtler consequence for policy analysis: individuals detected early via routine testing (e.g., during the proliferation phase) tend to be those with faster proliferation and hence higher eventual peaks. If this selection effect is ignored, na\"ive calculations of remaining infectiousness conditional on early detection will underestimate the peak viral load and therefore the duration of the infectious period. The hierarchical correlation structure accounts for this effect when computing the policy-relevant derived quantities in Section~\ref{sec:policy}.

\subsection{Inferring infectious virus from RNA-only data} \label{sec:inferred_pfu}

\begin{landscape}
\begin{figure}[p]
    \centering
    \includegraphics[width=\linewidth]{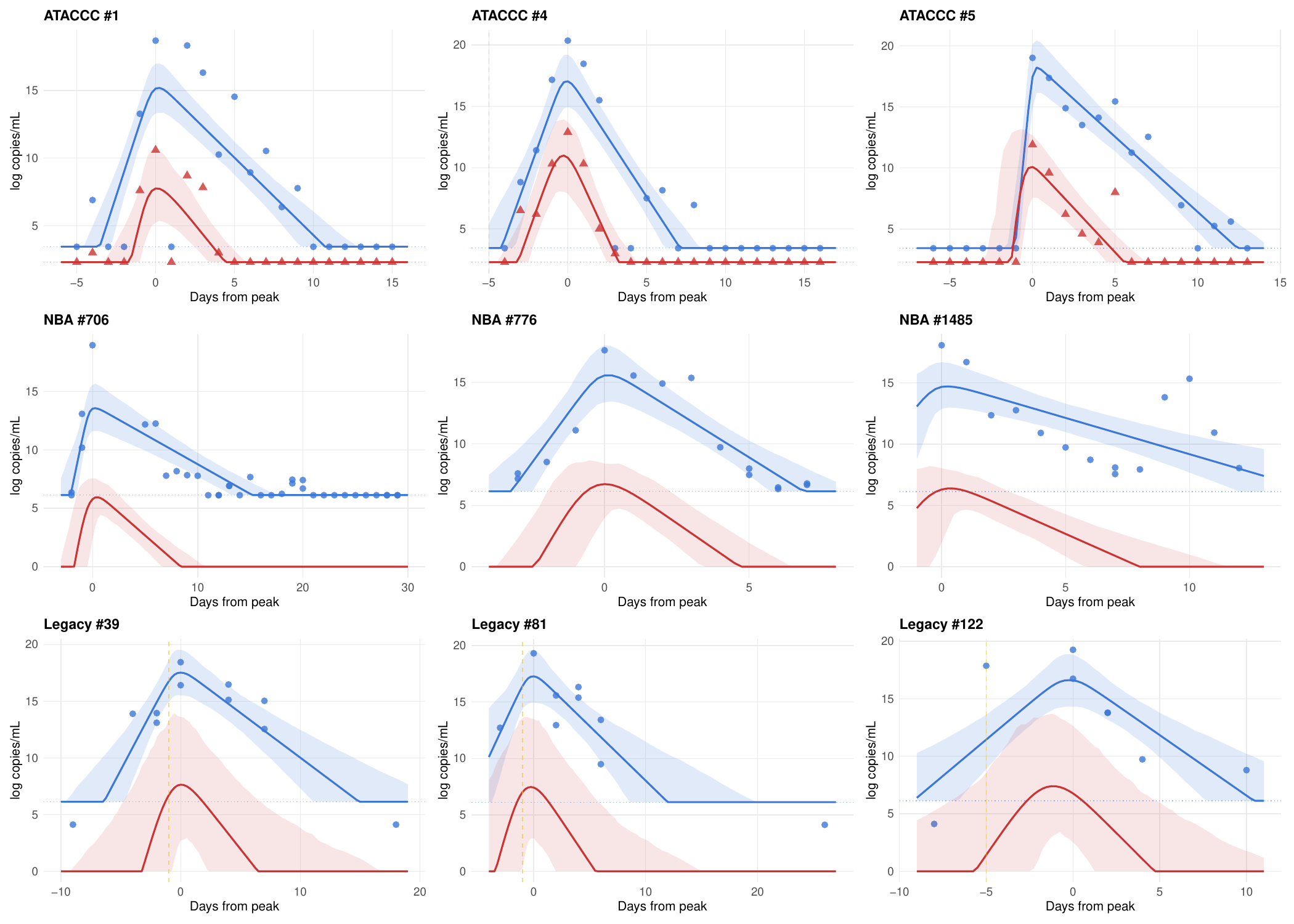}
    \caption{Inferred infectious virus (PFU) trajectories across data regimes. Blue: RNA trajectory (line/ribbon) and observed qPCR (points). Red: PFU trajectory (line/ribbon) and observed culture data (triangles, top row only). Gold dashed line: symptom onset. \textbf{Top:} ATACCC (PFU directly observed). \textbf{Middle:} NBA (RNA only). \textbf{Bottom:} Legacy (RNA + symptoms). PFU credible intervals widen from top to bottom, reflecting decreasing direct constraint on the infectious virus trajectory.}
    \label{fig:inferred_pfu}
\end{figure}
\end{landscape}

A central advantage of the joint model is its ability to impute the latent infectious virus trajectory for individuals who contribute only RNA data (or RNA and symptoms), without any direct culture measurements. This imputation operates through two mechanisms: (i)~the population-level log-affine transformation that maps RNA kinetic parameters to PFU kinetic parameters, and (ii)~the hierarchical prior on individual random effects. For individuals with culture data (ATACCC, UIUC, HCT), the PFU trajectory is further constrained by person-specific PFU random effects; for individuals without culture data (NBA, Legacy), the PFU trajectory reflects the population-level relationship plus the individual-specific RNA deviation propagated through the transformation.

Figure~\ref{fig:inferred_pfu} illustrates this across three data regimes. In the top row, ATACCC participants with dense culture data show well-constrained PFU trajectories (median 95\% credible interval width at peak: 5.4 log PFU/mL) that closely track observed culture measurements. In the middle row, NBA participants with RNA-only data show PFU trajectories inferred entirely from the estimated RNA trajectory, with moderately wider credible intervals (median width at peak: 5.7 log PFU/mL). In the bottom row, Legacy participants with RNA and symptom data show the widest PFU uncertainty (median width at peak: 14.2 log PFU/mL), reflecting both the sparser sampling cadence of this cohort and the absence of individual-level PFU random effects. Despite this wider uncertainty, the posterior median PFU trajectories maintain biologically plausible shapes --- peaking near the RNA peak and clearing more rapidly, consistent with the estimated sub-proportional clearance elasticity ($a_{1,\omega_r} = 0.66$).

The ability to infer individual-level infectious virus trajectories from RNA data alone is practically significant: viral culture is expensive, requires BSL-3 facilities, and is rarely available in clinical or surveillance settings. By linking culture-rich cohorts (ATACCC, HCT) with PCR-only cohorts (NBA, Legacy) through the joint hierarchical model, the population-level RNA-to-PFU transformation is estimated with the precision afforded by the culture data and then applied to the much larger RNA-only cohorts, enabling imputation of infectiousness dynamics for nearly 2{,}000 individuals from whom culture data were never collected. This imputation underpins the policy analyses in Section~\ref{sec:policy}, which require individual-level PFU trajectories to compute the probability of ongoing infectiousness.

\subsection{Applications to testing and isolation policy} \label{sec:policy}
A central advantage of the joint model is that any derived quantity can be computed by marginalizing over the posterior predictive distribution, propagating both parameter uncertainty and inter-individual heterogeneity. To demonstrate this, we draw 200 posterior samples and, for each, generate 50 independent trajectories per covariate profile by sampling new individual random effects from the fitted hierarchical distribution (10 profiles $\times$ 50 replicates $\times$ 200 draws = 100{,}000 trajectories). Each trajectory is then aligned to one of three clinically meaningful landmark events --- first positive PCR test, first positive lateral flow test, or symptom onset --- and aggregated to produce population-level probability curves and isolation duration statistics. Throughout this section, $t = 0$ refers to the landmark event, and all probabilities reflect the marginal posterior predictive distribution over both trajectory and measurement uncertainty.

\begin{figure}[t]
    \centering
    \includegraphics[width=\linewidth]{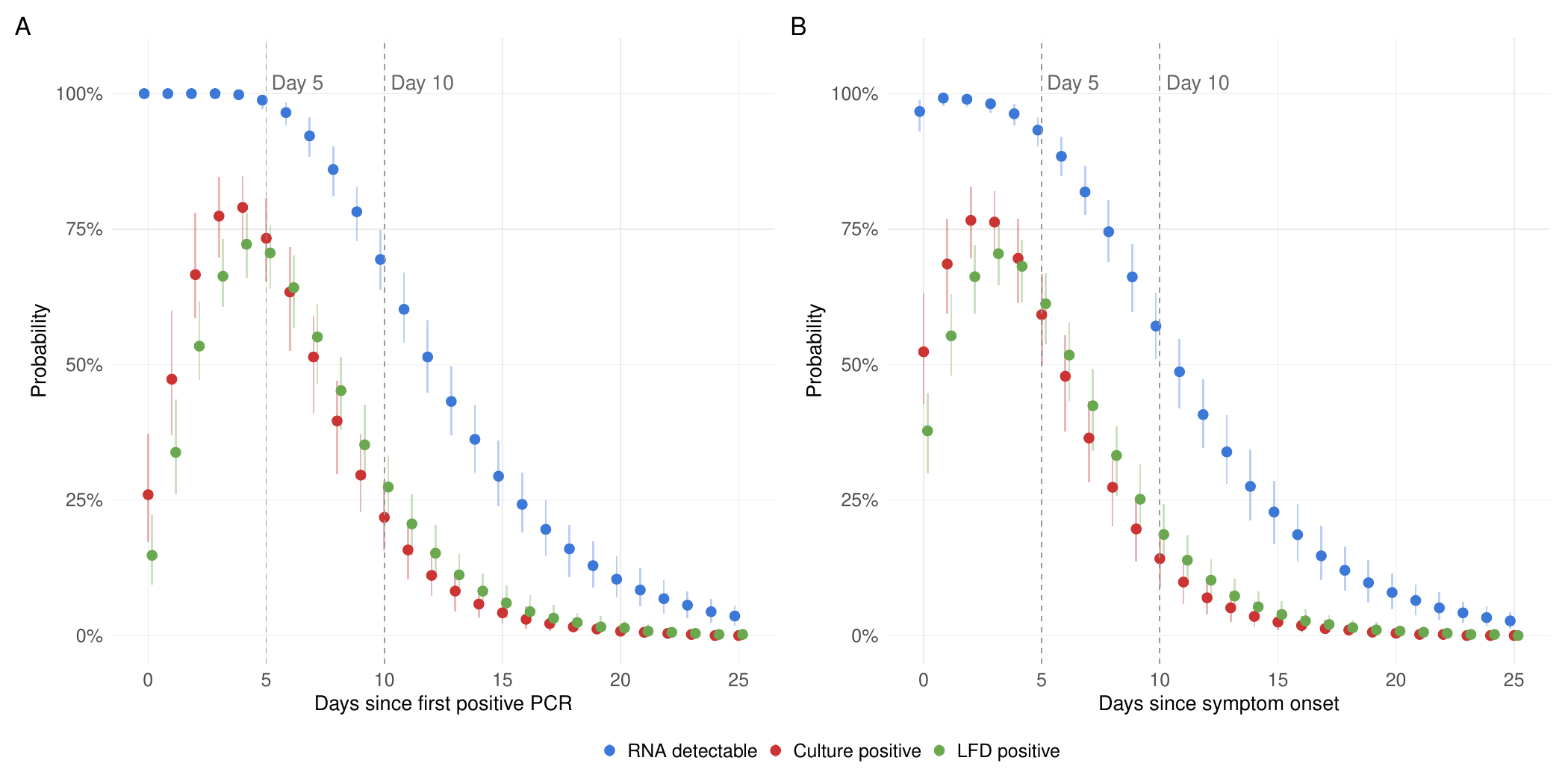}
    \caption{Posterior predictive probability of remaining culture-positive, LFD-positive, and RNA-detectable over days since a landmark event. Points: posterior medians; bars: 95\% credible intervals. \textbf{(A)}~Aligned to first positive PCR. The gap between RNA and culture positivity quantifies the false-positive window. \textbf{(B)}~Aligned to symptom onset. Conditioning on observed test results (e.g., LFD outcomes) is addressed via Bayesian filtering in Section~\ref{sec:filtering}.}
    \label{fig:probability_curves}
\end{figure}

Figure~\ref{fig:probability_curves} shows the probability of remaining culture-positive (infectious), LFD-positive, and RNA-detectable as a function of days since the landmark.

\paragraph{Time since first positive PCR (Figure~\ref{fig:probability_curves}A).} RNA detectability persists substantially longer than culture positivity, with a median delay between culture negativity and RNA negativity of approximately 5 days. This gap quantifies the ``false-positive window'' during which an individual would test positive by PCR but is no longer shedding replication-competent virus. LFD positivity tracks culture positivity more closely than RNA detectability, declining to approximately 25\% by day 10, which supports the use of antigen tests over PCR for informing isolation release decisions. By day 5 after the first positive PCR, 74.0\% (95\% CrI: 66.2--80.8\%) of individuals are still estimated to be culture-positive; by day 10, this falls to 24.1\% (17.6--31.0\%).

\paragraph{Time since symptom onset (Figure~\ref{fig:probability_curves}B).} Aligning to symptom onset rather than first positive test shifts the curves, because symptom onset typically lags peak viral load. The probability of remaining culture-positive at day 5 after symptom onset is lower than at day 5 after first positive PCR, reflecting the fact that individuals have been infectious for longer by the time symptoms emerge.

These population-averaged curves do not condition on an individual's observed test history. In Section~\ref{sec:filtering}, we show how conditioning on specific LFD results via Bayesian filtering dramatically updates the probability of ongoing infectiousness --- for example, a single negative LFD at day~5 reduces day-10 residual risk from 9.4\% to 2.7\%, while three consecutive positive LFDs increase it to 29.4\%.

\subsection{Isolation duration and residual infectiousness} \label{sec:isolation}
\begin{landscape}
\begin{figure}[p]
    \centering
    \includegraphics[width=\linewidth]{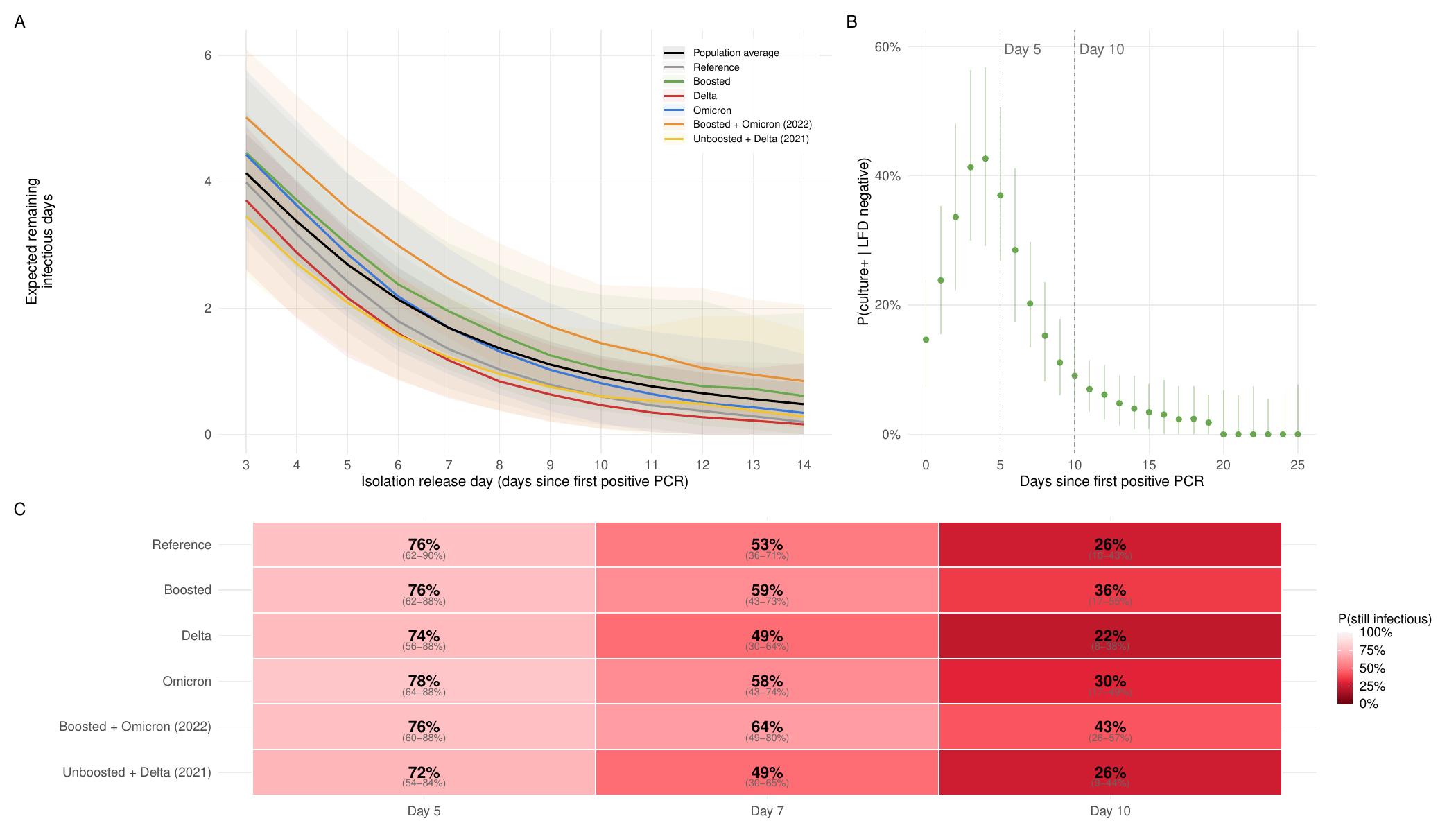}
    \caption{Policy-relevant quantities from the joint posterior predictive distribution. \textbf{(A)}~Expected remaining infectious days if released on day $d$ since first positive PCR, by covariate profile (lines: medians; ribbons: 95\% CrI). \textbf{(B)}~False reassurance rate: P(culture$+$ $|$ LFD$-$) on day $d$, showing the residual risk of test-to-release. \textbf{(C)}~Probability of remaining culture-positive at days 5, 7, and 10 by covariate profile (posterior medians with 95\% CrI).}
    \label{fig:isolation_policy}
\end{figure}
\end{landscape}

Figure~\ref{fig:isolation_policy} translates the probability curves into quantities directly relevant to isolation policy design.

\paragraph{Residual infectiousness (Figure~\ref{fig:isolation_policy}A).} For each candidate release day $d$ (days since first positive PCR), we compute the expected number of remaining days that an individual would still be shedding infectious virus if released on day $d$. This quantity captures the diminishing marginal benefit of each additional day of isolation. The marginal (population-averaged) curve shows rapid decline from day 3 to day 7, with modest additional benefit beyond day 10. Covariate stratification reveals meaningful heterogeneity: individuals with prior infection show 1.2 fewer expected residual infectious days at day 5 compared to the reference profile (1.6 vs.\ 2.8 days), consistent with the faster clearance associated with anamnestic immunity. The combined Boosted + Omicron (2022) profile shows the highest residual burden (3.7 days at day 5), reflecting the prolonged proliferation phase associated with both vaccination and Omicron that offsets the benefits of lower peak shedding.

\paragraph{Test-to-release analysis (Figure~\ref{fig:isolation_policy}B).} Among individuals who test LFD-negative on day $d$, a fraction may still be culture-positive --- representing the ``false reassurance rate'' of test-based release. This rate is computed as P(culture$+$ $|$ LFD$-$, day $d$) from the joint trajectory draws. The false reassurance rate is 33.5\% (95\% CrI: 22.9--50.7\%) at day 3, reflecting the high prevalence of infectiousness early in the course, and falls to 27.9\% (18.3--39.3\%) by day 7. This means that a test-to-release policy requiring a negative rapid test is substantially safer than fixed-duration isolation alone, particularly in the day 5--7 window where the two strategies diverge most.

\paragraph{Isolation table (Figure~\ref{fig:isolation_policy}C).} The heatmap shows the probability of remaining infectious at days 5, 7, and 10 for six representative covariate profiles. Under a fixed 5-day isolation policy, 79\% (95\% CrI: 62--90\%) of individuals in the reference category are still culture-positive at release, compared to 67\% (47--80\%) for those with prior infection. By day 10, residual infectiousness falls below 43\% across all profiles, with the Reinfection profile reaching 20\% (5--38\%). These results suggest that a combined policy --- 5 days of isolation plus a negative LFD for release --- could achieve residual risk levels comparable to 10-day fixed isolation while substantially reducing the total days of isolation.

\paragraph{Serial rapid testing by symptom onset.} To evaluate the incremental value of serial rapid testing, Figure~\ref{fig:conditional_symptom} (Appendix) shows the probability of remaining culture-positive conditional on lateral flow test outcomes, aligned to the day of symptom onset rather than first positive PCR, for three covariate profiles (Reinfection, Reference, Boosted + Omicron). In addition to the single-test conditions (LFD$+$ and LFD$-$), we compute the probability of culture positivity conditional on \emph{two consecutive} negative LFD results (i.e., both the current day and the preceding day negative). Across all three profiles, two consecutive negative results provide substantially stronger evidence against ongoing infectiousness than a single negative test, particularly in the day 5--10 window. For the Reference profile, a single negative LFD at day 7 post-symptom onset is associated with a residual culture-positive probability of approximately 25--30\%, while two consecutive negatives reduce this to below 15\%. These results provide quantitative support for serial testing strategies in test-to-release policies.

\subsection{Personalized infectiousness estimation via Bayesian filtering} \label{sec:filtering}

The probability curves in Sections~\ref{sec:policy}--\ref{sec:isolation} are population-averaged: they reflect the marginal posterior predictive distribution over all possible trajectories for a new individual drawn from the fitted hierarchical distribution. In clinical practice, however, a clinician typically has access to a specific patient's test history --- a PCR Ct value at diagnosis, followed by one or more lateral flow results --- which substantially narrows the set of plausible trajectories. To bridge this gap, we implement importance-sampling-based Bayesian filtering~\cite{gelman2013bayesian} that conditions the population of posterior predictive trajectories on an observed test history, reweighting each trajectory by the likelihood of the observed data under the fitted observation models.

Formally, let $\mathbf{D}^{\text{obs}} = (D^{\text{obs}}_1, \ldots, D^{\text{obs}}_K)$ denote a sequence of test results at times $t_1, \ldots, t_K$ relative to diagnosis, and let $(V_t^{(m)}, R_t^{(m)})$ denote the latent viral trajectories of the $m$-th posterior predictive draw. The importance weight for trajectory $m$ is
\begin{equation}
    w_m = \prod_{k=1}^K f(D^{\text{obs}}_k \mid V_{t_k}^{(m)}, R_{t_k}^{(m)}; \theta),
\end{equation}
where the likelihood terms use the same observation models as the main fit (Normal with LOD censoring for RNA; Bernoulli for LFD). The filtered probability of remaining infectious at time $t$ is then $\hat{P}(V_t > \text{LOD} \mid \mathbf{D}^{\text{obs}}) = \sum_m \tilde{w}_m \cdot \mathbf{1}[V_t^{(m)} > \text{LOD}]$, where $\tilde{w}_m = w_m / \sum_j w_j$ are the self-normalized importance weights.

\begin{landscape}
\begin{figure}[p]
    \centering
    \includegraphics[width=\linewidth]{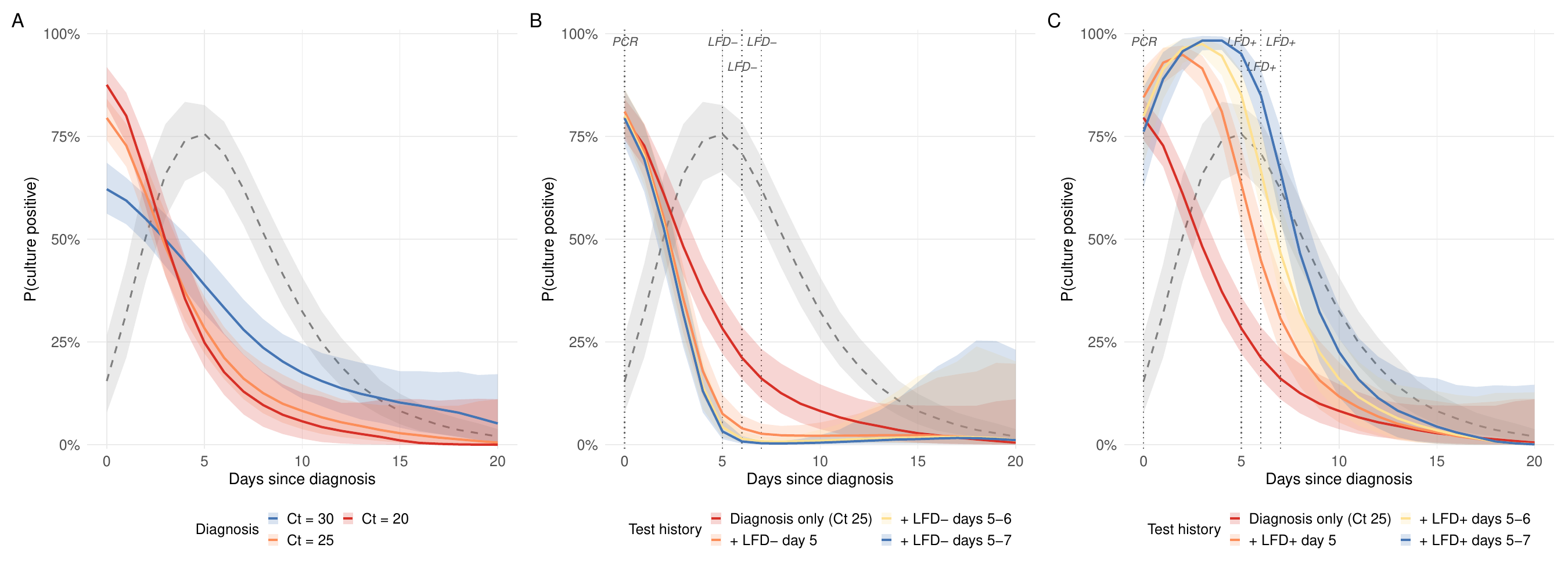}
    \caption{Personalized probability of remaining culture-positive, conditioned on observed test history via importance-sampling-based Bayesian filtering (100{,}000 posterior predictive trajectories). Grey dashed line and band: population-averaged baseline (median and 95\% CrI). \textbf{(A)}~Conditioning on a single PCR result at diagnosis with Ct values of 20 (red), 25 (orange), or 30 (blue). Lower Ct (higher viral load) corresponds to higher initial probability of infectiousness but faster decline, while higher Ct results in a longer tail of residual risk. \textbf{(B)}~Sequential updating with negative LFD results for a patient diagnosed at Ct~25. Dotted vertical lines mark the timing of each conditioning event (PCR at diagnosis; negative LFD on days 5, 6, 7). \textbf{(C)}~Sequential updating with positive LFD results.}
    \label{fig:bayesian_filtering}
\end{figure}
\end{landscape}

We generate 100{,}000 posterior predictive trajectories (200 posterior draws $\times$ 500 replicates) for a reference agent profile (Omicron, boosted, age 30--49, first infection) and apply importance-sampling filters for nine scenarios organized into three groups (Figure~\ref{fig:bayesian_filtering}).

\paragraph{Effect of Ct value at diagnosis (Figure~\ref{fig:bayesian_filtering}A).} Conditioning on a single PCR result at diagnosis dramatically narrows the distribution of plausible trajectories compared to the population baseline. At diagnosis (day 0), the posterior probability of being culture-positive is 86\% for Ct~20, 78\% for Ct~25, and 61\% for Ct~30, compared to 15\% for a randomly drawn individual from the population. The ordering reverses at later time points: by day 10, the culture-positive probability is 6.9\% for Ct~20 but 19.0\% for Ct~30. This crossover reflects the model's inference that a low Ct (high viral load) at diagnosis indicates the individual is near peak shedding and will decline rapidly, whereas a high Ct is ambiguous --- it could reflect early proliferation, late clearance, or a lower-amplitude trajectory --- leading to a longer tail of uncertainty.

\paragraph{Sequential LFD updating (Figure~\ref{fig:bayesian_filtering}B).} Starting from a Ct~25 diagnosis, each additional negative LFD result provides substantial new information (vertical dotted lines mark conditioning events). A negative LFD on day 5 reduces the day-10 culture-positive probability from 9.4\% to 2.7\%; adding a second negative on day 6 reduces it further to 1.1\%; and three consecutive negatives (days 5--7) yield a residual risk of 0.6\%. The effective sample sizes remain healthy (median ESS 197--489 across scenarios), confirming that the importance weights are not dominated by a small number of trajectories. These results demonstrate that the joint model's observation models --- connecting the latent trajectory to both PCR and LFD observations --- enable coherent fusion of heterogeneous test data into a single personalized infectiousness estimate.

\paragraph{Positive LFD confirmation (Figure~\ref{fig:bayesian_filtering}C).} Conversely, positive LFD results on days 5--7 select for trajectories with higher and more sustained infectiousness, shifting the posterior upward relative to diagnosis alone. A positive LFD on day 5 increases the day-5 probability from 31.4\% (diagnosis only) to 68.0\%; two consecutive positives (days 5--6) increase it to 88.5\%; and three (days 5--7) to 96.3\%. By day 10, three consecutive positive LFDs increase the residual risk from 9.4\% to 29.4\%, a three-fold increase. The effective sample sizes are lower for the positive-LFD scenarios (median ESS 95--232) than for the corresponding negative scenarios, reflecting the fact that positive LFDs select for longer-tailed trajectories that represent a smaller fraction of the prior predictive pool. Together, Panels B and C illustrate the bidirectional informativeness of rapid antigen tests: negative results rapidly exclude ongoing infectiousness, while positive results confirm it, with each additional test providing diminishing but still meaningful incremental information.

\section{Discussion} \label{sec:discussion}

We have developed a Bayesian joint model for the within-host viral kinetics of SARS-CoV-2 that integrates longitudinal data on viral RNA, infectious virus, lateral flow test results, and symptom onset from approximately 2{,}000 infections across five prospective cohorts. By modeling the joint distribution of these proxies conditional on a shared latent trajectory of infectiousness, the model provides a coherent framework for inferring unobserved or difficult-to-measure markers of viral shedding from routinely available data.

\paragraph{Summary of findings.} The population-level viral kinetics exhibit the characteristic asymmetric profile of acute respiratory viral infections, with rapid proliferation (median 5.3 days) followed by slower clearance (median 13.7 days). The peak RNA concentration (median 16.2 log copies/mL) is consistent with previous estimates from densely sampled cohorts~\cite{ke2022daily,hay2022quantifying}. The log-affine transformation from RNA to infectious virus reveals sub-proportional clearance elasticity ($a_{1,\omega_r} = 0.66$), providing quantitative support for the well-known observation that infectious virus clears faster than viral RNA~\cite{cevik2021sars,wolk2020sars}. Covariate effects are broadly consistent with the existing literature~\cite{puhach2022infectious,kissler2021viral}: Delta is associated with higher peak shedding, vaccination reduces peak load while prolonging proliferation, and prior infection accelerates clearance. The correlated individual random effects reveal interpretable structure, with negative correlations between proliferation rate and both peak timing and clearance rate. The policy analysis demonstrates that a test-to-release strategy using LFD results reduces the estimated residual infectiousness at release compared to fixed-duration isolation alone, and that vaccination status and variant substantially modulate the optimal isolation duration. Furthermore, importance-sampling-based Bayesian filtering demonstrates that conditioning on a patient's observed test history --- even a single PCR result --- dramatically narrows the uncertainty in residual infectiousness, and that serial rapid tests provide bidirectional information: negative LFDs can reduce the estimated probability of ongoing culture positivity to below 1\%, while positive LFDs confirm sustained infectiousness and increase the estimated risk three-fold. These derived quantities illustrate the practical value of the joint posterior: rather than treating each assay in isolation, the model enables coherent, uncertainty-quantified predictions of clinically relevant outcomes, both at the population level and for individual patients.

\paragraph{Methodological contributions.} The primary methodological contribution is the formulation of a generative model that jointly characterizes multiple proxies of viral shedding within a single Bayesian framework. Previous approaches have typically modeled individual proxies in isolation or used simplified linkage functions. Our approach allows each observation type to borrow strength from the others through the shared latent trajectory, which is particularly valuable when individual-level data are sparse or heterogeneous. The non-centered parameterization of the hierarchical random effects, combined with the smooth trajectory approximation and the Cholesky factorization of the correlation matrix, enables efficient posterior sampling for a model with over 10{,}000 latent parameters.

The model's factorization into an infectious virus trajectory, a derived RNA trajectory, a symptom onset hazard, and observation models is biologically motivated and facilitates interpretation: each component maps onto a distinct aspect of within-host viral dynamics or the measurement process. The log-affine transformation linking RNA to PFU parameters provides a parsimonious yet flexible specification that nests proportional models and permits estimation of the elasticity of each kinetic parameter.

\paragraph{Limitations.} Several limitations should be noted. First, the model assumes that the trajectories of viral RNA and infectious virus follow the same functional form (smooth piecewise exponential), differing only in their parameters. While this approximation captures the dominant features of the kinetic profile, it may not adequately represent more complex dynamics such as biphasic clearance or viral rebound. Second, the PFU individual random effects are restricted to the 275 individuals with viral culture data, and the limited number of culture observations per individual constrains the precision of the RNA-to-PFU transformation parameters. Third, the model does not currently include covariate effects on the PFU transformation parameters or the LFD observation model; these extensions are fully implemented in the Stan code but were not activated in the current fit to limit model complexity. Fourth, the symptom onset model is limited by the availability and consistency of symptom diary data across cohorts, with some cohorts providing only aggregate onset times rather than daily diaries, and ``symptom onset'' itself is variably defined across studies (first any symptom, first respiratory symptom, or patient-reported onset), introducing heterogeneity that the individual random effect can accommodate but not eliminate. Fifth, while Section~\ref{sec:filtering} demonstrates importance-sampling-based Bayesian filtering for personalized infectiousness estimation, this approach reweights a fixed set of trajectories rather than performing fully online sequential updating; for patients with many test results or highly informative observations, sequential Monte Carlo methods~\cite{doucet2000sequential} could provide more efficient inference. Sixth, the current model treats each observation as a single RNA measurement, averaging across anatomical sites (nose, throat, saliva) where site-specific specimens were collected (HCT, UIUC); this discards site-specific kinetic information (see Supplementary Section~\ref{sec:supp_site} for a proposed extension). Seventh, and more broadly, all cohorts in this analysis rely on upper respiratory tract swabs, which may not fully capture the viral shedding relevant to transmission via aerosols or exhaled breath. Emerging sampling approaches such as aerosol or breath-based collection devices may provide different and potentially more transmission-relevant measures of infectiousness; incorporating such data into the joint model framework is a natural extension as these datasets become available.

\paragraph{Extensions and future work.} Several natural extensions emerge from this work. First, the model can generate plausible viral kinetic trajectories for synthetic populations, enabling agent-based transmission modeling that incorporates realistic within-host dynamics and heterogeneity. Second, the model could be extended to include viral rebound dynamics, time-varying covariates (e.g., antiviral treatment), or additional pathogen targets beyond SARS-CoV-2. Third, the model could be extended to account for anatomical site-specific viral kinetics and the dilution physics of swab pooling, leveraging the site-specific nasal and throat measurements available from the HCT and UIUC cohorts; the mathematical framework for this extension is developed in Supplementary Section~\ref{sec:supp_site}. Fourth, the symptom onset model currently treats onset as a binary event; richer symptom data --- such as daily severity scores, individual symptom profiles, or symptom-specific onset times --- could support more detailed modeling of the relationship between viral dynamics and symptom presentation~\cite{hakki2026symptoms}. Finally, correlated PFU individual random effects (mirroring the RNA correlation structure) could be added to capture dependencies among the infectious virus trajectory parameters, though this requires sufficient culture data to identify the additional parameters.

\paragraph{Conclusion.} Accurate characterization of within-host viral dynamics is essential for informing public health policy during infectious disease outbreaks. By jointly modeling multiple proxies of viral shedding, our approach provides a principled framework for inferring infectiousness from routinely available data. The model's ability to predict individual-level trajectories and to estimate the probability of ongoing infectiousness at any point during infection has direct applications to isolation and testing policy. The Bayesian framework naturally quantifies uncertainty in these predictions, providing decision-makers with probabilistic guidance rather than point estimates.

\printbibliography

\clearpage
\newcommand{\suppprefix}{S}
\newcommand{\suppfigpath}{figures}
\setcounter{page}{0}

\begin{appendix}

    \renewcommand{\thefigure}{\suppprefix\arabic{figure}}
    \setcounter{figure}{0}

    \renewcommand{\thetable}{\suppprefix\arabic{table}}
    \setcounter{table}{0}

    \renewcommand{\theequation}{\suppprefix\arabic{equation}}
    \setcounter{equation}{0}

    \renewcommand{\thesection}{\suppprefix\arabic{section}}
    \setcounter{section}{0}

    \newpage
    \begin{center}
        {\Large \textbf{Supplementary Materials}} \\[0.5em]
        {\large Inferring infectiousness: a joint model of the within-host viral kinetics of SARS-CoV-2} \\[0.5em]
        \today
    \end{center}
    \vspace{1em}

    \noindent This supplement contains prior predictive checks (Section~\ref{sec:supp_prior}), a comparison of the semi-mechanistic trajectory model to the target cell limited ODE (Section~\ref{sec:supp_tcl}), a description of the flat-top trajectory extension (Section~\ref{sec:supp_flattop}), convergence diagnostics (Section~\ref{sec:supp_convergence}), individual-level trajectory fits for each cohort (Section~\ref{sec:supp_fits}), posterior predictive checks (Section~\ref{sec:supp_ppc}), the parameter recovery simulation study (Section~\ref{sec:supp_recovery}), additional parameter estimates including covariate effects, the RNA-to-PFU transformation, and the individual random effect correlation structure (Section~\ref{sec:supp_params}), and covariate-stratified probability curves for testing and isolation policy (Section~\ref{sec:supp_strat}).

    \section{Prior predictive checks} \label{sec:supp_prior}

\begin{figure}[p]
        \centering
        \includegraphics[width=\linewidth]{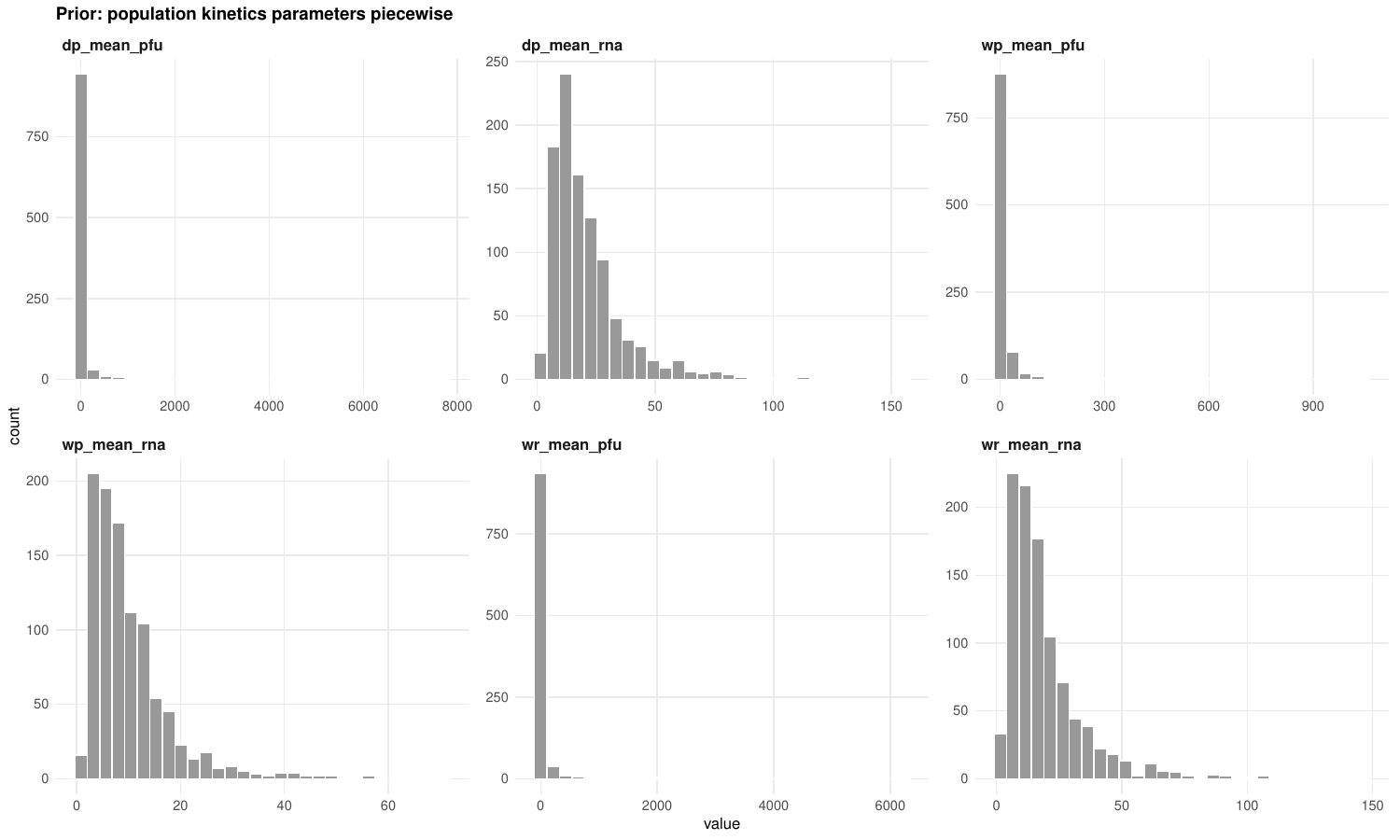}
        \caption[Prior predictive: RNA trajectory parameters]{Prior predictive check for the piecewise exponential trajectory parameters. Each panel shows 500 draws from the marginal prior distribution of a single kinetic parameter (peak height $\delta$, proliferation duration $\omega_p$, clearance duration $\omega_r$, peak timing $t_p$), illustrating the range of values admitted by the prior before observing data. Gray shading indicates the approximate range observed in the literature.}
        \label{fig:prior_pe}
    \end{figure}

\begin{figure}[p]
        \centering
        \includegraphics[width=\linewidth]{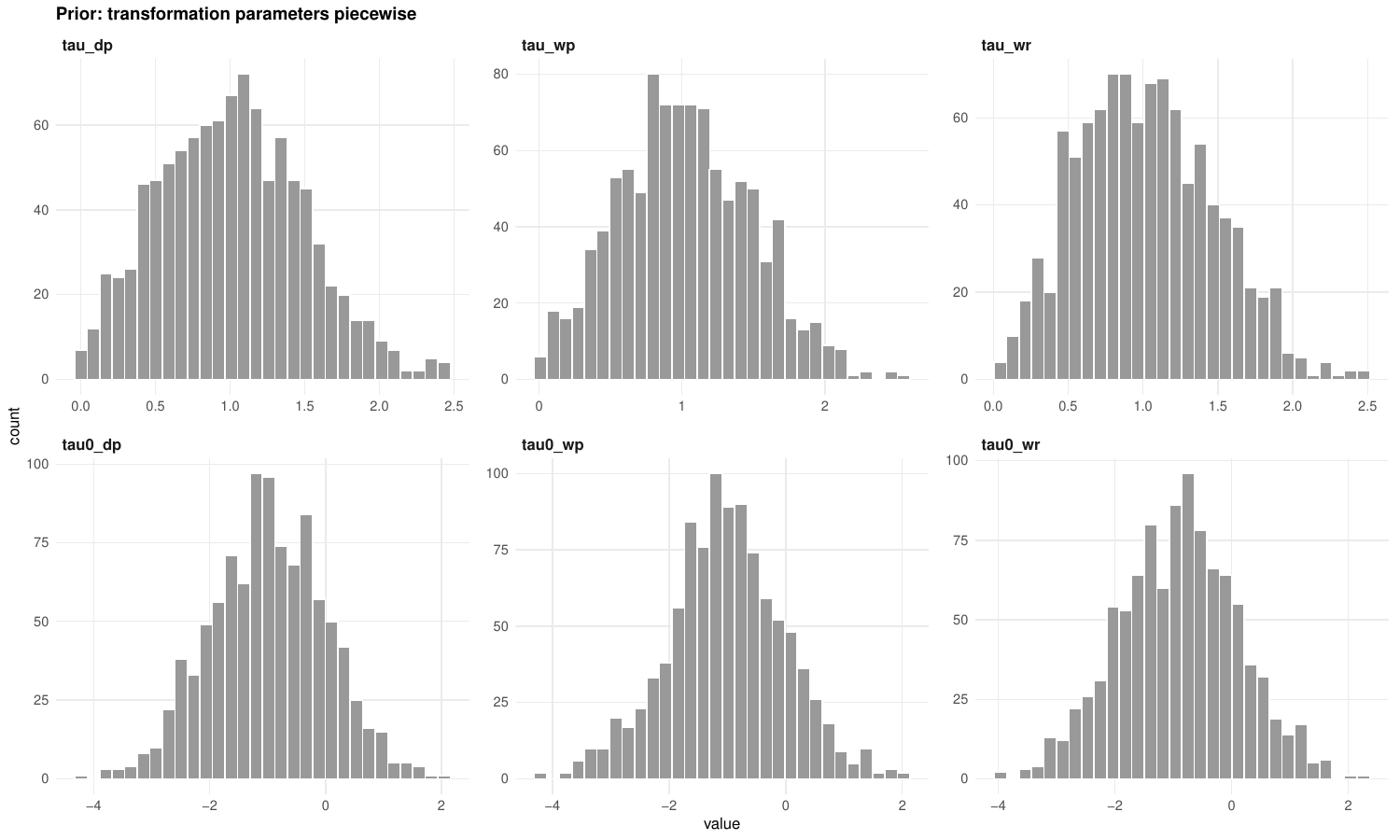}
        \caption[Prior predictive: RNA-to-PFU transformation]{Prior predictive check for the log-affine RNA-to-PFU transformation parameters. Each panel shows draws from the joint prior of the intercept ($a_0$) and elasticity ($a_1$) parameters linking infectious virus trajectory parameters to viral RNA trajectory parameters. The elasticity prior is centered at 1 (proportional scaling) with positivity enforced by truncation.}
        \label{fig:prior_trans}
    \end{figure}

\begin{figure}[p]
        \centering
        \includegraphics[width=\linewidth]{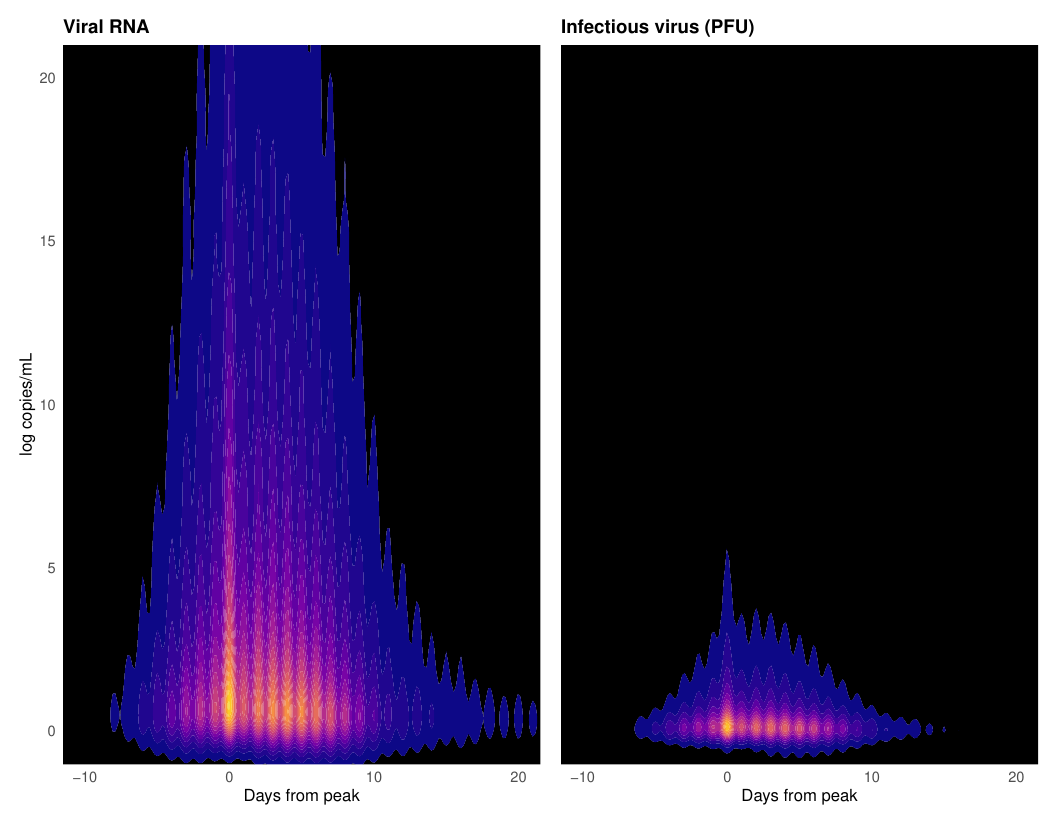}
        \caption[Prior predictive: trajectory draws]{Prior predictive trajectories. Two-dimensional kernel density estimate of simulated viral trajectories drawn from the joint prior distribution of all kinetic parameters, including population-level parameters, individual random effects, and the log-affine RNA-to-PFU transformation. Left panel: viral RNA; right panel: infectious virus (PFU). The y-axis is constrained to 0--25 log copies/mL to match the observed data range. The density illustrates the range and concentration of kinetic profiles implied by the prior before conditioning on data.}
        \label{fig:prior_traj}
    \end{figure}

\begin{figure}[p]
        \centering
        \includegraphics[width=\linewidth]{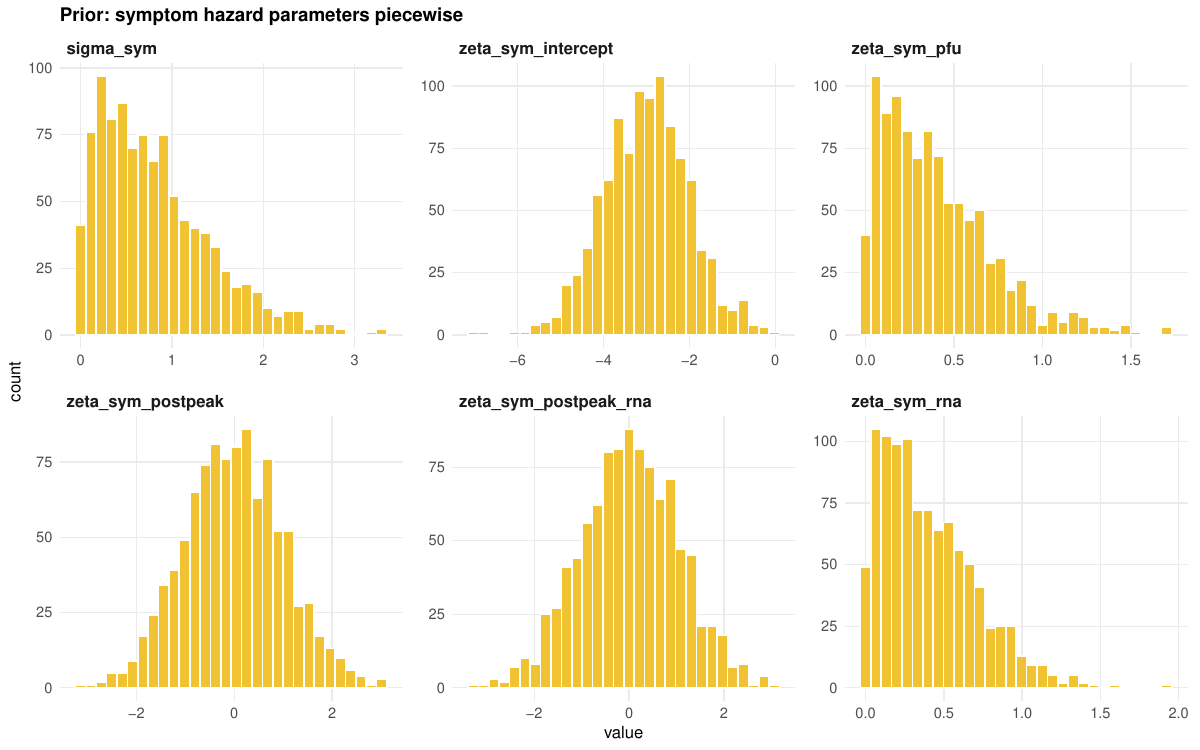}
        \caption[Prior predictive: symptom onset model]{Prior predictive check for the discrete-time cloglog symptom onset model. Prior draws of the daily hazard of symptom onset as a function of days since infection, illustrating the range of onset timing and cumulative incidence curves consistent with the prior specification. The prior intercept is centered at $\zeta_0 = -3$ (corresponding to a ${\sim}5\%$ baseline daily hazard), with positive viral load effects $\zeta_1, \zeta_2 \sim N^+(0, 0.5)$ and post-peak sigmoid coefficients $\zeta_3, \zeta_4 \sim N(0, 1)$.}
        \label{fig:prior_sym}
    \end{figure}

    Figures~\ref{fig:prior_pe}--\ref{fig:prior_sym} display draws from the joint prior distribution of model parameters before conditioning on data, illustrating the range of trajectories and derived quantities consistent with the prior specification described in Section~\ref{sec:priors} of the main text.

    \section{Comparison with target cell limited model} \label{sec:supp_tcl}

\begin{figure}[p]
        \centering
        \includegraphics[width=\linewidth]{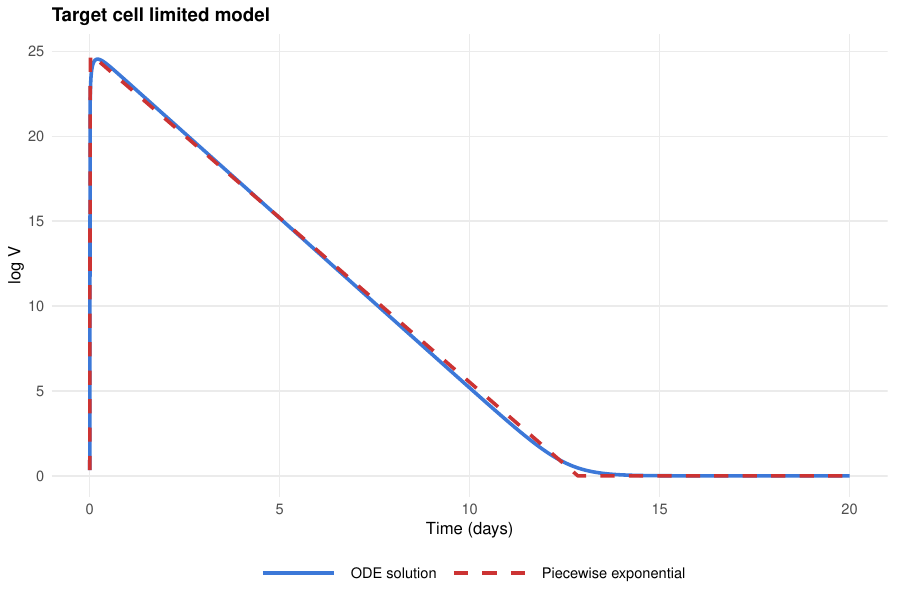}
        \caption[TCL comparison: standard model]{Comparison between the standard target cell limited ODE solution (solid blue line; $\beta = 2.4 \times 10^{-5}$, $\delta = 2.0$, $\pi = 1700$, $c = 10$, $T_0 = 4 \times 10^8$, $V_0 = 0.4$) and a smooth piecewise exponential approximation fitted by least squares (dashed red line). The piecewise model closely tracks the exponential proliferation and clearance phases; the main discrepancy is a slight geometric mismatch near the rounded ODE peak versus the smooth log-sum-exp cap of the piecewise model.}
        \label{fig:tcl_1}
    \end{figure}

\begin{figure}[p]
        \centering
        \includegraphics[width=\linewidth]{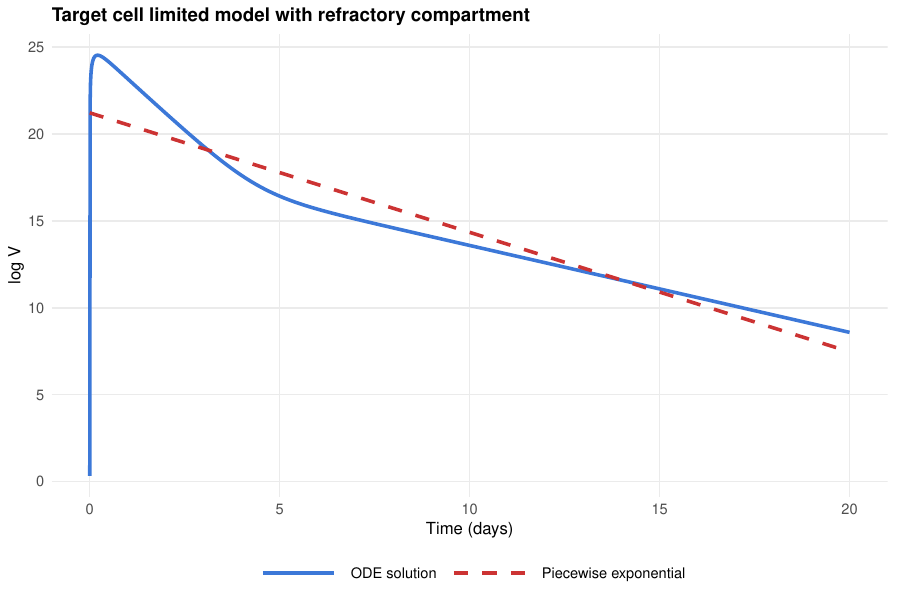}
        \caption[TCL comparison: refractory compartment]{As Figure~\ref{fig:tcl_1}, but for an extended TCL model that includes a refractory cell compartment ($\phi = 1$, $\rho = 0.5$; all other parameters identical). The innate immune response modeled by the refractory state produces a blunted, asymmetric trajectory with a faster initial decline followed by a prolonged tail. The piecewise exponential approximation captures the dominant features of this more complex ODE solution, supporting its use as a tractable reduced-form model even when the underlying biology includes additional compartments.}
        \label{fig:tcl_2}
    \end{figure}

    The semi-mechanistic piecewise exponential trajectory model (Section~3 of the main text) is motivated by the phenomenological similarity between its trajectory shape and the numerical solution of the target cell limited (TCL) ODE. Figures~\ref{fig:tcl_1}--\ref{fig:tcl_2} illustrate this correspondence by overlaying a least-squares--fitted smooth piecewise exponential approximation (Equation~3 of the main text) on the ODE solution for two model configurations, demonstrating that the semi-mechanistic model captures the salient features---exponential growth, peak, and exponential decline---while directly parameterizing the quantities of clinical interest.

    \section{Flat-top trajectory extension} \label{sec:supp_flattop}

    The smooth trajectory function used in the main text (Section~3.1) assumes that the viral load profile transitions immediately from proliferation to clearance at the peak. In reality, some individuals may exhibit an approximate plateau near peak viral load, during which infectious virus production and immune-mediated clearance are roughly balanced. To accommodate this possibility, we consider a generalized smooth trajectory that introduces a \emph{flat-top duration} parameter $\omega_f \geq 0$:
    \begin{equation}
        g_s(t; \theta) = \delta + \log \frac{a + b}{b \exp(-a(t - t_p)) + a \exp(b(t - (t_p + \omega_f)))} \label{eq:flattop}
    \end{equation}
    where $a = \delta / \omega_p$ and $b = \delta / \omega_r$ are the proliferation and clearance rates, as in the main text. When $\omega_f = 0$, this reduces to the standard smooth two-arm envelope used in our primary analysis. For $\omega_f > 0$, the rising and falling exponential arms are separated by $\omega_f$ days, producing an approximate plateau of width $\omega_f$ around the peak.

    However, when $\omega_f > 0$, the log-sum-exp envelope can overshoot the peak value $\delta$---an inherent artifact of merging two separated exponential arms. To prevent this, we apply a soft cap:
    \begin{equation}
        \tilde{g}_s(t; \theta) = g_s(t; \theta) - \frac{1}{\kappa}\log\!\left(1 + \exp\!\left(\kappa \left(g_s(t; \theta) - \delta\right)\right)\right) \label{eq:softcap}
    \end{equation}
    where $\kappa = 10$ controls the sharpness of the cap. This ensures $\tilde{g}_s \leq \delta$ everywhere while maintaining smooth gradients.

    When the flat-top extension is used, $\omega_f$ is shared between the RNA and PFU trajectories, and the parameter vector becomes $\theta = (\delta, \omega_p, \omega_r, t_p, \omega_f)$ for RNA and $\theta' = (\delta', \omega'_p, \omega'_r, t'_p, \omega_f)$ for PFU. Individual and source-level effects on $\omega_f$ are modeled independently (not included in the correlated random effect block) to keep the dimensionality of the Cholesky factor manageable.

    \paragraph{Model comparison.} Our implementation includes a compile-time toggle (\texttt{use\_wf}) that enables or disables the flat-top extension. In preliminary fits with $\omega_f$ enabled, the posterior for $\omega_f$ concentrated near zero, indicating that the data do not support a plateau phase beyond what is already captured by the smooth peak geometry. Formally, the model with $\omega_f = 0$ (the specification used in the main text) was preferred on the basis of WAIC, and the remaining parameter estimates were substantively unchanged. We therefore use the simpler specification without the flat-top parameter throughout the main analysis.

    \section{Convergence diagnostics} \label{sec:supp_convergence}

    \begin{figure}[p]
        \centering
        \includegraphics[width=\linewidth]{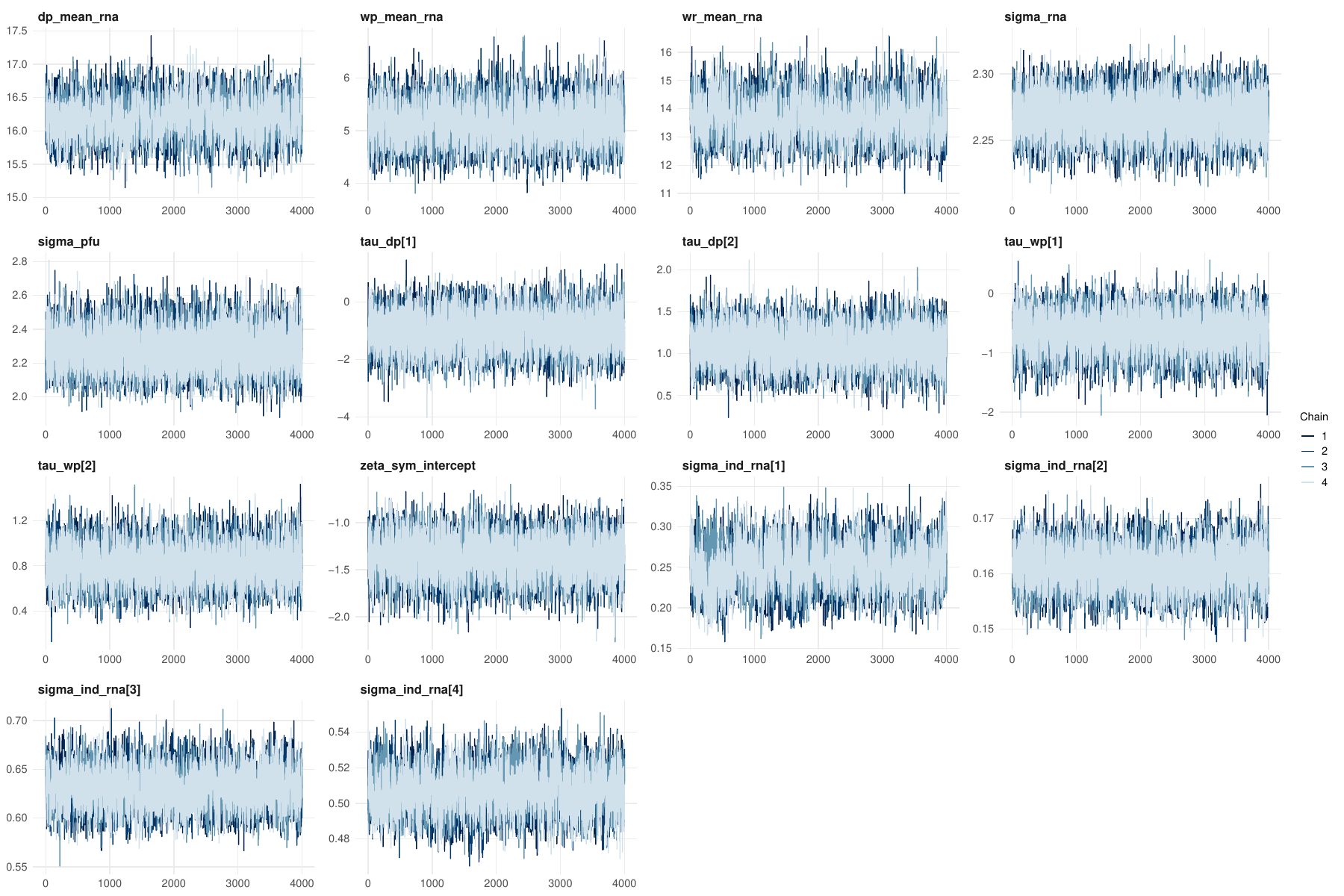}
        \caption[MCMC trace plots]{MCMC trace plots for key population-level parameters across all four chains. Each panel shows the sampled values for one parameter as a function of iteration number, with chains distinguished by color. All parameters show good mixing with no visible trends, multimodality, or stuck chains, consistent with the split-$\hat{R}$ values reported in Section~\ref{sec:model_checking} of the main text.}
        \label{fig:trace}
    \end{figure}

    \section{Individual trajectory fits} \label{sec:supp_fits}

    Figures~\ref{fig:fit_ataccc}--\ref{fig:fit_legacy} show individual-level trajectory fits for all participants in each of the five cohorts. For each individual, the posterior median viral RNA trajectory (blue line) with 95\% credible interval (blue ribbon) is overlaid on observed qPCR-derived log RNA concentrations (blue points). Where available, the posterior median infectious virus (PFU) trajectory (red line) with 95\% credible interval (red ribbon) is shown with observed culture-based measurements (red points). Predicted lateral flow device (LFD) positivity probabilities are displayed as shaded tiles (green = positive, gray = negative), and observed symptom onset is indicated by markers. The limit of detection for each assay is shown as a horizontal dashed line.

    \begin{landscape}
        \begin{figure}[p]
            \centering
            \includegraphics[width=\linewidth]{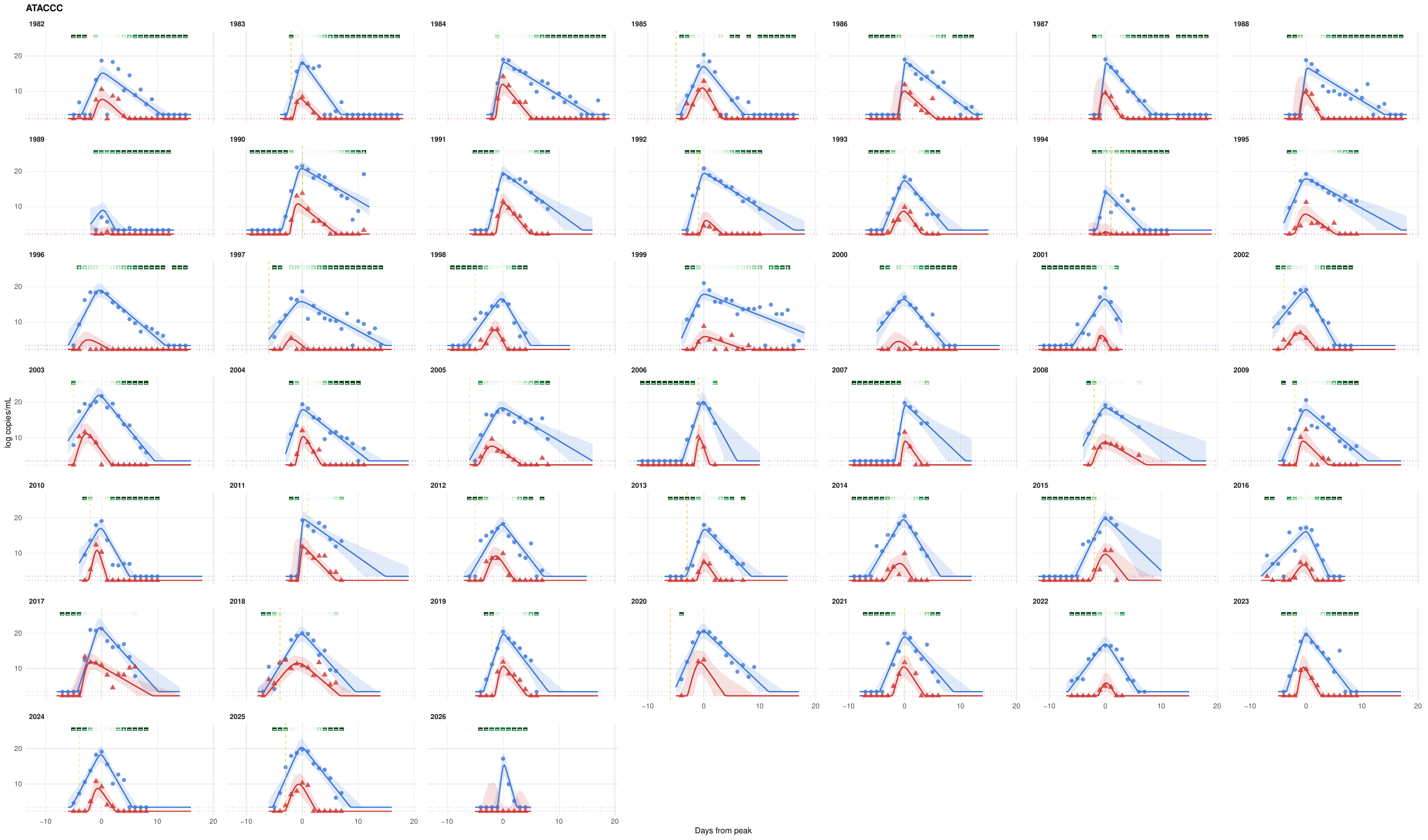}
            \caption[ATACCC individual fits]{Individual trajectory fits for all 57 ATACCC participants. This cohort provides the most complete multi-modal data, with paired qPCR, viral culture, LFD, and symptom diary observations at each time point, enabling direct comparison of all four observation models within each individual.}
            \label{fig:fit_ataccc}
        \end{figure}

        \begin{figure}[p]
            \centering
            \includegraphics[width=\linewidth]{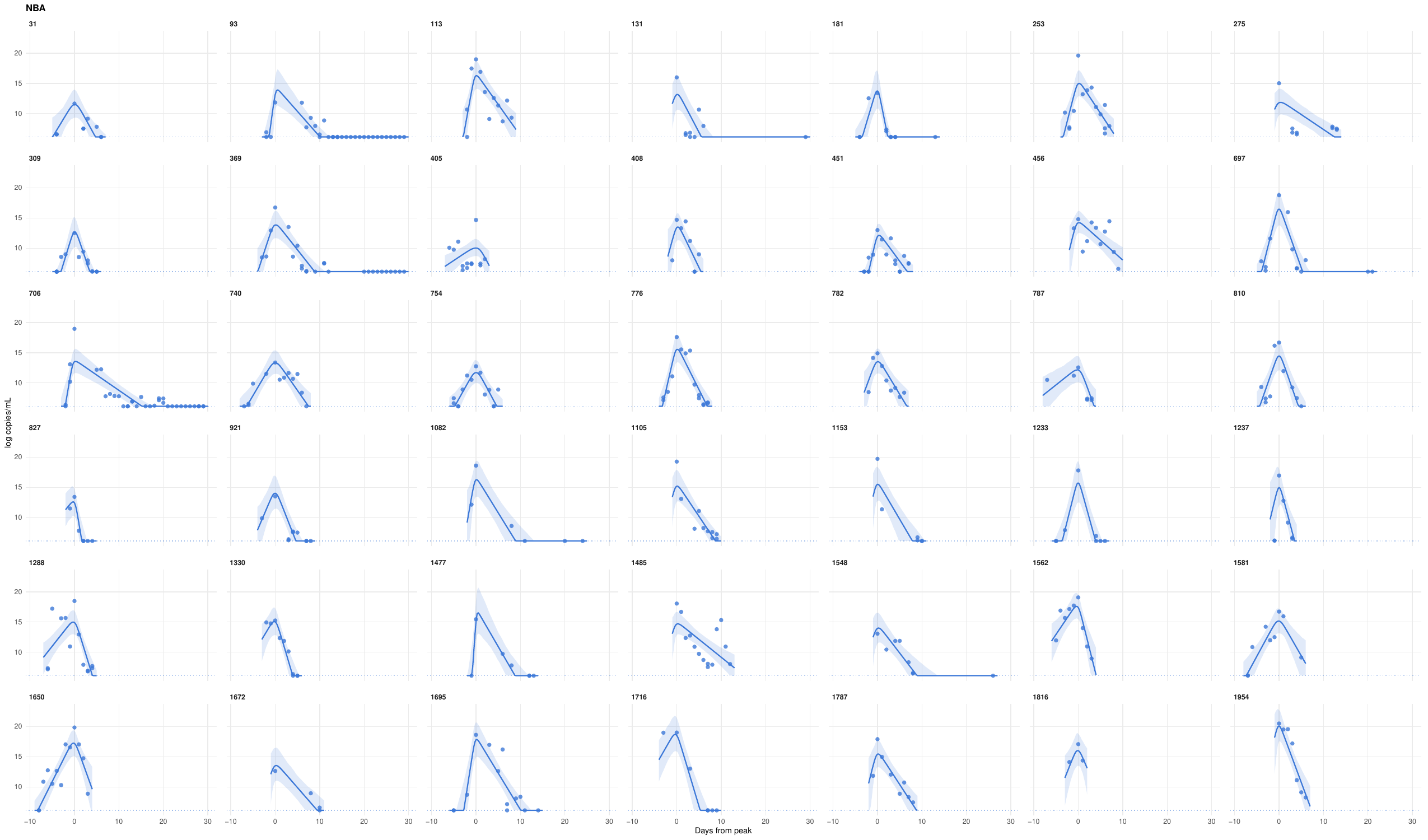}
            \caption[NBA individual fits]{Individual trajectory fits for selected NBA participants. This cohort contributes only qPCR data; the infectious virus trajectory (red) is inferred entirely through the hierarchical RNA-to-PFU transformation and the population prior on PFU kinetics, illustrating the model's ability to impute unobserved infectious virus dynamics from RNA data alone.}
            \label{fig:fit_nba}
        \end{figure}

        \begin{figure}[p]
            \centering
            \includegraphics[width=\linewidth]{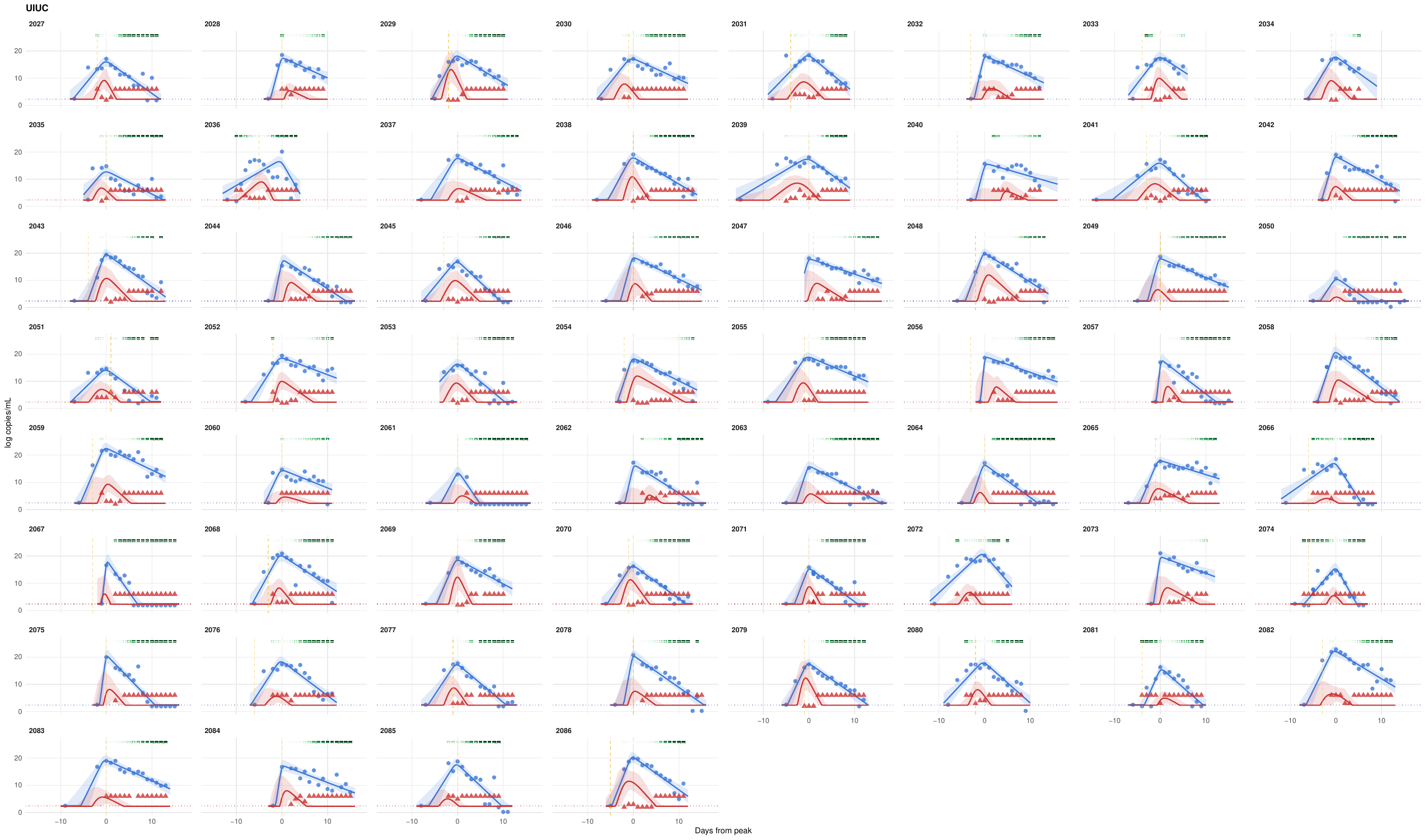}
            \caption[UIUC individual fits]{Individual trajectory fits for all 60 UIUC participants. Like ATACCC, this cohort has multi-modal data (qPCR, TCID$_{50}$ culture, LFD, symptoms), with the TCID$_{50}$ culture results modeled via the mechanistic interval-censored normal observation model described in Section~3.4 of the main text.}
            \label{fig:fit_uiuc}
        \end{figure}

        \begin{figure}[p]
            \centering
            \includegraphics[width=\linewidth]{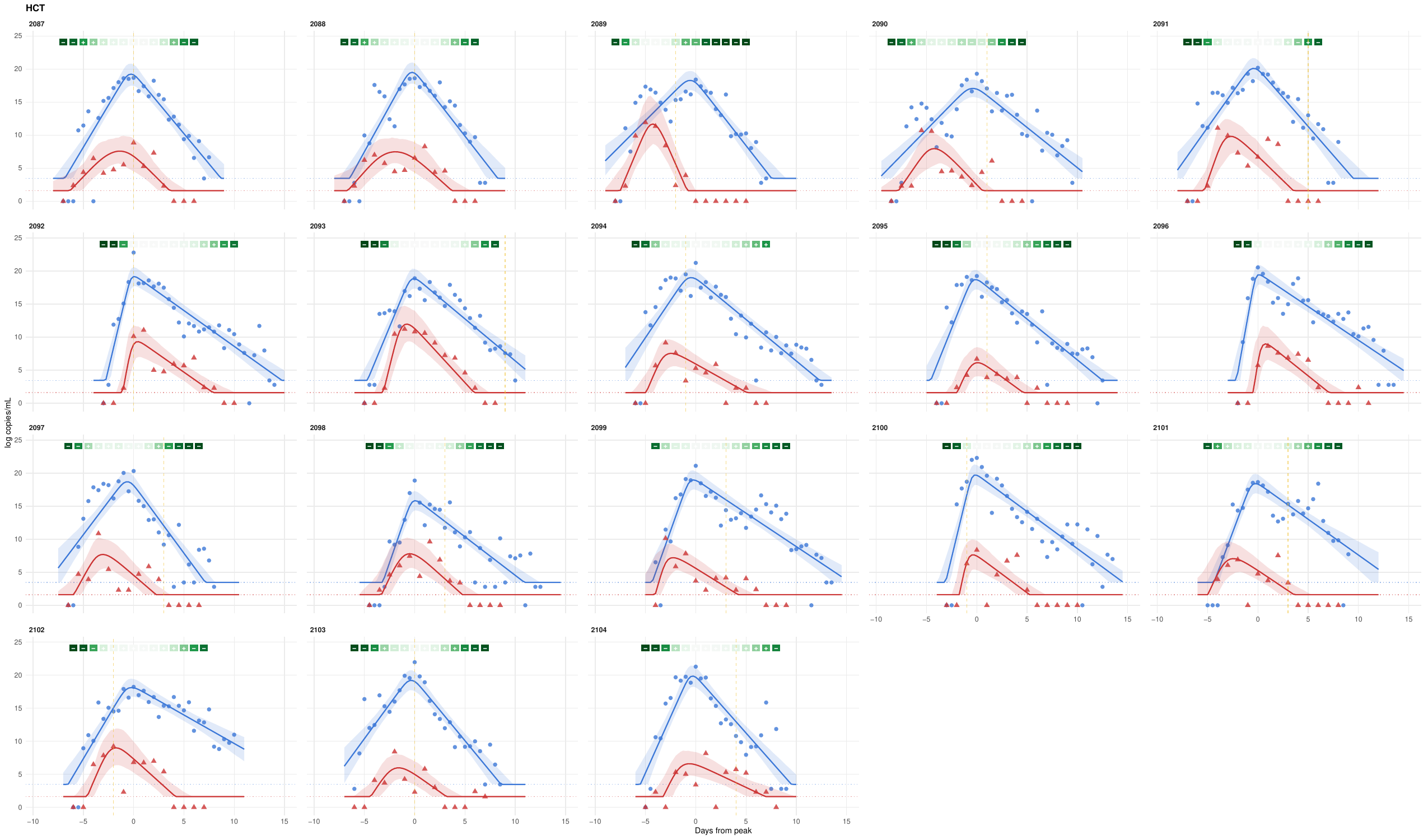}
            \caption[HCT individual fits]{Individual trajectory fits for all 18 Human Challenge Trial participants. The known inoculation time ($t=0$) provides a fixed reference point absent in observational cohorts, enabling precise estimation of the proliferation phase duration and timing. PFU and FFA culture measurements are available for all time points.}
            \label{fig:fit_hct}
        \end{figure}

        \begin{figure}[p]
            \centering
            \includegraphics[width=\linewidth]{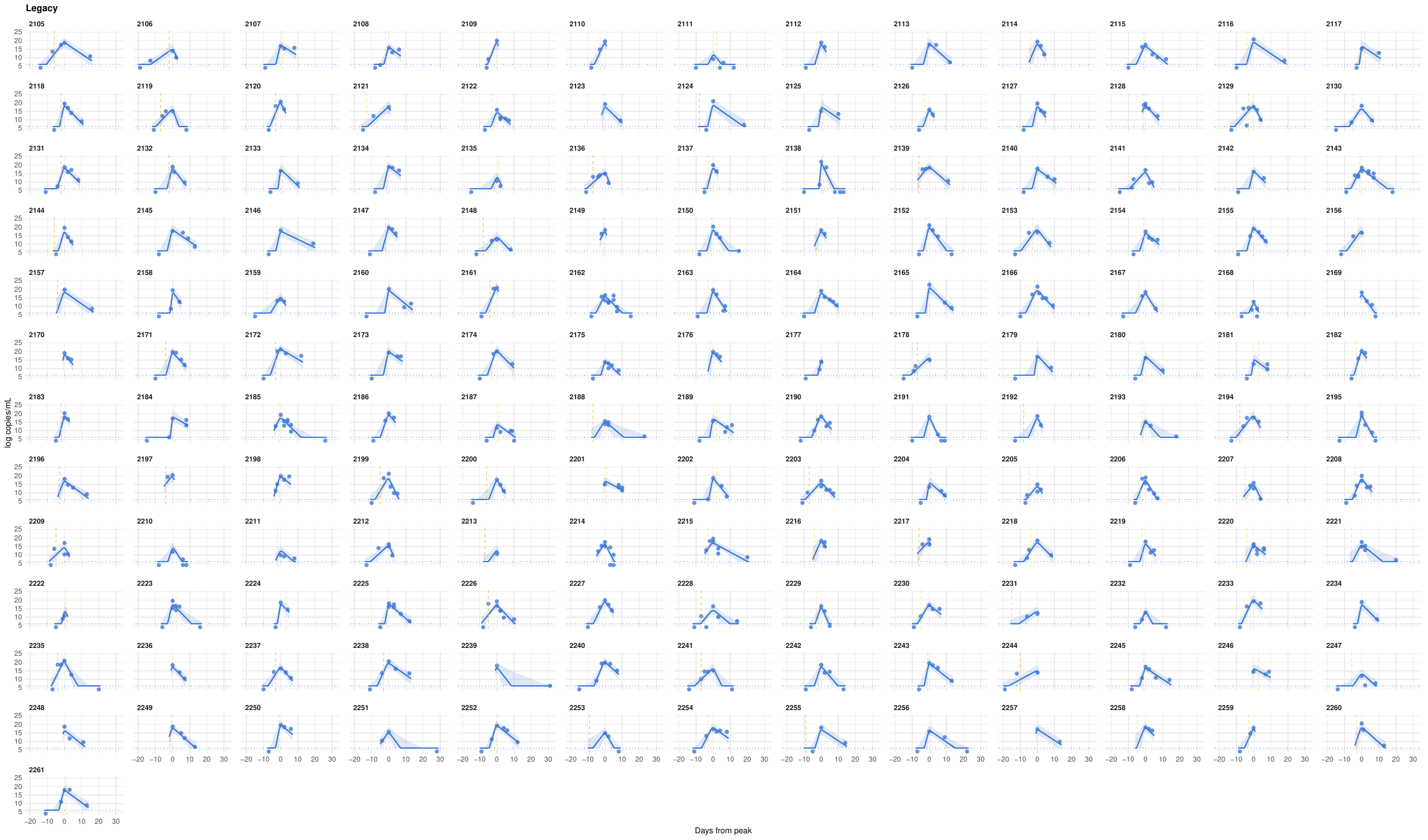}
            \caption[Legacy individual fits]{Individual trajectory fits for all 157 Crick Legacy participants. This cohort contributes qPCR and symptom onset data but no viral culture or LFD results. The model uses the joint structure and hierarchical prior to impute the infectious virus trajectory from RNA data and the estimated population-level RNA-to-PFU transformation. All participants are vaccinated (2 or 3 doses), and infections span the Delta and Omicron waves.}
            \label{fig:fit_legacy}
        \end{figure}
    \end{landscape}

    \section{Posterior predictive checks} \label{sec:supp_ppc}

\begin{figure}[p]
        \centering
        \includegraphics[width=\linewidth]{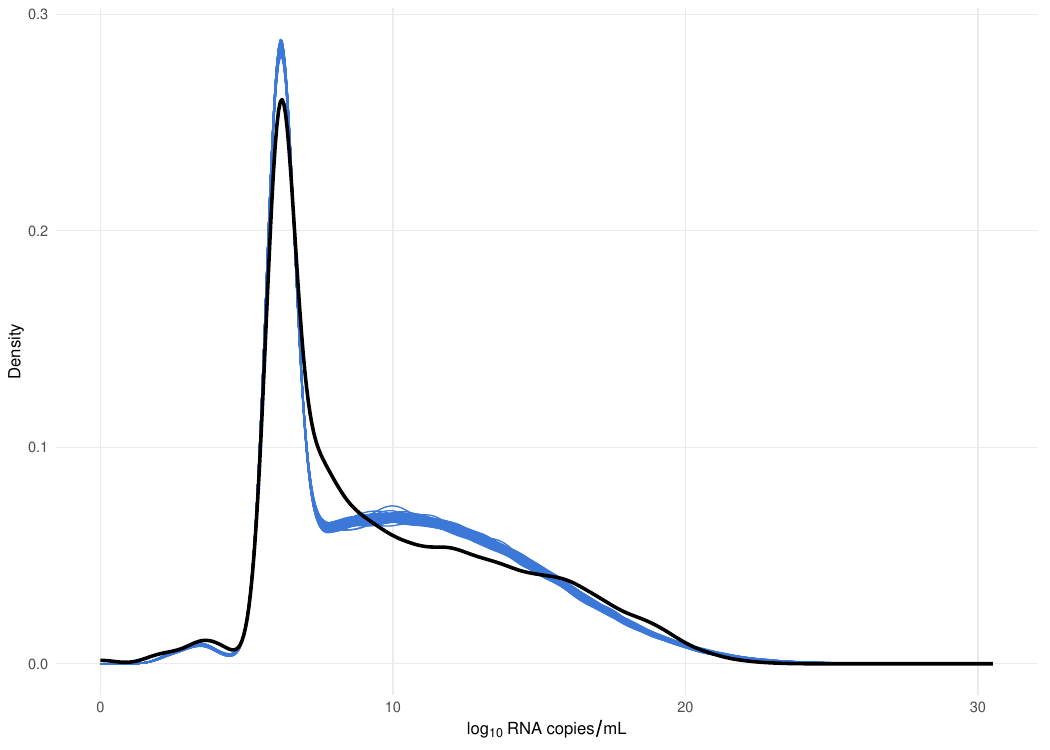}
        \caption[PPC: viral RNA]{Posterior predictive check for viral RNA (qPCR) observations. The distribution of observed log RNA concentrations (black density curve) is compared to 100 replicated datasets drawn from the posterior predictive distribution (coloured density curves). Agreement between observed and replicated distributions indicates adequate model fit for the RNA observation model.}
        \label{fig:ppc_rna}
    \end{figure}

\begin{figure}[p]
        \centering
        \includegraphics[width=\linewidth]{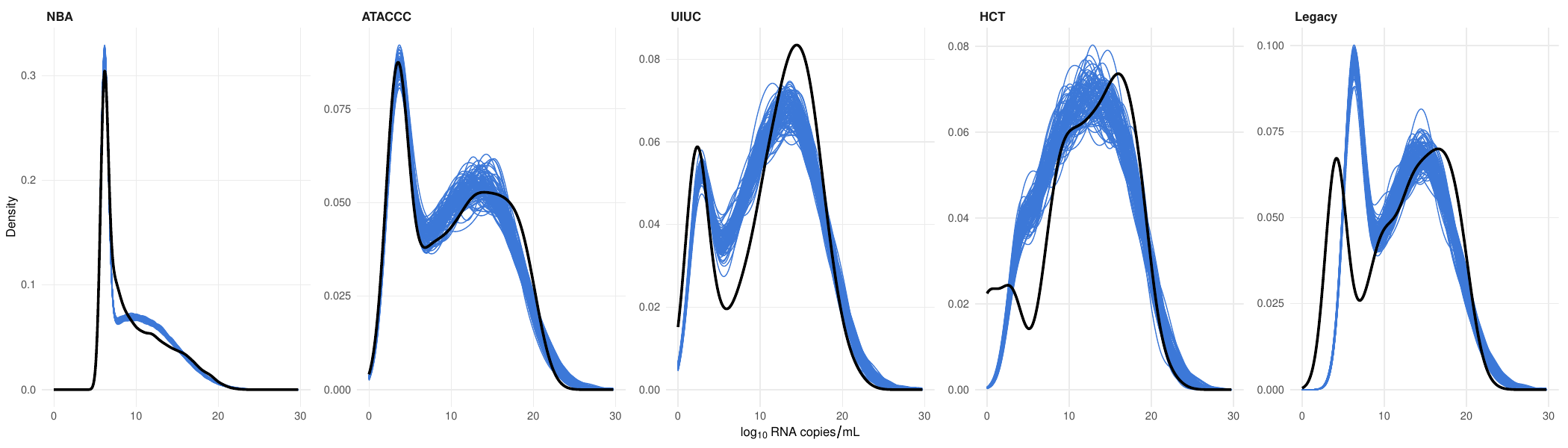}
        \caption[PPC: viral RNA by cohort]{Posterior predictive check for viral RNA, stratified by cohort. Each panel shows the observed versus replicated RNA distribution for a single cohort. The model captures differences in the marginal RNA distributions across cohorts, including the broader distribution in the NBA cohort (which has the most individuals) and the more concentrated distributions in the smaller cohorts.}
        \label{fig:ppc_rna_cohort}
    \end{figure}

\begin{figure}[p]
        \centering
        \includegraphics[width=\linewidth]{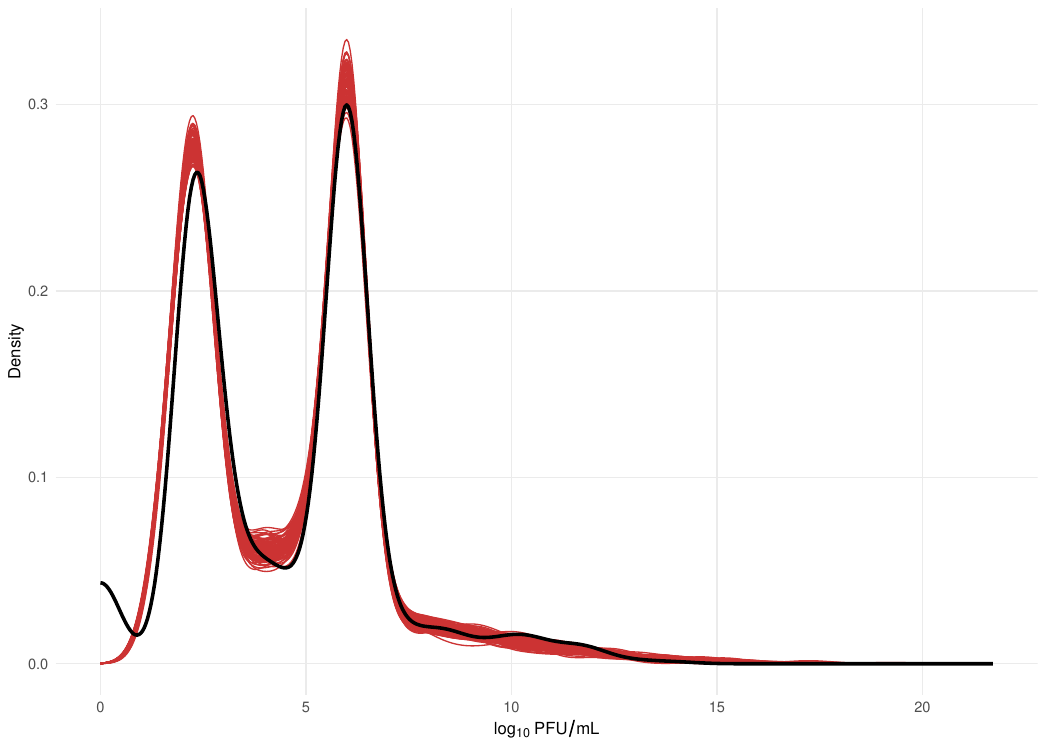}
        \caption[PPC: infectious virus]{Posterior predictive check for infectious virus (PFU/TCID$_{50}$/FFA) observations. The observed distribution of log$_{10}$ PFU/mL values is shown as a black density curve, overlaid with density curves from 100 replicated datasets drawn from the posterior predictive distribution. Close overlap between the observed and replicated densities indicates that the PFU observation model adequately captures the location, spread, and shape of the culture-based measurements.}
        \label{fig:ppc_pfu}
    \end{figure}

\begin{figure}[p]
        \centering
        \includegraphics[width=\linewidth]{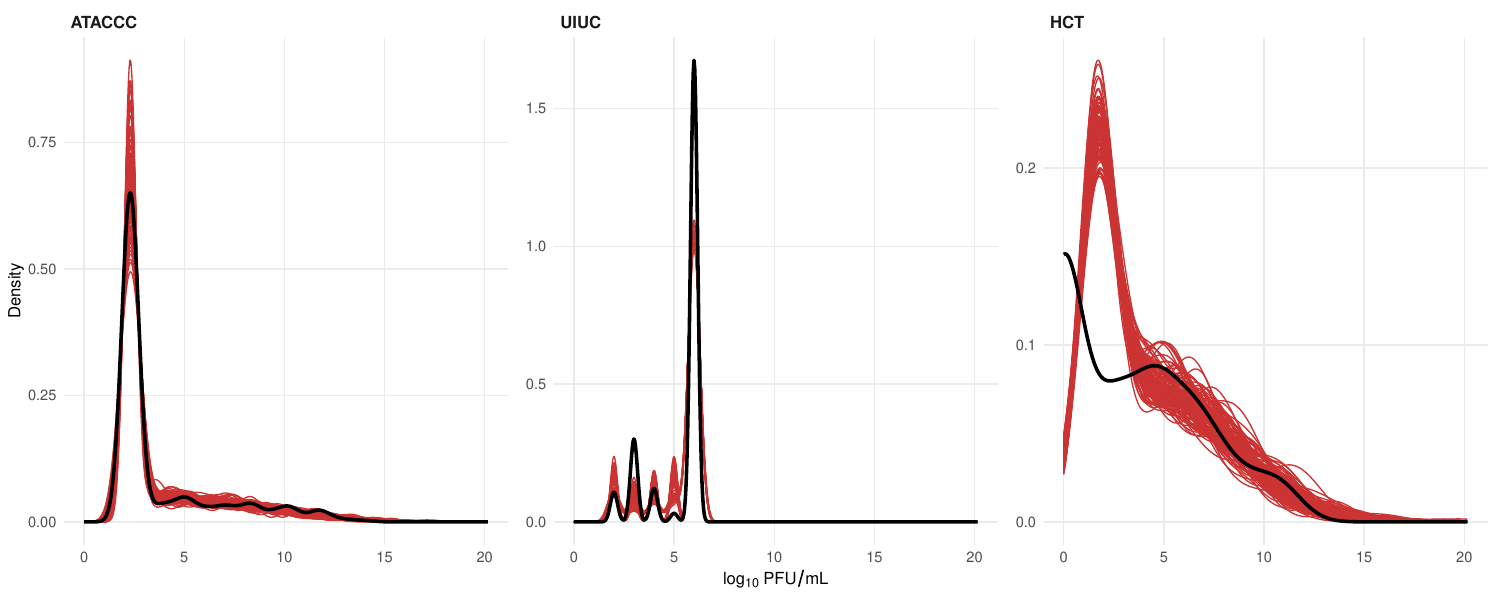}
        \caption[PPC: infectious virus by cohort]{Posterior predictive check for infectious virus, stratified by cohort. Panels show results for the three cohorts with culture data (ATACCC, UIUC, HCT). The model accommodates differences in PFU distributions across cohorts arising from different culture assay types (viral culture, TCID$_{50}$, PFU/FFA).}
        \label{fig:ppc_pfu_cohort}
    \end{figure}

\begin{figure}[p]
        \centering
        \includegraphics[width=\linewidth]{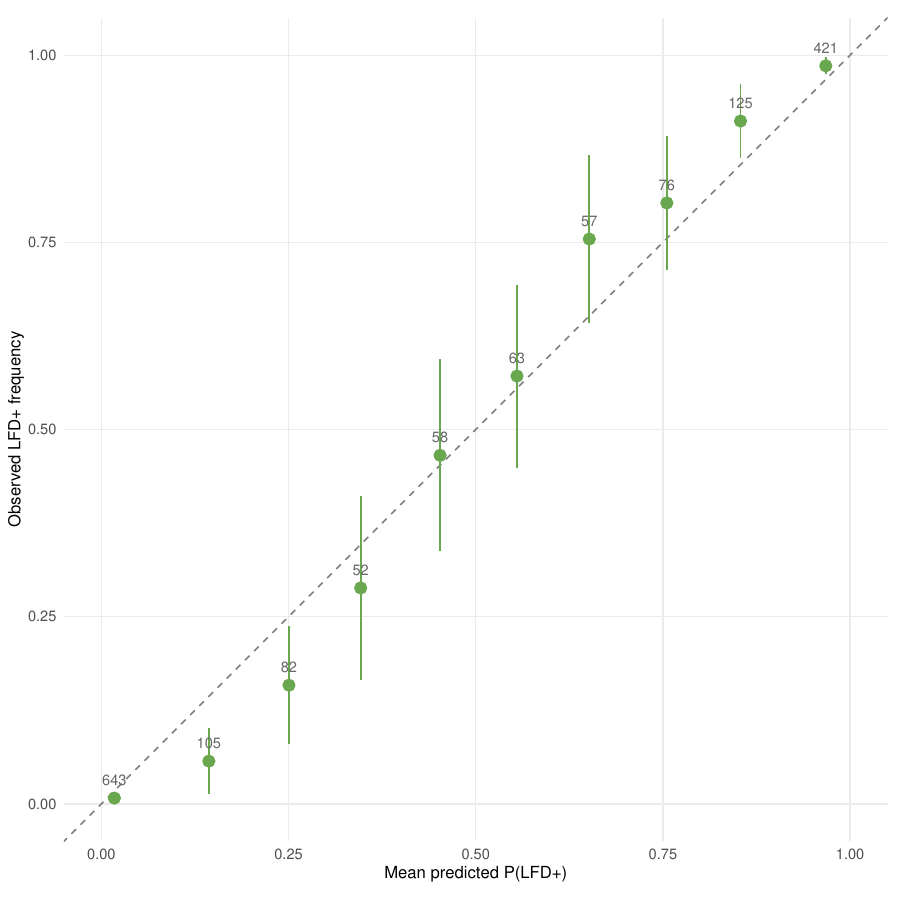}
        \caption[PPC: LFD calibration]{Calibration plot for the lateral flow device (LFD) observation model. Predicted LFD positivity probabilities are binned into deciles; within each bin, the observed LFD-positive frequency ($y$-axis) is plotted against the mean predicted probability ($x$-axis), with 95\% binomial confidence intervals (vertical bars) and the number of observations per bin labeled above each point. The dashed diagonal line represents perfect calibration. Points falling near the diagonal indicate that the logistic LFD model is well calibrated across the full range of predicted positivity.}
        \label{fig:ppc_lfd}
    \end{figure}

\begin{figure}[p]
        \centering
        \includegraphics[width=\linewidth]{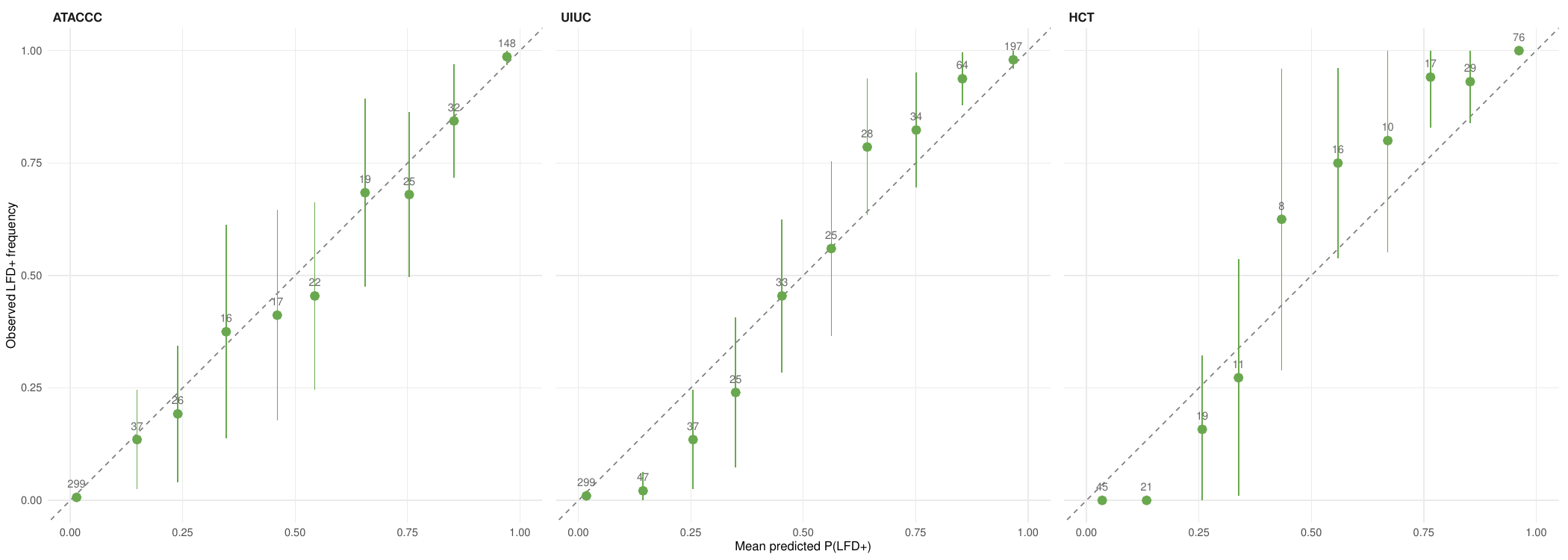}
        \caption[PPC: LFD calibration by cohort]{Calibration plot for the LFD observation model, stratified by cohort. Panels show results for the three cohorts with LFD data (ATACCC, UIUC, HCT). Calibration is maintained across cohorts despite differences in sampling frequency and LFD assay conditions.}
        \label{fig:ppc_lfd_cohort}
    \end{figure}

\begin{figure}[p]
        \centering
        \includegraphics[width=\linewidth]{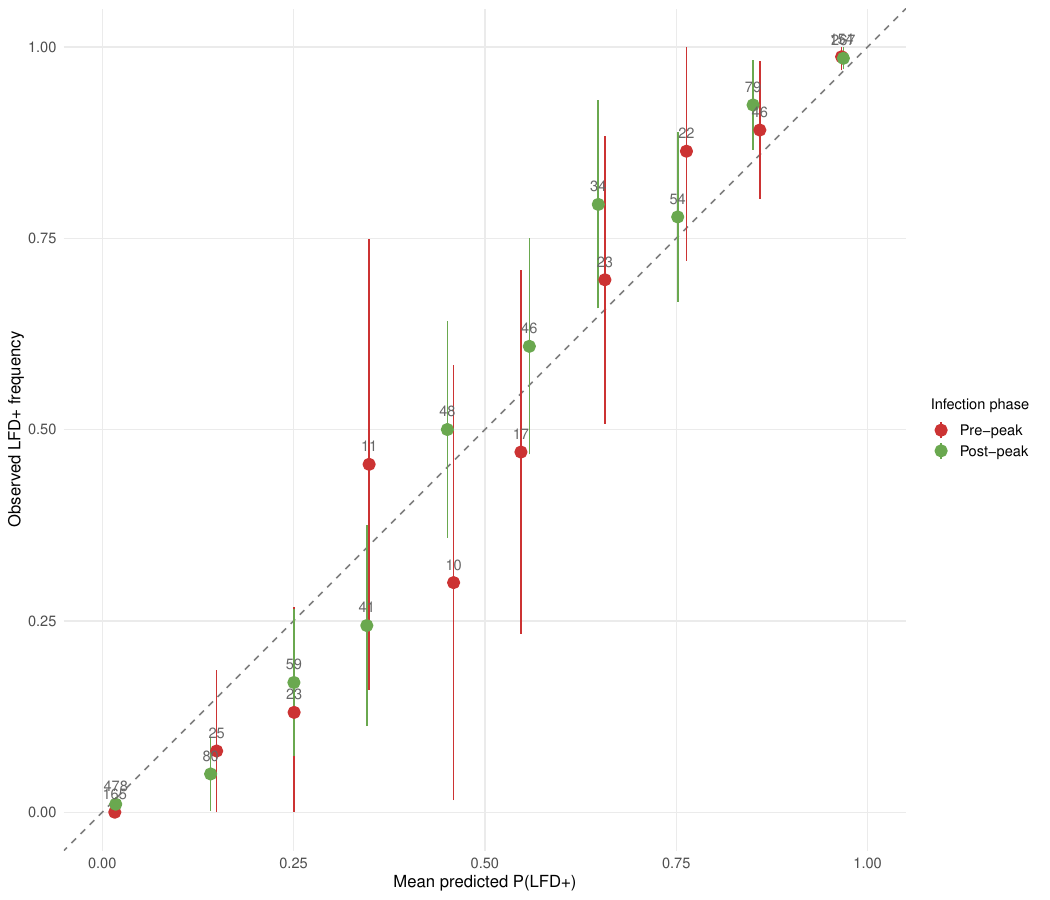}
        \caption[PPC: LFD calibration by infection phase]{Calibration plot for the LFD observation model, stratified by infection phase (pre-peak versus post-peak viral RNA). Pre-peak observations (time $\leq 0$ relative to observed RNA peak) show lower observed LFD positivity than predicted at comparable predicted probability levels, consistent with the lag between viral replication and antigen accumulation to the LFD detection threshold reported by~\cite{hakki2022onset}. Post-peak observations are well calibrated along the diagonal. This asymmetry reflects a biologically expected feature: during viral proliferation, RNA levels rise faster than antigen protein concentrations, leading to lower LFD sensitivity for a given predicted positivity level than during the clearance phase.}
        \label{fig:ppc_lfd_phase}
    \end{figure}

\begin{figure}[p]
        \centering
        \includegraphics[width=\linewidth]{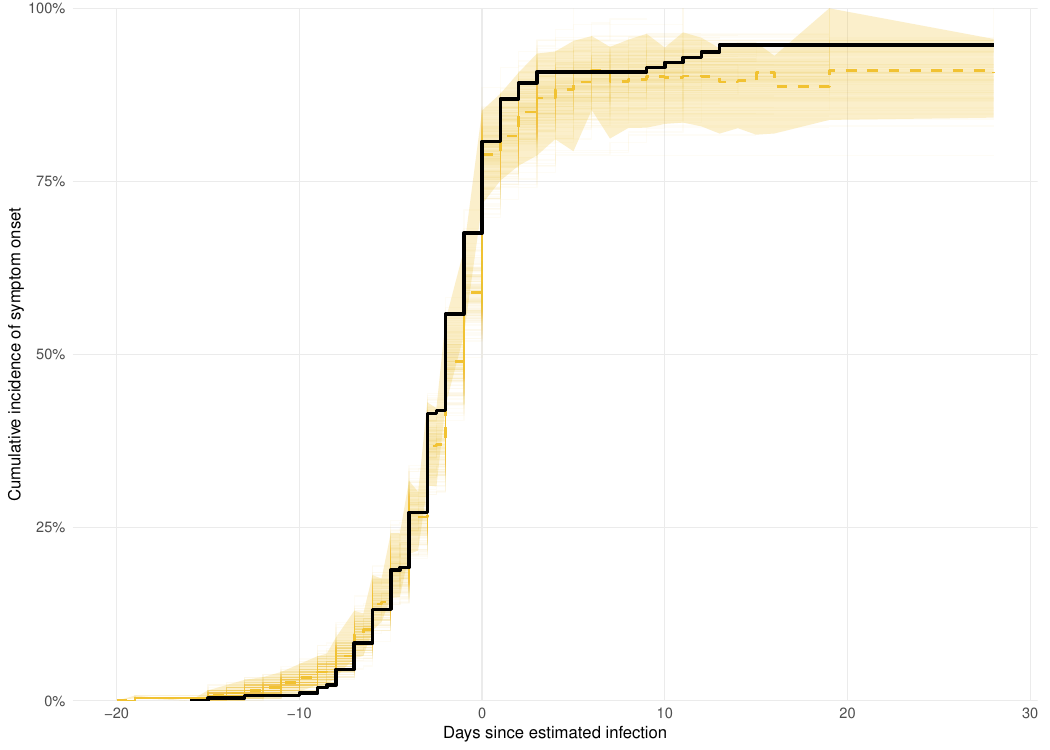}
        \caption[PPC: symptom onset]{Posterior predictive check for symptom onset. The observed Kaplan--Meier cumulative incidence curve of symptom onset (black step function) is compared to cumulative incidence curves from 200 replicated datasets drawn from the posterior predictive distribution of the discrete-time complementary log-log hazard model (coloured lines, with 95\% pointwise credible band shaded). For each replicated dataset, symptom onset is simulated by drawing Bernoulli outcomes at each at-risk day using the posterior predictive hazard probability. Agreement between the observed and replicated cumulative incidence indicates that the symptom onset model adequately captures the timing and proportion of symptomatic infections.}
        \label{fig:ppc_sym}
    \end{figure}

    Figures~\ref{fig:ppc_rna}--\ref{fig:ppc_sym} display posterior predictive checks comparing simulated replicated data from the fitted model to the observed data across all four assay types, both in aggregate and stratified by cohort. Replicated datasets are generated by drawing from the posterior predictive distribution at each observed time point for each individual.

    \begin{figure}[p]
        \centering
        \includegraphics[width=\linewidth]{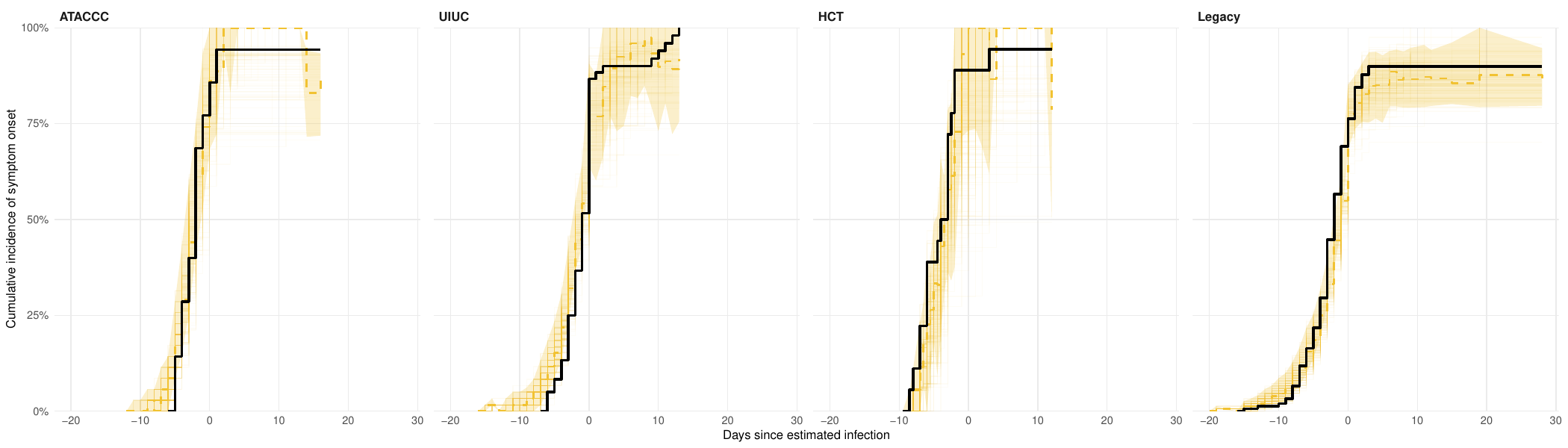}
        \caption[PPC: symptom onset by cohort]{Posterior predictive check for symptom onset, stratified by cohort. Panels show cumulative incidence of symptom onset for the four cohorts with symptom data (ATACCC, UIUC, HCT, Legacy). Per-cohort agreement between observed and replicated cumulative incidence curves indicates that the symptom model captures cohort-level differences in symptom onset timing and overall symptomatic fraction.}
        \label{fig:ppc_sym_cohort}
    \end{figure}

    \section{Parameter recovery} \label{sec:supp_recovery}

    \begin{figure}[p]
        \centering
        \includegraphics[width=\linewidth]{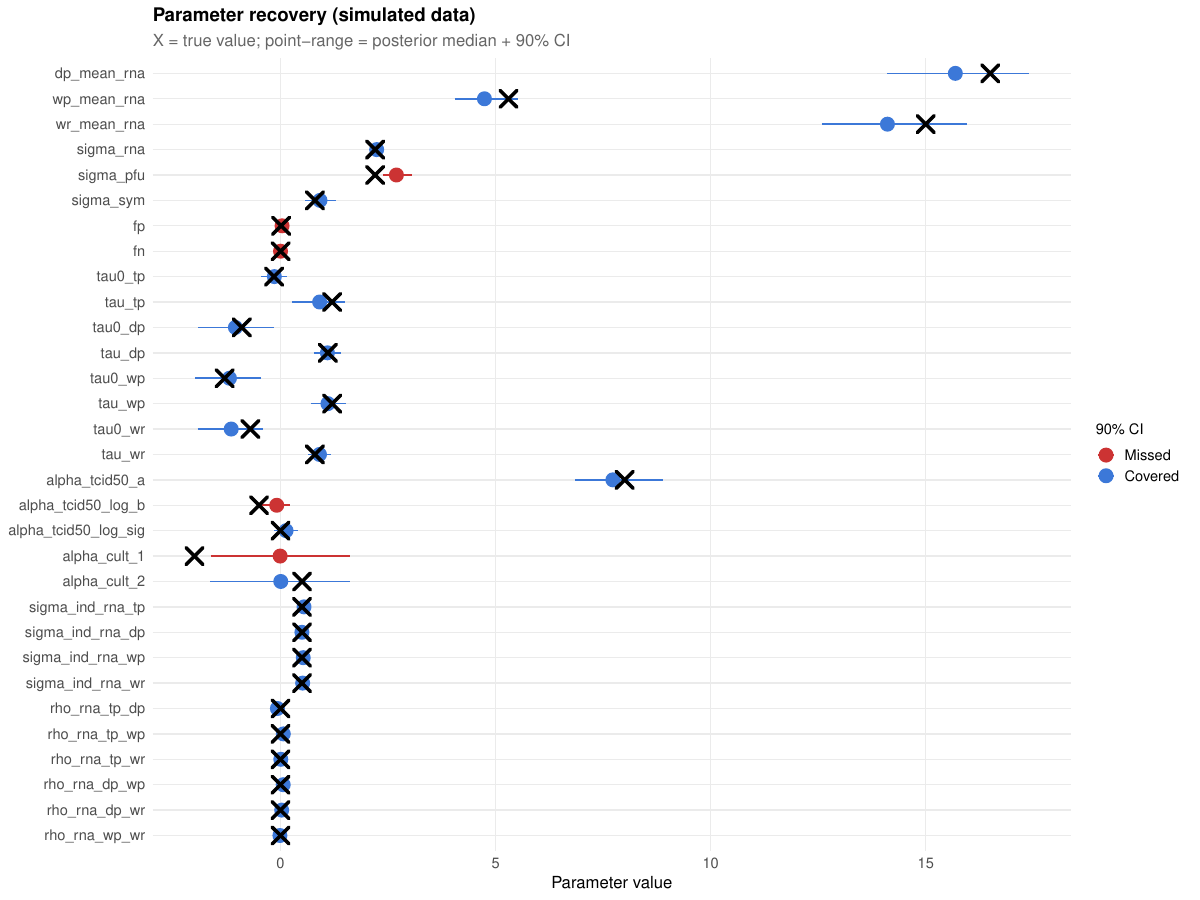}
        \caption[Parameter recovery]{Parameter recovery from simulated data. A synthetic dataset was generated from the model's prior predictive distribution using a 50\% random subsample of the observed covariate and missingness structure, and the model was refit to the simulated data. For each of 34 assessed parameters, the true generating value (horizontal line) is compared to the posterior 95\% credible interval from the recovery fit (vertical bar with point at the median). Covered parameters (true value within the 95\% CrI) are shown in blue; non-covered parameters are shown in red. Overall coverage is 76.5\% (26/34). Non-covered parameters ($\sigma_{\text{pfu}}$, $\hat\lambda_{\text{fp}}$, $\hat\lambda_{\text{fn}}$, $\alpha_{\text{cult},1}$, and three others) are discussed in Section~\ref{sec:model_checking} of the main text.}
        \label{fig:recovery}
    \end{figure}

    \section{Additional parameter estimates} \label{sec:supp_params}

    \subsection{Covariate effects on RNA kinetics}

    Tables~\ref{tab:cov_peak}--\ref{tab:cov_clearance} report the full posterior estimates of covariate effects on each RNA kinetic parameter. Effects are expressed as $\exp(\beta)$, representing the multiplicative fold change in the kinetic parameter relative to the reference category (unvaccinated, immunologically na\"ive, 0--30 year old with pre-Alpha variant). Values greater than 1 indicate an increase relative to the reference; values less than 1 indicate a decrease. These results are summarized visually in Figure~\ref{fig:covariates} of the main text.

\begin{table}[p]
    \centering
    \caption{Posterior estimates of covariate effects on peak viral RNA concentration. Each row reports the posterior median multiplicative effect $\exp(\hat\beta)$ and 95\% credible interval (CrI) for the indicated covariate level relative to the reference category. The reference value row reports the estimated population-mean peak concentration for the reference category.}
    \label{tab:cov_peak}
    \begin{tabular}{lccc}
    \toprule
     & \multicolumn{2}{c}{Peak value} \\
    Characteristic & $\exp(\beta)$ & 95\% CrI\\
    \midrule
         Age: [30,50) & 0.99 & (0.97, 1.01)\\
         Age: [50,100) & 0.99 & (0.97, 1.02)\\
         Recurrence: Yes & 0.96 & (0.94, 0.99)\\
         Variant: Alpha & 1.03 & (0.98, 1.08)\\
         Variant: Delta & 1.17 & (1.12, 1.22)\\
         Variant: Omicron & 1.06 & (1.02, 1.09)\\
         Variant: BA.4/BA.5 & 1.12 & (1.06, 1.18)\\
         Variant: Other & 0.98 & (0.94, 1.03)\\
         History: Vaccinated boosted & 0.87 & (0.84, 0.90)\\
         History: Vaccinated unboosted & 0.88 & (0.84, 0.92)\\
         History: Vaccinated unreported & 0.84 & (0.79, 0.88)\\
         History: Unreported & 0.85 & (0.80, 0.91)\\
         History: Boosted unreported primary & 0.88 & (0.85, 0.92)\\
         \midrule
         Reference value, log [RNA] per ml & 16.18 & (15.64, 16.76)\\
     \bottomrule
    \end{tabular}
\end{table}

\begin{table}[p]
    \caption{Posterior estimates of covariate effects on proliferation phase duration. Format as in Table~\ref{tab:cov_peak}. Values $>1$ indicate longer proliferation (slower rise to peak) relative to the reference category; values $<1$ indicate faster proliferation. Reference values are the estimated population-mean proliferation duration (in days) for the reference category, reported both for the full trajectory (from infection onset to peak) and from the limit of detection to peak.}
    \label{tab:cov_prolif}
    \centering
    \begin{tabular}{lcc}
     \toprule
     & \multicolumn{2}{c}{Proliferation duration } \\
     Characteristic & $\exp(\beta)$ & 95\% CrI\\
     \midrule
         Age: [30,50) & 1.01 & (0.92, 1.10)\\
         Age: [50,100) & 1.06 & (0.93, 1.20)\\
         Recurrence: Yes & 0.83 & (0.73, 0.95)\\
         Variant: Alpha & 0.85 & (0.68, 1.07)\\
         Variant: Delta & 0.76 & (0.62, 0.93)\\
         Variant: Omicron & 1.10 & (0.93, 1.29)\\
         Variant: BA.4/BA.5 & 1.22 & (0.87, 1.71)\\
         Variant: Other & 1.27 & (1.04, 1.55)\\
         History: Vaccinated boosted & 1.49 & (1.23, 1.80)\\
         History: Vaccinated unboosted & 1.42 & (1.16, 1.74)\\
         History: Vaccinated unreported & 1.31 & (1.02, 1.69)\\
         History: Unreported & 1.25 & (0.94, 1.68)\\
         History: Boosted unreported primary & 1.40 & (1.16, 1.68)\\
         \midrule
         Reference value, days & 5.12 & (4.42, 5.92)\\
         Reference value (lod), days & 3.18 & (2.69, 3.75)\\
     \bottomrule
     \end{tabular}
\end{table}

\begin{table}[p]
    \centering
    \caption{Posterior estimates of covariate effects on clearance phase duration. Format as in Table~\ref{tab:cov_peak}. Values $>1$ indicate slower clearance (longer time from peak to limit of detection) relative to the reference category; values $<1$ indicate faster clearance.}
    \label{tab:cov_clearance}
    \begin{tabular}{lcc}
     \toprule
     & \multicolumn{2}{c}{Clearance duration} \\
     Characteristic & $\exp(\beta)$ & 95\% CrI\\
     \midrule
         Age: [30,50) & 1.02 & (0.96, 1.08)\\
         Age: [50,100) & 1.20 & (1.11, 1.29)\\
         Recurrence: Yes & 0.76 & (0.70, 0.83)\\
         Variant: Alpha & 1.00 & (0.86, 1.16)\\
         Variant: Delta & 1.04 & (0.92, 1.17)\\
         Variant: Omicron & 0.96 & (0.87, 1.07)\\
         Variant: BA.4/BA.5 & 0.91 & (0.77, 1.08)\\
         Variant: Other & 1.05 & (0.93, 1.20)\\
         History: Vaccinated boosted & 0.82 & (0.73, 0.93)\\
         History: Vaccinated unboosted & 0.68 & (0.60, 0.78)\\
         History: Vaccinated unreported & 0.69 & (0.59, 0.81)\\
         History: Unreported & 0.90 & (0.74, 1.08)\\
         History: Boosted unreported primary & 0.78 & (0.70, 0.88)\\
         \midrule
         Reference value, days & 13.67 & (12.34, 15.12)\\
         Reference value (lod), days & 8.49 & (7.50, 9.59)\\
     \bottomrule
     \end{tabular}
\end{table}

    \subsection{RNA-to-PFU transformation and correlation structure}

\begin{figure}[p]
        \centering
        \includegraphics[width=\linewidth]{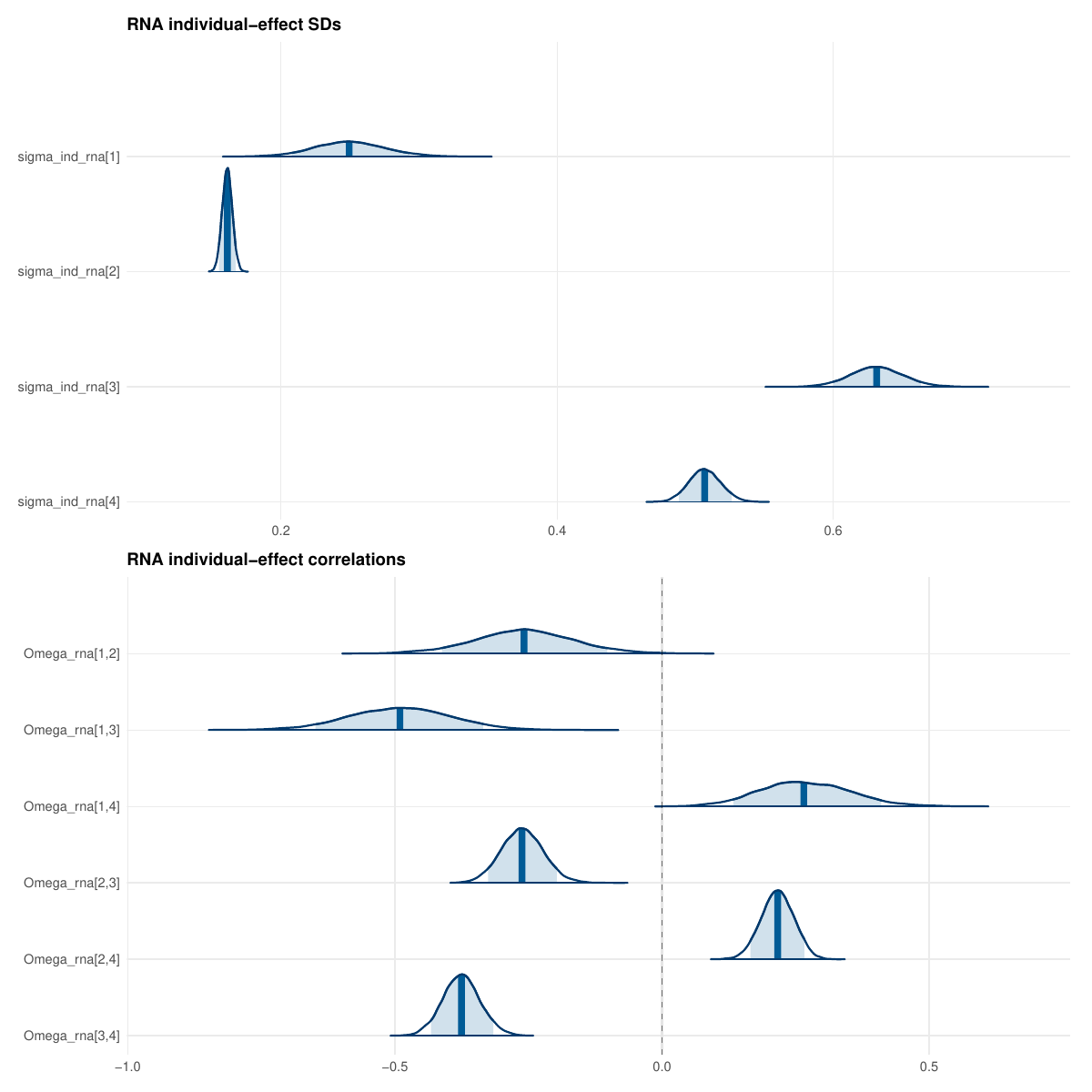}
        \caption[RNA RE correlation densities]{Marginal posterior densities for each pairwise correlation among the four RNA individual random effects (peak timing $t_p$, peak height $\delta$, proliferation duration $\omega_p$, clearance duration $\omega_r$). Each panel shows the posterior density (filled curve) with the prior density (dashed line) overlaid for comparison. The shift between prior and posterior indicates the direction and degree of data learning for each correlation.}
        \label{fig:corr_densities}
    \end{figure}

\begin{figure}[p]
        \centering
        \includegraphics[width=\linewidth]{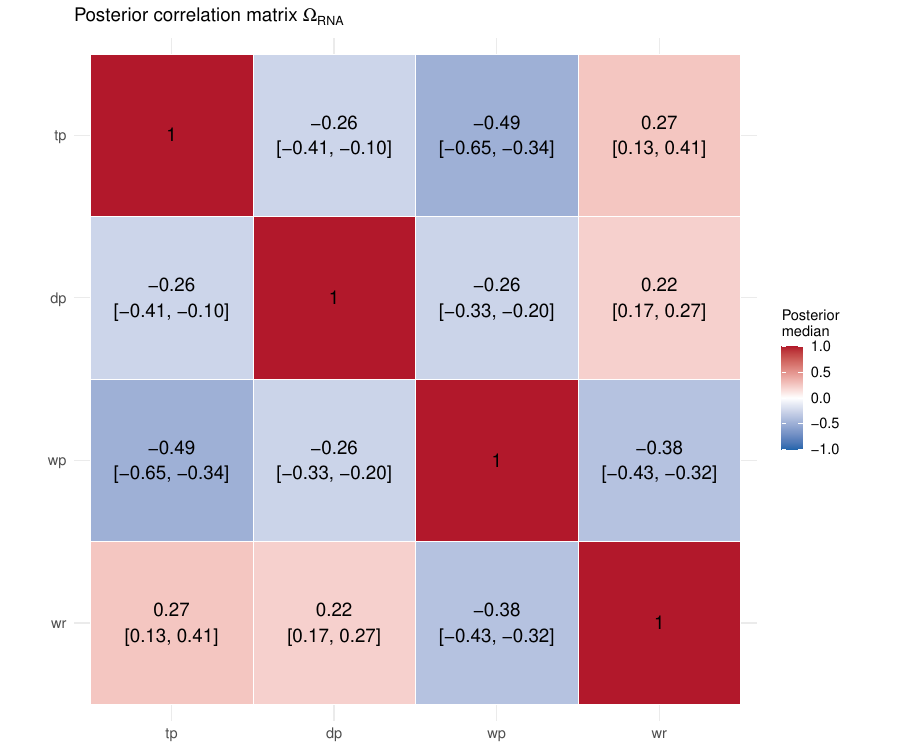}
        \caption[RNA RE correlation matrix]{Posterior median correlation matrix of the RNA individual random effects, displayed as a heatmap. Colors indicate the sign and magnitude of each pairwise correlation: blue tones represent negative correlations and red tones represent positive correlations. Numeric values show the posterior median; 95\% credible intervals are reported in Section~8.5 of the main text.}
        \label{fig:corr_matrix}
    \end{figure}

    Figure~\ref{fig:corr_densities} shows the marginal posterior densities for each element of the RNA individual random effect correlation matrix, and Figure~\ref{fig:corr_matrix} displays the full posterior median correlation matrix with uncertainty. These supplement the summary in Section~8.5 and Figure~\ref{fig:correlations} of the main text.

\begin{table}[p]
    \centering
    \caption{RNA-to-PFU transformation: posterior estimates of the log-affine parameters linking infectious virus trajectory parameters to viral RNA trajectory parameters. The transformation is $\theta'_k = \exp(a_{0,k} + a_{1,k} \log \theta_k)$ for peak height, proliferation duration, and clearance duration, where $\theta_k$ is the RNA parameter and $\theta'_k$ is the corresponding PFU parameter. The elasticity $a_1$ represents the percent change in the PFU parameter for a 1\% change in the RNA parameter; values $<1$ indicate sub-proportional scaling. The peak time transformation is affine: $t'_p = a_{0,tp} + a_{1,tp} \cdot t_p$.}
    \label{tab:transformation}
    \begin{tabular}{lcccc}
     \toprule
     Parameter & Intercept ($a_0$) & 95\% CrI & Elasticity ($a_1$) & 95\% CrI\\
     \midrule
     Peak ($\delta'$) & $-1.00$ & ($-2.23$, $0.25$) & 1.09 & (0.67, 1.52)\\
     Proliferation ($\omega_p'$) & $-0.64$ & ($-1.31$, $-0.03$) & 0.81 & (0.50, 1.16)\\
     Clearance ($\omega_r'$) & $0.06$ & ($-0.50$, $0.59$) & 0.63 & (0.42, 0.83)\\
     \midrule
     & Offset ($a_0$) & 95\% CrI & Scaling ($a_1$) & 95\% CrI\\
     \midrule
     Peak time ($t_p'$) & $-0.55$ & ($-0.97$, $-0.11$) & 3.39 & (1.94, 4.80)\\
     \bottomrule
     \end{tabular}
\end{table}

    \subsection{Individual random effect standard deviations}

    Table~\ref{tab:re_sd} reports the posterior estimates of individual-level random effect standard deviations for both the RNA and PFU trajectories. Larger values indicate greater between-individual variability in that kinetic parameter.

\begin{table}[p]
    \centering
    \caption{Posterior estimates of individual random effect standard deviations for the RNA and PFU trajectories. RNA random effects are correlated (Section~8.5 of the main text); PFU random effects are independent with a half-normal(0, 0.3) prior. Each row reports the posterior median and 95\% credible interval for the standard deviation of individual-level deviations from the population mean on the log scale.}
    \label{tab:re_sd}
    \begin{tabular}{lcc}
     \toprule
     Parameter & SD & 95\% CrI\\
     \midrule
         RNA: Peak time ($t_p$) & 0.249 & (0.199, 0.301)\\
         RNA: Peak height ($\delta$) & 0.161 & (0.154, 0.168)\\
         RNA: Proliferation ($\omega_p$) & 0.631 & (0.594, 0.670)\\
         RNA: Clearance ($\omega_r$) & 0.507 & (0.485, 0.530)\\
         PFU: Peak time ($t_p'$) & 0.899 & (0.548, 1.204)\\
         PFU: Peak height ($\delta'$) & 0.296 & (0.231, 0.377)\\
         PFU: Proliferation ($\omega_p'$) & 0.455 & (0.268, 0.674)\\
         PFU: Clearance ($\omega_r'$) & 0.371 & (0.294, 0.464)\\
         Symptom: Susceptibility ($u_i$) & 0.65 & (0.32, 1.01)\\
     \bottomrule
     \end{tabular}
\end{table}

    \subsection{Observation model parameters}

    Tables~\ref{tab:lfd} and~\ref{tab:symptom} report the posterior estimates of the LFD and symptom onset observation model parameters, respectively. Both models include smooth post-peak sigmoid interaction terms that allow the relationship between viral load and the outcome to differ between the proliferation and clearance phases: the LFD model uses a logistic link (Section~3.3 of the main text) and the symptom model uses a discrete-time complementary log-log hazard (Section~3.5).

\begin{table}[p]
    \centering
    \caption{Posterior estimates of LFD logistic observation model parameters. The exponentiated coefficients $\exp(\gamma)$ represent odds ratios. The post-peak indicator $\mathbbm{1}(t \geq t_p)$ and its interaction with log RNA capture the asymmetry between the proliferation and clearance phases of antigen accumulation. The intercept is on the logit scale.}
    \label{tab:lfd}
    \begin{tabular}{lcc}
     \toprule
     Predictor & $\exp(\gamma)$ & 95\% CrI\\
     \midrule
     log RNA copies ($\gamma_1$) & 1.14 & (1.02, 1.27)\\
     log PFU culturable virus ($\gamma_2$) & 1.92 & (1.67, 2.27)\\
     Post-peak indicator ($\gamma_3$) & 3.29 & (0.84, 12.94)\\
     Post-peak $\times$ log RNA ($\gamma_4$) & 1.09 & (0.98, 1.21)\\
     \midrule
     & Coefficient & 95\% CrI\\
     \midrule
     Intercept ($\gamma_0$) & -4.46 & (-5.59, -3.36)\\
     \bottomrule
     \end{tabular}
\end{table}

\begin{table}[p]
    \centering
    \caption{Posterior estimates of discrete-time complementary log-log symptom onset model parameters. The hazard of symptom onset on day $t$ is $h_t = 1 - \exp\bigl(-\exp(\zeta_0 + \zeta_1 \log V_t/s + \zeta_2 \log R_t/s + \zeta_3 \mathbbm{1}(t \geq t_p) + \zeta_4 \mathbbm{1}(t \geq t_p) \log R_t/s + u_i)\bigr)$, where $V_t$ and $R_t$ are the latent infectious virus and viral RNA trajectories, $\mathbbm{1}(t \geq t_p)$ is a post-peak indicator capturing the immune activation lag, and $u_i \sim N(0, \sigma_{\text{sym}}^2)$ is an individual random effect. Exponentiated coefficients $\exp(\zeta)$ represent the multiplicative change in the complementary log-log hazard per unit increase in the predictor.}
    \label{tab:symptom}
    \begin{tabular}{lcc}
     \toprule
     Predictor & $\exp(\zeta)$ & 95\% CrI\\
     \midrule
     log PFU culturable virus ($\zeta_1$) & 2.27 & (1.32, 4.22)\\
     log RNA copies ($\zeta_2$) & 2.80 & (1.51, 5.42)\\
     Post-peak indicator ($\zeta_3$) & 0.47 & (0.17, 1.30)\\
     Post-peak $\times$ log RNA ($\zeta_4$) & 1.58 & (0.52, 4.90)\\
     \midrule
     & Coefficient & 95\% CrI\\
     \midrule
     Intercept ($\zeta_0$) & -1.36 & (-1.79, -0.96)\\
     Individual RE SD ($\sigma_{\text{sym}}$) & 0.65 & (0.32, 1.01)\\
     \bottomrule
     \end{tabular}
\end{table}

    \subsection{Stratified probability curves by covariate profile}
    \label{sec:supp_strat}

\begin{landscape}
    \begin{figure}[p]
        \centering
        \includegraphics[width=\linewidth]{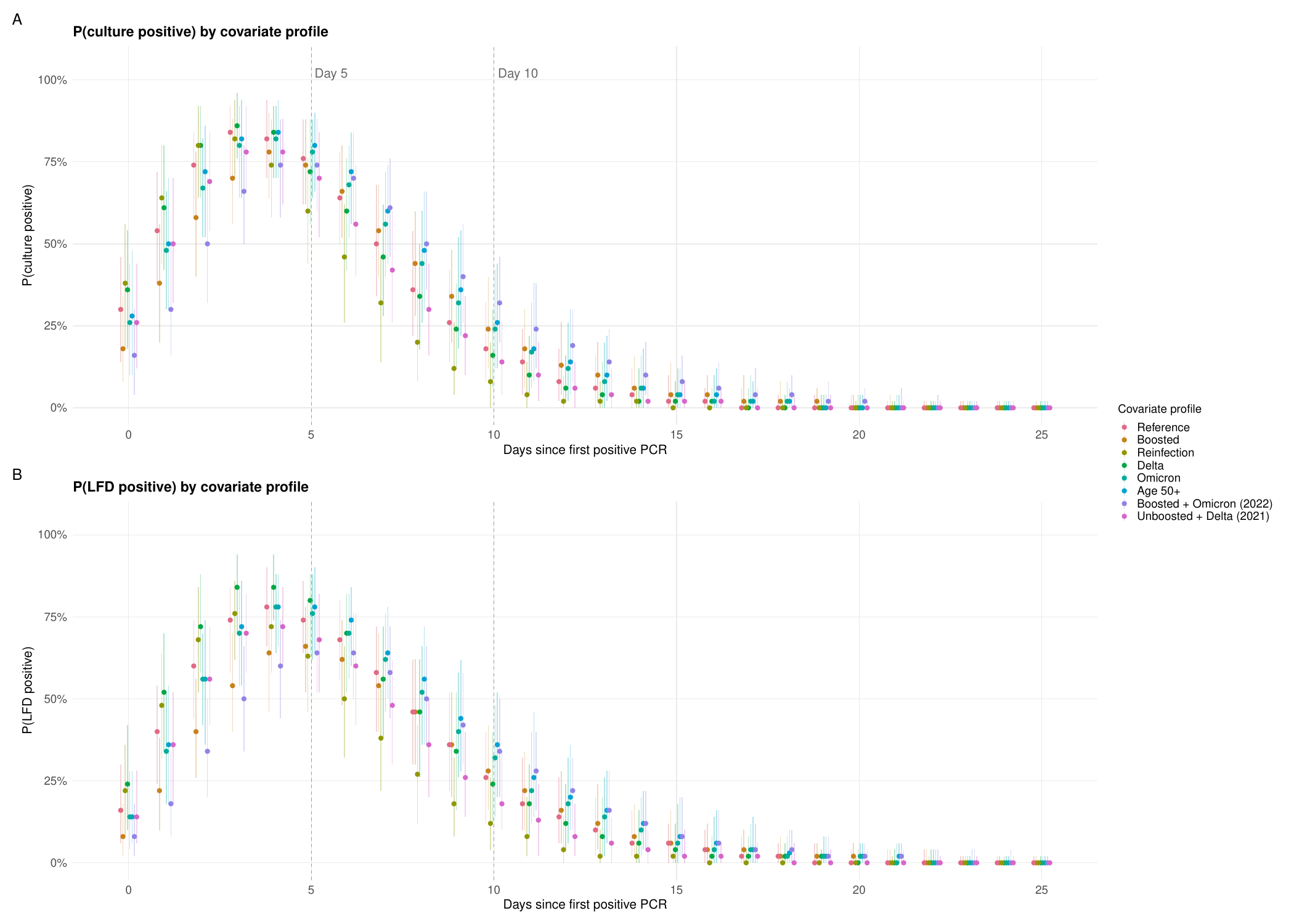}
        \caption[Stratified probability curves]{Probability of remaining culture-positive \textbf{(A)} and LFD-positive \textbf{(B)} as a function of days since first positive PCR test, stratified by covariate profile. Points show posterior median probabilities at each integer day; vertical bars show 95\% credible intervals. Profiles are horizontally dodged for readability. The Reinfection profile clears fastest, consistent with anamnestic immunity accelerating viral clearance. The Boosted + Omicron (2022) profile shows the most prolonged culture positivity, reflecting the longer proliferation phase associated with both vaccination and Omicron. Culture and LFD positivity curves track each other closely across profiles, supporting the use of LFD as a proxy for infectious status regardless of covariate profile.}
        \label{fig:strat_curves}
    \end{figure}
    \end{landscape}

    Figure~\ref{fig:strat_curves} shows the probability of remaining culture-positive and LFD-positive as a function of days since first positive PCR, stratified by covariate profile. These curves supplement the marginal (population-averaged) probability curves presented in Figure~\ref{fig:probability_curves} of the main text by illustrating how variant, vaccination status, and infection history modulate the duration of infectiousness. Profiles are selected to maximise visual contrast and include the Reference (unvaccinated, pre-Alpha, first infection, age $<$30), single-factor profiles (Boosted, Reinfection, Delta, Omicron, Age 50+), and composite profiles (Boosted + Omicron 2022, Unboosted + Delta 2021).

    \subsection{Serial rapid testing conditional on symptom onset}
    \label{sec:supp_serial_lfd}

\begin{landscape}
    \begin{figure}[p]
        \centering
        \includegraphics[width=\linewidth]{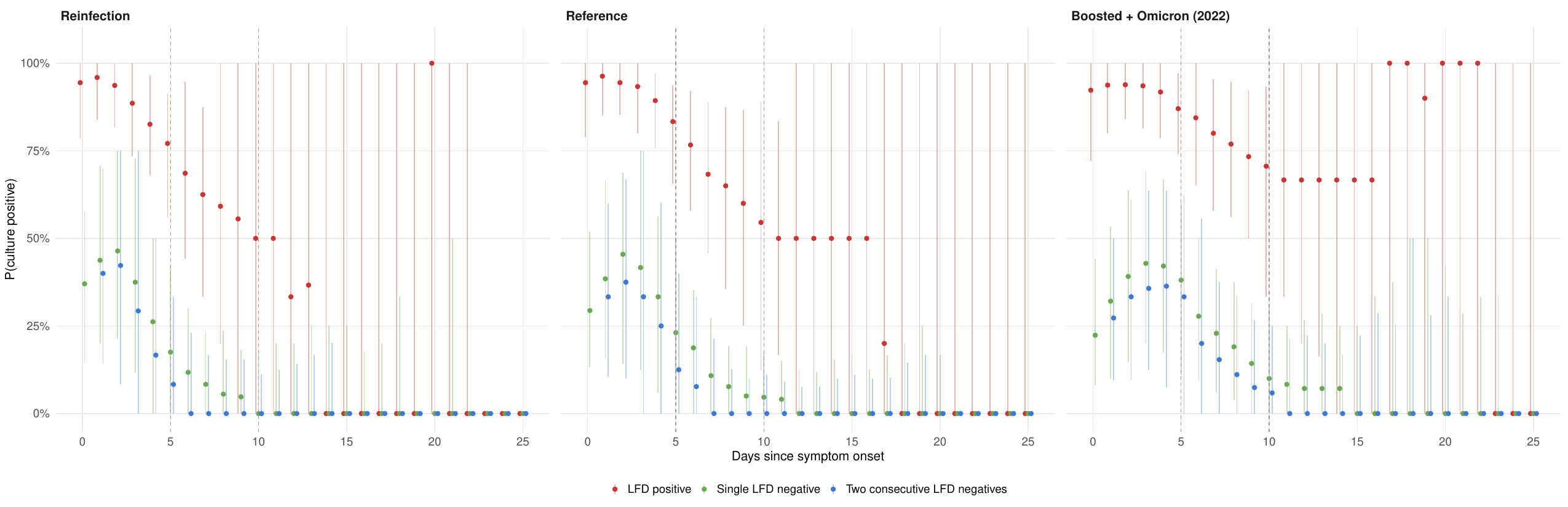}
        \caption[Conditional culture positivity by LFD outcome and symptom onset]{Probability of remaining culture-positive conditional on lateral flow device (LFD) test outcomes, as a function of days since symptom onset, for three representative covariate profiles. Points show posterior medians; vertical bars show 95\% credible intervals. Three conditions are compared: LFD positive (red), single LFD negative (blue), and two consecutive LFD negatives --- current day plus preceding day (green). Two consecutive negative results provide substantially greater reassurance than a single negative, particularly in the day 5--10 window where the single-negative false reassurance rate remains appreciable. Dashed vertical lines indicate days 5 and 10.}
        \label{fig:conditional_symptom}
    \end{figure}
    \end{landscape}

    Figure~\ref{fig:conditional_symptom} shows the probability of remaining culture-positive conditional on lateral flow device (LFD) test outcomes, aligned to the day of symptom onset, for three representative covariate profiles (Reinfection, Reference, Boosted + Omicron 2022). In addition to conditioning on a single positive or negative LFD result, the figure includes a series conditioning on two consecutive negative LFD results (current day plus preceding day), quantifying the incremental reassurance provided by serial testing.

    \section{Anatomical site-specific sampling and pooling model}
    \label{sec:supp_site}

    \subsection{Motivation}

    The five cohorts in this study differ in their upper respiratory tract sampling protocols. Three cohorts (ATACCC, NBA, Legacy) combine nasal and oropharyngeal swabs into a single collection tube prior to PCR, yielding one RNA concentration measurement per time point that reflects a \emph{pooled} sample from both anatomical sites. Two cohorts (HCT, UIUC) collect site-specific specimens --- nasal and throat swabs (HCT) or nasal and saliva samples (UIUC) --- that are processed separately, providing distinct RNA concentration measurements from each site.

    In the current model, site-specific measurements from HCT and UIUC are combined prior to model fitting by taking the arithmetic mean on the linear scale: $R_{\text{obs}} = \log\bigl(\tfrac{1}{2}e^{R_n} + \tfrac{1}{2}e^{R_t}\bigr)$, where $R_n$ and $R_t$ denote the log RNA concentrations from the nasal and throat sites, respectively. While this produces a single summary value per time point, the averaging has two drawbacks. First, it discards information about site-specific kinetic differences --- for example, whether the nose or throat leads in proliferation or clears virus faster. Second, it introduces a subtle inconsistency with the PFU likelihood, which uses \emph{throat-only} culture data (the only site from which viable virus was assayed in HCT).

    Figure~\ref{fig:site_comparison}A illustrates this issue using data from three representative HCT participants. The nasal and throat RNA trajectories can differ substantially in timing and magnitude, yet the current model receives only their average. The throat-only PFU measurements (orange triangles) align with the throat trajectory but may diverge from the nose trajectory, particularly during the early proliferation phase when viral replication in the nose may precede the throat.

    \subsection{Chemistry of swab pooling}

    When two swabs at concentrations $C_n$ and $C_t$ (copies/mL) are pooled into a single tube, the resulting measured concentration is a weighted sum:
    \begin{equation}
        C_{\text{pool}} = w_n C_n + w_t C_t, \quad w_n + w_t = 1,
        \label{eq:pool_linear}
    \end{equation}
    where $w_n$ and $w_t$ reflect the relative collection efficiencies (encompassing swab absorption capacity, elution volume, and extraction efficiency at each anatomical site). On the natural-log scale, this becomes:
    \begin{equation}
        R_{\text{pool}} = \operatorname{log\_sum\_exp}\!\bigl(R_n + \alpha_n,\; R_t + \alpha_t\bigr),
        \label{eq:pool_log}
    \end{equation}
    where $\alpha_s = \log w_s$ for $s \in \{n, t\}$ are log-scale collection-efficiency weights. This is numerically stable and can be computed directly in Stan using the built-in \texttt{log\_sum\_exp} function.

    The current averaging procedure implicitly assumes equal weights ($w_n = w_t = 0.5$, i.e.\ $\alpha_n = \alpha_t = \log 0.5$). A site-specific extension would estimate $\alpha_n$ and $\alpha_t$ from data, allowing the model to learn whether one site dominates the pooled signal.

    \subsection{Site-specific trajectory model}

    A natural extension is to model each anatomical site $s \in \{n, t\}$ as having its own latent RNA trajectory, derived from the individual's shared infection process but offset by site-specific parameters:
    \begin{align}
        \tau_{p,s} &= \tau_p + \delta_\tau^{(s)}, &
        d_{p,s} &= d_p + \delta_d^{(s)}, \nonumber \\
        \omega_{p,s} &= \omega_p + \delta_{\omega_p}^{(s)}, &
        \omega_{r,s} &= \omega_r + \delta_{\omega_r}^{(s)},
        \label{eq:site_offsets}
    \end{align}
    where $\delta_\cdot^{(s)}$ are population-level site offset parameters (with throat as the reference site, so $\delta_\cdot^{(t)} = 0$). The site-specific trajectory $R_s(t)$ is then computed using the same piecewise exponential function as the main model but evaluated at the site-shifted parameters.

\begin{figure}[p]
        \centering
        \includegraphics[width=\linewidth]{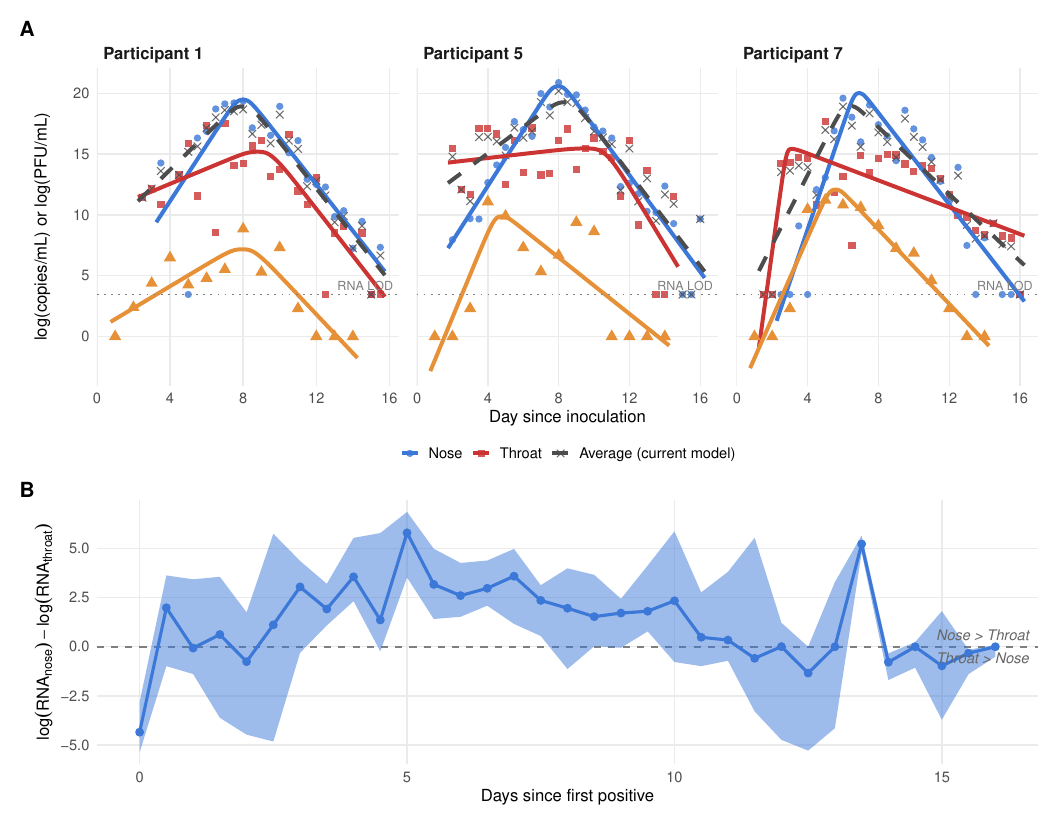}
        \caption[Site-specific RNA trajectories in the HCT cohort]{\textbf{(A)} Nasal (blue) and throat (red) RNA concentration data (points) for three representative participants in the SARS-CoV-2 Human Challenge Study, with the arithmetic mean used in the current model shown as crosses. Smooth curves show smoothed piecewise exponential fits to each series independently: solid blue (nose), solid red (throat), dashed grey (average), and solid orange (PFU). Orange triangles show PFU measurements from throat-only viral culture. The two anatomical sites exhibit distinct kinetic profiles, with the nasal site often reaching higher peak concentrations and the PFU trajectory aligning more closely with the throat RNA than with the averaged signal. The current model's averaging obscures these site-specific differences. \textbf{(B)} Population-level median (shaded IQR) difference $\log(\text{RNA}_{\text{nose}}) - \log(\text{RNA}_{\text{throat}})$ across all HCT participants as a function of days since first positive test. The difference is not constant over time, supporting a model with site-specific kinetic offsets rather than a simple multiplicative correction. Bins with fewer than three observations are excluded.}
        \label{fig:site_comparison}
    \end{figure}

    Figure~\ref{fig:site_comparison}B provides empirical support for this structure. The median log-difference between nasal and throat RNA ($R_n - R_t$) varies systematically over the course of infection, suggesting that the two sites do not simply differ by a constant offset but exhibit distinct kinetic profiles.

    \subsection{Likelihood structure}

    Under the site-specific extension, the RNA likelihood would depend on the cohort's sampling protocol:
    \begin{itemize}
        \item \textbf{Site-specific cohorts} (HCT, UIUC): Each site contributes an independent observation:
        \begin{equation}
            R_{s,\text{obs}} \sim \operatorname{Normal}\!\bigl(R_s(t),\; \sigma_{\text{rna}}\bigr), \quad s \in \{n, t\}.
        \end{equation}
        This doubles the number of RNA likelihood terms for these cohorts, providing substantially more information per time point.
        \item \textbf{Pooled cohorts} (ATACCC, NBA, Legacy): The observed pooled measurement follows from Equation~\eqref{eq:pool_log}:
        \begin{equation}
            R_{\text{obs}} \sim \operatorname{Normal}\!\bigl(R_{\text{pool}}(t),\; \sigma_{\text{rna}}\bigr),
        \end{equation}
        where $R_{\text{pool}}(t)$ is computed from the two site-specific trajectories and the estimated pooling weights.
    \end{itemize}
    The PFU likelihood would use the throat-specific trajectory $R_t(t)$ directly, eliminating the current inconsistency between averaged RNA and throat-only PFU data.

    \subsection{Implementation notes}

    This extension is fully compatible with the existing model architecture and could be controlled by a \texttt{use\_site\_model} flag (analogous to the existing \texttt{use\_smooth} and \texttt{use\_wf} flags). When disabled ($= 0$), the site offsets collapse to zero and the pooling weights to $w_n = w_t = 0.5$, recovering the current model exactly. When enabled ($= 1$), the model would estimate up to four site offset parameters ($\delta_\tau, \delta_d, \delta_{\omega_p}, \delta_{\omega_r}$) plus one free pooling weight parameter (since $\alpha_n + \alpha_t$ is determined by the softmax constraint). These are population-level parameters identifiable from the $\sim$18 HCT and $\sim$60 UIUC individuals with site-specific data.

    The data pipeline would expand site-specific observations into separate rows (one per site per time point) and add a site index to the Stan data. Site-specific limits of detection would also be required, since the LOD may differ between anatomical sites due to differences in swab type, collection technique, or assay calibration.

\section{Relationship between log-affine transformation and convolution}
\label{sec:supp_logaffine}

The log-affine transformation that maps RNA kinetic parameters to PFU kinetic parameters (Section~3.2 of the main text) can be motivated from the underlying biology. In the target cell limited model, viral RNA accumulates according to $dR/dt = qV_t - eR_t$, where $q$ is the RNA production rate per infectious virion and $e$ is the RNA degradation rate. The solution is the convolution
\begin{equation*}
    R_t = q \int_0^t e^{-e(t-s)} V_s\,ds.
\end{equation*}
For a piecewise exponential infectious virus trajectory $\log V_t = g(t; \theta)$ with proliferation rate $a = \delta/\omega_p$ and clearance rate $b = \delta/\omega_r$, the convolution can be evaluated in closed form on each arm.

\paragraph{Pre-peak ($t \leq t_p$).} On the rising arm, $V_t = V_0\,e^{at}$ and the convolution yields
\begin{equation*}
    R_t = \frac{q}{e + a}\,V_t
\end{equation*}
which is a pure scaling of the virus trajectory. Taking logarithms, $\log R_t = \log(q/(e+a)) + \log V_t$, which has the same functional form as the rising arm of a piecewise exponential with a log-additive offset to the peak height. The log-affine transformation captures this relationship exactly.

\paragraph{Post-peak ($t > t_p$).} On the falling arm, $V_t = V_p\,e^{-b(t-t_p)}$, and the convolution gives
\begin{equation*}
    R_t = \frac{q}{e - b}\,V_t + \frac{q(a + b)}{(e + a)(e - b)}\,V_p\,e^{-e(t-t_p)}
\end{equation*}
(assuming $e \neq b$). The first term tracks the virus with a scaling factor; the second is a ``memory'' term that decays at the RNA clearance rate $e$ rather than the viral clearance rate $b$. When $e > b$ (RNA clears faster than virus declines), the memory term decays rapidly and the log-affine transformation is an excellent approximation. When $e < b$ (RNA persists longer than virus, the typical biological scenario), the memory term dominates at late times, producing the characteristic delayed and broadened RNA clearance observed empirically.

\paragraph{Smooth trajectory.} For the smooth log-sum-exp trajectory $g_s(t)$, the convolution integral does not admit an elementary closed form---it reduces to a hypergeometric-type integral because the virus trajectory is a nonlinear function of two competing exponentials. However, the closed-form expressions for the piecewise trajectory are asymptotically correct far from the peak (where one exponential arm dominates), and the discrepancy is localized to a narrow region around the peak where both the virus and RNA concentrations are at their maximum and diagnostically least ambiguous. The log-affine transformation therefore provides a computationally tractable approximation that captures the dominant kinetic features---delayed peak and prolonged clearance of RNA relative to infectious virus---without requiring numerical quadrature or ODE integration inside the MCMC sampler (see Figure~\ref{fig:antigen_schematic}c--d).

\section{LFD observation model: indicator interaction and alternative approaches}
\label{sec:supp_lfd_deriv}

\subsection{Antigen dynamics and the proliferation--clearance asymmetry}

Lateral flow devices detect viral nucleocapsid protein (antigen), whose concentration $A_t$ follows the same production--clearance ODE as viral RNA (Section~\ref{sec:supp_logaffine}), but with virus rather than infected cells as the source:
\begin{equation*}
    \frac{dA}{dt} = \alpha_1 V_t - \alpha_2 A_t.
\end{equation*}
The ratio $A_t / V_t$ is not constant over the infection. On the rising arm, $A_t / V_t \approx \alpha_1 / (\alpha_2 + a)$; on the falling arm, $A_t / V_t \approx \alpha_1 / (\alpha_2 - b) + \text{memory term}$. Since $\alpha_2 + a > \alpha_2 - b$ (assuming $a, b > 0$ and $\alpha_2 > b$), the antigen-to-virus ratio is lower during proliferation than during clearance. This means that at a given viral load, antigen concentration is lower pre-peak than post-peak---the proliferation--clearance asymmetry that motivates the model extension.

\subsection{Adopted approach: smooth post-peak sigmoid interaction}

The LFD observation model uses a smooth sigmoid interaction to allow the relationship between viral load and LFD sensitivity to differ between the rising and falling phases of infection:
\begin{equation*}
    \text{logit}\,P(L_t^* = 1) = \gamma_0 + \gamma_1 \log R_t + \gamma_2 \log V_t + \gamma_3 \,\sigma_\kappa(t - t_p) + \gamma_4 \,\sigma_\kappa(t - t_p) \log R_t
\end{equation*}
where $t_p$ is the individual's estimated RNA peak time and $\sigma_\kappa(x) = \text{logit}^{-1}(\kappa x)$ is a smooth sigmoid with steepness $\kappa$ (fixed at 5, giving a transition width of approximately one day). The effective model for LFD sensitivity is approximately:
\begin{equation*}
    \text{logit}\,P(L_t^* = 1) \approx \begin{cases}
        \gamma_0 + \gamma_1 \log R_t + \gamma_2 \log V_t & \text{pre-peak} \\
        (\gamma_0 + \gamma_3) + (\gamma_1 + \gamma_4) \log R_t + \gamma_2 \log V_t & \text{post-peak}
    \end{cases}
\end{equation*}
This is a natural extension of the regression framework already used throughout the model. It requires no additional functions or numerical computation: the peak time $t_p$ is already estimated for each individual as part of the RNA trajectory. Because $t_p$ is a latent parameter estimated jointly with all other model parameters via Hamiltonian Monte Carlo, the smooth sigmoid is preferred over a hard indicator $\mathbbm{1}(t \geq t_p)$: the latter creates a discontinuity that zeroes out the gradient with respect to $t_p$, impeding the leapfrog integrator. With $\kappa = 5$, the sigmoid transitions from 0.01 to 0.99 over approximately one day, which is both biologically reasonable (the switch from proliferation to clearance is not instantaneous) and numerically well-behaved. Near the peak, LFD sensitivity is close to its maximum, so the precise transition shape is inconsequential for the binary LFD outcome.

\subsection{Alternative approaches considered}

We considered several alternative approaches to capturing the asymmetry (summarized in Table~\ref{tab:lfd_approaches} and Figure~\ref{fig:antigen_schematic}d):

\begin{table}[h]
\centering
\caption{Comparison of approaches for modeling phase-asymmetric LFD sensitivity.}
\label{tab:lfd_approaches}
\begin{tabular}{lcccc}
\toprule
 & Convolution & Log-affine & Indicator & Derivative \\
\midrule
Additional parameters & 2 ($\alpha_1, \alpha_2$) & 8 (log-affine) & 2 ($\gamma_3, \gamma_4$) & 1 ($\gamma_3$) \\
Closed form & No & Approximate & Yes & Yes \\
Compatible with \texttt{reduce\_sum} & No\textsuperscript{a} & Yes & Yes & Yes \\
Identifiable from binary data & Partially & Poorly & Yes & Yes \\
Discontinuity at peak & No & No & Yes\textsuperscript{b} & No \\
\bottomrule
\multicolumn{5}{l}{\textsuperscript{a}Requires quadrature or discrete recursion at each observation.} \\
\multicolumn{5}{l}{\textsuperscript{b}Negligible: binary data makes this practically undetectable.}
\end{tabular}
\end{table}

\paragraph{Full convolution.} The most principled approach would model $A_t$ explicitly via the convolution $A_t = \alpha_1 \int_0^t e^{-\alpha_2(t-s)} V_s\,ds$. For the piecewise trajectory, this has a closed form (Section~\ref{sec:supp_logaffine}). For the smooth trajectory used in practice, no elementary closed form exists; the integral would require numerical quadrature at each observation within each MCMC iteration, which is incompatible with Stan's \texttt{reduce\_sum} parallelism.

\paragraph{Log-affine antigen layer.} One could add a second log-affine transformation mapping PFU parameters to antigen parameters, producing a piecewise exponential antigen trajectory with 8 additional parameters. This approximates the convolution well in the tails but misses the post-peak shoulder. More importantly, with only binary LFD outcomes (no quantitative antigen measurements), the 8 additional parameters would be poorly identified.

\paragraph{Derivative-based approach.} An alternative to the sigmoid interaction is to replace $\gamma_3 \,\sigma_\kappa(t - t_p) + \gamma_4 \,\sigma_\kappa(t - t_p) \log R_t$ with a single term $\gamma_3 \dot{g}_s(t)$, where $\dot{g}_s(t)$ is the time derivative of the smooth log-RNA trajectory. The derivative transitions smoothly from positive values during proliferation through zero at the peak to negative values during clearance, providing a continuous phase indicator with a single parameter. However, as a first-order correction that is additive in the logit, the derivative term cannot modulate the \emph{slope} of LFD sensitivity with respect to viral load---it only shifts the intercept. The sigmoid interaction is more flexible and fits naturally within the regression-based framework used throughout the model.

\subsection{Derivative of the smooth trajectory}

\begin{figure}[p]
    \centering
    \includegraphics[width=\linewidth]{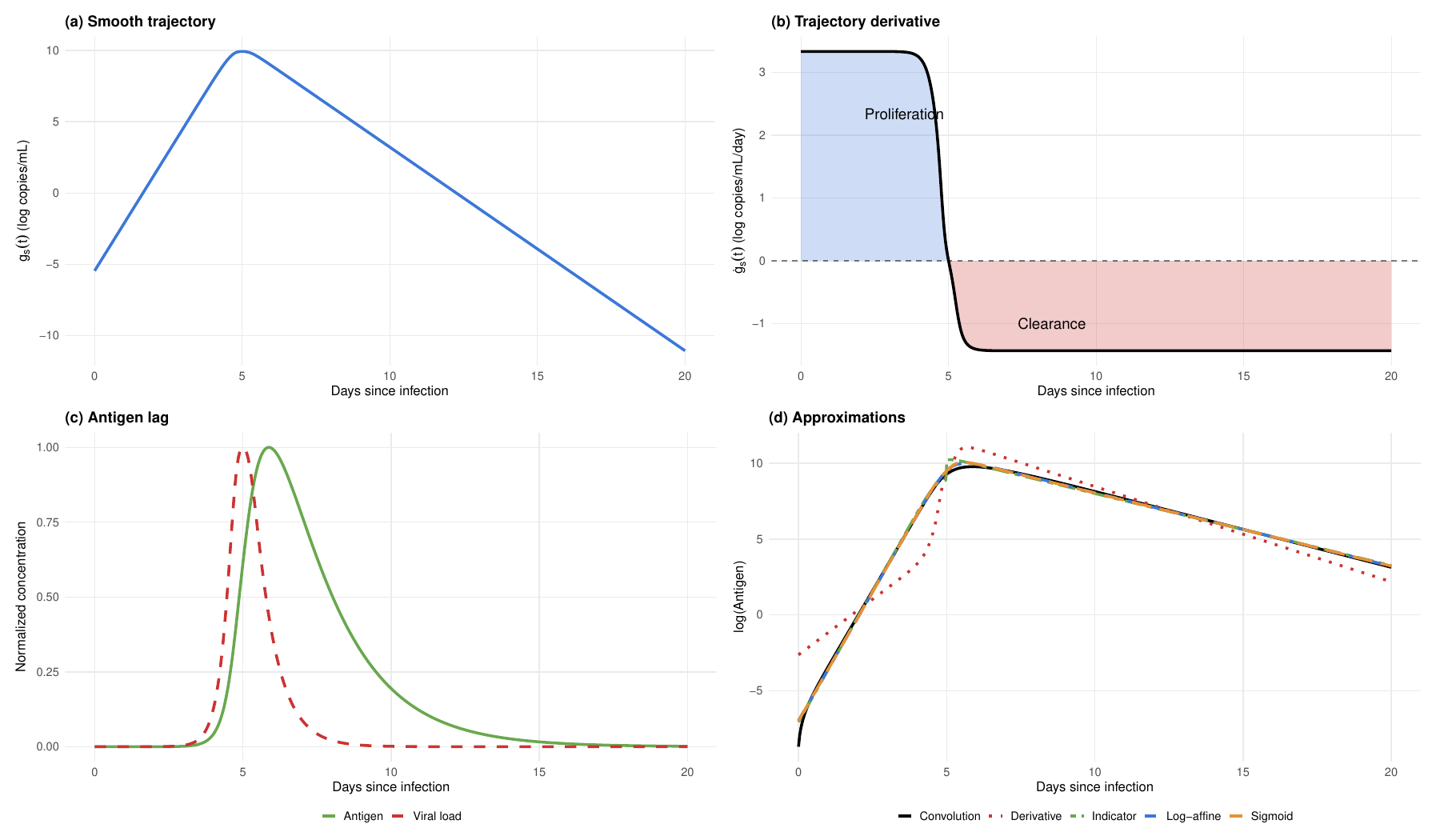}
    \caption[Trajectory representations for the LFD observation model]{Comparison of trajectory representations for the LFD observation model, computed using representative parameters ($\delta = 10$, $\omega_p = 3$, $\omega_r = 7$, $t_p = 5$). \textbf{(a)} Smooth log-viral-load trajectory $g_s(t)$. \textbf{(b)} Time derivative $\dot{g}_s(t)$: positive during viral proliferation (blue shading) and negative during clearance (red shading), providing a smooth indicator of infection phase. \textbf{(c)} Antigen concentration $A_t$ (solid) from numerical convolution of the production--clearance ODE, overlaid on viral load $V_t = e^{g_s(t)}$ (dashed), both normalized to unit peak. The antigen curve lags behind viral load during proliferation and persists above it during clearance, illustrating the mechanism underlying asymmetric LFD sensitivity. \textbf{(d)} Comparison of approximation approaches on the log scale: the log-affine piecewise exponential (blue dashed) closely tracks the full convolution (black solid); the smooth sigmoid interaction (orange long-dash) captures the phase-dependent slope change with a differentiable transition at the peak; the hard indicator interaction (green dash-dot) is similar but with a discontinuous step at $t_p$; the derivative term (red dotted) provides a smooth first-order correction. All four approximations are fit to the convolution by least squares for illustration.}
    \label{fig:antigen_schematic}
\end{figure}

For reference, the time derivative of the smooth piecewise exponential trajectory is
\begin{equation*}
    \dot{g}_s(t) = \left[1 - \sigma\!\left(\kappa(\text{raw} - \delta)\right)\right] \cdot \frac{ab\left(e^{-a\tau} - e^{b\tau}\right)}{b\,e^{-a\tau} + a\,e^{b\tau}}
\end{equation*}
where $\tau = t - t_p$, $a = \delta/\omega_p$, $b = \delta/\omega_r$, $\sigma(\cdot)$ is the logistic sigmoid, $\kappa = 10$ is the soft-cap sharpness, and $\text{raw} = \delta + \log\!\left((a+b)/(b\,e^{-a\tau} + a\,e^{b\tau})\right)$ is the uncapped trajectory value. On the arms, this simplifies to $\dot{g}_s(t) \approx a$ for $t \ll t_p$ and $\dot{g}_s(t) \approx -b$ for $t \gg t_p$, recovering the piecewise derivative (Figure~\ref{fig:antigen_schematic}b). The derivative functions remain implemented in the Stan model code for potential future use.

\end{appendix}

\end{document}